\documentclass{aa}

\usepackage{graphics,epsfig}
\usepackage{natbib}
\def\bibfiles{1688maas.bib}
\def\aareferences{\longrefs=0  \bibliographystyle{aa}
            \bibliography{1688aa,\bibfiles}}
\newcount\longrefs
\def\aap{\ifnum\longrefs=1 {Astron.\ Astrophys.}\else 
                           {A\hbox{\rm \&}A}\fi}
\def\aapr{\ifnum\longrefs=1 {Astron.\ Astrophys.\ Rev.}\else 
                            {A\hbox{\rm \&}AR}\fi}
\def\aaps{\ifnum\longrefs=1 {Astron.\ Astrophys.\ Suppl.}\else 
                            {A\hbox{\rm \&}AS}\fi}
\def\aj{\ifnum\longrefs=1 {Astron.\ J.}\else 
                          {AJ}\fi} 
\def\ao{\ifnum\longrefs=1 {Applied Optics}\else 
                           {Appl.\ Opt.}\fi} 
\def\aspcs{\ifnum\longrefs=1 {Astron.\ Soc.\ Pacific Conf. Series}\else 
                           {ASP Conf.\ Ser.}\fi} 
\def\apj{\ifnum\longrefs=1 {Astrophys.\ J.}\else 
                           {ApJ}\fi} 
\def\apjl{\ifnum\longrefs=1 {Astrophys.\ J. Lett.}\else 
                            {ApJ}\fi} 
\def\aplett{\ifnum\longrefs=1 {Astrophys.\ J. Lett.}\else 
                            {ApJ}\fi} 
\def\apjs{\ifnum\longrefs=1 {Astrophys.\ J. Suppl.}\else 
                            {ApJS}\fi}
\def\apss{\ifnum\longrefs=1 {Astrophys.\ and Space Science}\else 
                            {Ap\hbox{\rm \&}SS}\fi}
\def\araa{\ifnum\longrefs=1 {Ann.\ Rev.\ Astron.\ Astrophys.}\else 
                            {ARA\hbox{\rm \&}A}\fi}
\def\azh{\ifnum\longrefs=1 {Astronomicheskii Zhurnal}\else 
                            {Astron.\ Zhur.}\fi}
\def\baas{\ifnum\longrefs=1 {Bull.\ Am.\ Astron.\ Soc.}\else 
                            {BAAS}\fi}
\def\bain{\ifnum\longrefs=1 {Bull.\ Astronom.\ Institutes Netherlands}\else
                            {Bull.\ Astr.\ Inst.\ Neth.}\fi}
\def\gca{\ifnum\longrefs=1 {Geochim.\ Cosmochim.\ Acta}\else 
                           {Geochim.\ Cosmochim.\ Acta}\fi}
\def\grl{\ifnum\longrefs=1 {Geophys.\ Res.\ Lett.}\else 
                           {Geoph.\ Res.\ Lett.}\fi}
\def\iaucirc{\ifnum\longrefs=1 {IAU Circulars}\else 
                          {IAU Circ.}\fi}
\def\ip{\ifnum\longrefs=1 {in press}\else 
                          {in press}\fi}
\def\jgr{\ifnum\longrefs=1 {J.\ Geophys.\ Res.}\else 
                           {J.\ Geophys.\ Res.}\fi}  
\def\jrasc{\ifnum\longrefs=1 {J.\ Royal Astron.\ Soc.\ Canada}\else 
                           {JRAS Can.}\fi}  
\def\mnras{\ifnum\longrefs=1 {Mon.\ Not.\ Roy.\ Astron.\ Soc.}\else 
                             {MNRAS}\fi} 
\def\nat{\ifnum\longrefs=1 {Nature}\else 
                           {Nat}\fi}
\def\pasj{\ifnum\longrefs=1 {Pub.\ Astron.\ Soc.\ Japan}\else 
                            {PASJ}\fi} 
\def\pasp{\ifnum\longrefs=1 {Pub.\ Astron.\ Soc.\ Pacific}\else 
                            {PASP}\fi} 
\def\physscr{\ifnum\longrefs=1 {Physica Scripta}\else 
                            {Phys.\ Scrip.}\fi} 
\def\planss{\ifnum\longrefs=1 {Planetary \& Space Science}\else 
                            {Plan. \& Space Sci.}\fi} 
\def\procspie{\ifnum\longrefs=1 {Proc.\ SPIE}\else 
                            {Proc.\ SPIE}\fi} 
\def\qjras{\ifnum\longrefs=1 {Quarterly J.\ Royal Astron.\ Soc.}\else 
                            {QJRAS}\fi} 
\def\sa{\ifnum\longrefs=1 {Soviet Astron..}\else 
                               {Sov.\ Astron.}\fi}
\def\skytel{\ifnum\longrefs=1 {Sky \& Telescope}\else 
                            {Sky \& Tel.}\fi} 
\def\solphys{\ifnum\longrefs=1 {Solar Phys.}\else 
                               {Solar Phys.}\fi}
\def\ssr{\ifnum\longrefs=1 {Space Science Rev.}\else 
                               {Space\ Sci.\ Rev.}\fi}

\def\bibfiles{/data1/lte/thomas/latex/bibtex/bibliofile}   %% bibfiles

\def\aareferences{\longrefs=0  \bibliographystyle{/data1/lte/thomas/latex/bibtex/aa}
            \bibliography{/data1/lte/thomas/latex/bibtex/aajour,\bibfiles}}
\def\dutch{\def\refname{Referenties}\def\abstractname{Samenvatting}%
  \def\bibname{Bibliografie}\def\chaptername{Hoofdstuk}%
  \def\appendixname{Bijlage}\def\contentsname{Inhoudsopgave}%
  \def\listfigurename{Lijst van figuren}\def\listtablename{Lijst van tabellen}%
  \def\indexname{Index}\def\figurename{Figuur}\def\tablename{Tabel}%
  \def\partname{Deel}\def\enclname{Bijlage(n)}\def\ccname{Ter attentie van}%
  \def\headtoname{Aan}\def\headpagename{Pagina}%
  \def\today{\number\day\space\ifcase\month\or januari\or februari\or maart\or%
     april\or mei\or juni\or juli\or augustus\or september\or oktober\or%
     november\or december\fi \space\number\year}%
  \typeout{
              >>>>> use hlatex209 for Dutch hyphenation <<<<< 
         }}
\newcounter{onefig} \newcounter{fignumber}
\newcount\nocaptions \newcount\nofigures \newcount\figwidth
\newcount\viewgraphs
  \def\paper{}  \def\figlabel{} 
\long\def\nextfig#1{\setcounter{figure}{\value{fignumber}}
  \addtocounter{fignumber}{1}
  \ifnum \viewgraphs=1 \newpage \pagestyle{empty} \fi 
  \ifnum\value{onefig}=0 #1 \fi                 
  \ifnum\value{onefig}=\value{fignumber} #1 \fi}
\def\figwidths#1#2{\ifnum \nocaptions=1 #2mm \else #1mm \fi}  
\def\paper#1{}  %% redefine for separate-figure identification line
\long\def\plotfig#1#2{\ifnum \nofigures=1 \else #2 \fi}
\long\def\captiontext#1{\ifnum \nofigures=1 \raggedright \fi 
   \ifnum \nocaptions=1 \paper
     \ifnum \viewgraphs=0 
       \newline  \mbox{}\hrulefill\mbox{} \newline 
       \newline label:~\{\figlabel\} 
     \fi 
     \else \ifnum \nofigures=0 \fi 
   #1 \fi}

\newcount\panelwidth \newcount\panelheight 
\newcount\bxmin \newcount\bymin \newcount\bxmax \newcount\bymax
\newcount\tbxmin \newcount\tbymin
\newcount\tpanelwidth \newcount\tpanelheight \newcount\tpdif
\def\panelsize #1,#2;{\panelwidth=#1 \panelheight=#2}  
\def\setbb #1,#2;#3,#4;#5,#6;{% UNITS: bp (from ghostview)
  \tbxmin=#1 \tbymin=#2    %% full box (axis titles) lower left corner
  \bxmin=#3 \bymin=#4      %% bare box (ticks only) lower left corner
  \bxmax=#5 \bymax=#6}     %% upper right corner
\def\barepanel #1{%
  \ifnum\panelheight=0 
    \tpdif=\bymax \advance\tpdif by -\bymin
    \multiply \tpdif by \panelwidth
    \tpanelheight=\tpdif
    \tpdif=\bxmax \advance\tpdif by -\bxmin
    \divide \tpanelheight by \tpdif
  \else \tpanelheight=\panelheight \fi
  \epsfig{file=#1,%
     bbllx=\bxmin bp,bblly=\bymin bp,bburx=\bxmax bp,bbury=\bymax bp,clip=,%
     width=\panelwidth mm,height=\tpanelheight mm}}
\def\labelypanel #1{% TeX permits only integer arithmetic, so bp and mm
  \ifnum\panelheight=0 
    \tpdif=\bymax \advance\tpdif by -\bymin
    \multiply \tpdif by \panelwidth
    \tpanelheight=\tpdif
    \tpdif=\bxmax \advance\tpdif by -\bxmin
    \divide \tpanelheight by \tpdif
  \else \tpanelheight=\panelheight \fi
  \tpdif=\bxmax \advance\tpdif by -\tbxmin
  \tpanelwidth=\panelwidth \multiply \tpanelwidth by \tpdif
  \tpdif=\bxmax \advance\tpdif by -\bxmin
  \divide \tpanelwidth by \tpdif
  \epsfig{file=#1,%
    bbllx=\tbxmin bp,bblly=\bymin bp,bburx=\bxmax bp,bbury=\bymax bp,%
    clip=,width=\tpanelwidth mm,height=\tpanelheight mm}}
\def\labelxpanel #1{%
  \ifnum\panelheight=0 
    \tpdif=\bymax \advance\tpdif by -\bymin
    \multiply \tpdif by \panelwidth
    \tpanelheight=\tpdif
    \tpdif=\bxmax \advance\tpdif by -\bxmin
    \divide \tpanelheight by \tpdif
  \else \tpanelheight=\panelheight \fi
  \tpdif=\bymax \advance\tpdif by -\tbymin
  \multiply \tpanelheight by \tpdif
  \tpdif=\bymax \advance\tpdif by -\bymin
  \divide \tpanelheight by \tpdif
  \epsfig{file=#1,%
    bbllx=\bxmin bp,bblly=\tbymin bp,bburx=\bxmax bp,bbury=\bymax bp,%
    clip=,width=\panelwidth mm,height=\tpanelheight mm}}
\def\labelxypanel #1{%
  \ifnum\panelheight=0 
    \tpdif=\bymax \advance\tpdif by -\bymin
    \multiply \tpdif by \panelwidth
    \tpanelheight=\tpdif
    \tpdif=\bxmax \advance\tpdif by -\bxmin
    \divide \tpanelheight by \tpdif
  \else \tpanelheight=\panelheight \fi
  \tpdif=\bxmax \advance\tpdif by -\tbxmin
  \tpanelwidth=\panelwidth \multiply \tpanelwidth by \tpdif
  \tpdif=\bxmax \advance\tpdif by -\bxmin
  \divide \tpanelwidth by \tpdif 
  \tpdif=\bymax \advance\tpdif by -\tbymin 
  \multiply \tpanelheight by \tpdif
  \tpdif=\bymax \advance\tpdif by -\bymin
  \divide \tpanelheight by \tpdif
  \epsfig{file=#1,%
    bbllx=\tbxmin bp,bblly=\tbymin bp,bburx=\bxmax bp,bbury=\bymax bp,%
    clip=,width=\tpanelwidth mm,height=\tpanelheight mm}}

\def\CC{\par \vspace*{-2ex} \footnotesize \baselineskip=8pt \begin{verbatim}}

\long\def\startignore #1\stopignore{}   %% use \startignore....\stopignore

\def\setlistparams{         
  \topsep=0.7ex                 %% ADAPT: parskip=0: 0.7;  parskip=1: -1.2ex
  \itemsep=0.7ex                %% space between items
  \leftmargini=3ex}             %% dashes at beginning of line 
\setlistparams                  %% recall after type changes 

\newcounter{alistindex}       %% problems: a)  b) etc

\newcounter{romenumnr}

\newlength{\minipagewidth}

\newsavebox{\boxcontent}
\newcommand{\ovalhead}[1]{
  \unitlength=1cm
  \sbox{\boxcontent}{\mbox{~~{#1}~~}}
  \begin{center}
    \ifdim\wd\boxcontent>6ex 
    \ifdim\wd\boxcontent<8cm 
    \begin{picture}(8,3) \thicklines     
      \put(4.0,0.8){\oval(8,1.6)} 
      \put(0.0,0.7){\parbox{8cm}{
         \begin{center} \usebox{\boxcontent} \end{center}}}
    \end{picture}
    \else \ifdim\wd\boxcontent<12cm 
    \begin{picture}(12,3) \thicklines     
        \put(6.0,0.8){\oval(12,1.6)} 
        \put(0.0,0.7){\parbox{12cm}{
           \begin{center} \usebox{\boxcontent} \end{center}}}
    \end{picture}
    \else
    \begin{picture}(16,3) \thicklines     
        \put(8.0,0.8){\oval(16,1.6)} 
        \put(0.0,0.7){\parbox{16cm}{
           \begin{center} \usebox{\boxcontent} \end{center}}}
    \end{picture}
    \fi \fi \fi
  \end{center}} 

\newcounter{headnr}            
\newcounter{subheadnr}[headnr]
\newcounter{subsubheadnr}[subheadnr]
\def\head #1\par{
  \stepcounter{headnr}                          %% sets subheadnr = 0 too 
  \vspace{2ex} \noindent                        %% 2ex = space above, no *
  {\bf \theheadnr~~~~#1}\\[1ex] \noindent}      %% 1ex = space below
\def\subhead #1\par{  
  \stepcounter{subheadnr}
  \vspace{1.3ex} \noindent
  {\bf \theheadnr.\arabic{subheadnr}~~~#1}\\[0.3ex] \noindent}
\def\subsubhead #1\par{
  \stepcounter{subsubheadnr}
  \vspace{1.0ex} \noindent
  {\bf \theheadnr.\arabic{subheadnr}.\arabic{subsubheadnr}~~~#1}\\ \noindent}

\font\dropfont= cmr12 scaled \magstep5
\def\dropcap#1#2{{\noindent
    \setbox0\hbox{\dropfont #1}\setbox1\hbox{#2}\setbox2\hbox{(}%
    \count0=\ht0\advance\count0 by\dp0\count1\baselineskip
    \advance\count0 by-\ht1\advance\count0by\ht2
    \dimen1=.5ex\advance\count0by\dimen1\divide\count0 by\count1
    \advance\count0 by1\dimen0\wd0
    \advance\dimen0 by.25em\dimen1=\ht0\advance\dimen1 by-\ht1
    \global\hangindent\dimen0\global\hangafter-\count0
    \hskip-\dimen0\setbox0\hbox to\dimen0{\raise-\dimen1\box0\hss}%
    \dp0=0in\ht0=0in\box0}#2}

\def\level #1 #2#3#4{$#1 \: ^{#2} \mbox{#3} ^{#4}$}   

\def\degree{\hbox{$^\circ$}}

\def\kms{\hbox{km$\;$s$^{-1}$}}

\def\mathstacksym#1#2#3#4#5{\def#1{\mathrel{\hbox to 0pt{\lower 
    #5\hbox{#3}\hss} \raise #4\hbox{#2}}}}

\mathstacksym\lta{$<$}{$\sim$}{1.5pt}{3.5pt} % less than approximately
\mathstacksym\gta{$>$}{$\sim$}{1.5pt}{3.5pt} % greater than approximately
\mathstacksym\lrarrow{$\leftarrow$}{$\rightarrow$}{2pt}{1pt} % equilibrium
\mathstacksym\lessgreat{$>$}{$<$}{3pt}{3pt} %% less or greater

\bibpunct{(}{)}{;}{a}{}{,}

\newcommand{\halfa}{H$\alpha$}
\begin{document}

%\thesaurus{06(08.01.1; 08.03.1; 08.05.3; 08.16.4)}

\title{Depletion in post-AGB stars with a dusty disc. II.
\thanks{based on observations collected at the European Southern Observatory
in Chile (64.L-117 and 67.D-0054) and at the Swiss Euler telescope at La Silla, Chile.}}

\author{Thomas Maas\inst{1,2}
\and Hans Van Winckel\inst{2}
\and Tom Lloyd Evans\inst{3}}

\offprints{Thomas Maas}
\mail{thomas@astro.as.utexas.edu}
\institute{Department of Astronomy, University of Texas, Austin, TX 78712
 \and Instituut voor Sterrenkunde, K.U.Leuven, Celestijnenlaan 200B,
B-3001 Leuven, Belgium \and
School of Physics and Astronomy, University of St Andrews, North Haugh,
St Andrews, Fife, Scotland KY16 9SS}
\date{Received  / Accepted}

%*****************************************************************************
%                      ABSTRACT
%*****************************************************************************

\abstract{We present a chemical abundance analysis on the basis of high signal-to-
noise and high resolution spectra for 12 stars of our newly defined sample.
The selection criterion for the stars of this sample was their position in the `RV\,Tauri' box
in the $[12]-[25]$, $[25]-[60]$ two-colour diagram. An additional requisite
was the presence of a near-IR excess.
 We found that the photospheres of nine stars are affected by the depletion process.
 Since the most likely origin  of the near-IR excess is 
a disc, this strongly suggests that the phenomenon of depletion and the
presence of a disc are linked. Among the depleted stars we find different
depletion patterns. In seven stars elements with a condensation temperature below
1500 K are not or only slightly affected. Large underabundances are
measured only for most elements with a condensation temperature  above 1500 K. 
}

\titlerunning{Depletion in post-AGB stars with a dusty disc. II.}
\authorrunning{T. Maas et al.}
\maketitle

%******************************************************************
%                    INTRODUCTION
%******************************************************************

\section{Introduction}
RV\,Tauri stars form a loosely defined class of variable stars with a
characteristic light curve showing alternating deep and shallow minima.
They have 'formal' periods (the time between two successive deep minima)
between 30 and 150 days and occupy the high luminosity end of the
Pop II Cepheids. \citet{1972ApJ...178..715G} and \citet{1985MNRAS.217..493E}
 pointed out that many
RV\,Tauri stars show near-IR excesses. \citet{1986ApJ...309..732J} classified RV\,Tauri
stars as post-AGB stars since many objects were shown to be surrounded by
considerable amounts of circumstellar dust as detected by IRAS.
\citet{1985MNRAS.217..493E} and \citet{1989MNRAS.238..945R} showed that the RV\,Tauri stars
are located in a  well-defined and relatively thinly-populated part of
the IRAS colour-colour diagram. The defining rectangle is $$
  \left\{\begin{array}{ll} [12]-[25] \equiv & 1.56 + 2.5 \log[F(25)/F(12)] = 1.0 - 1.55  \\ \,
[25]-[6
0] \equiv & 1.88 + 2.5 \log[F(60)/F(25)] = 0.20 - 1.0
   \end{array} \right. $$
\\
This is the second in a series of papers on a new sample of post-AGB
stars. They were selected by Lloyd Evans (1999 and in preparation) 
  on the basis of their IRAS colours,
which place them in the `RV\,Tauri' box in the $[12]-[25]$, $[25]-[60]$
 two-colour diagram. A few known RV\,Tauri stars fall in an extended region to the red, which
was less intensively searched for new examples. IRAS\,17038-4815 and 
IRAS\,20056+1834 fall in this extension (Lloyd Evans, in preparation).
An additional selection criterion is an observed
excess in the L-band, indicative of a dusty disc \citep{1997Ap&SS.251..239L}.
Both criteria lead to a typical spectral energy distribution with a broad IR excess
(see Fig.~\ref{sed}). 
In previous years we have been accumulating different data sets,
each probing different physical properties of the stars. Initial photometric monitoring 
in the optical and near-IR at SAAO were supplemented with radial velocity monitoring 
at La Silla. Beside that, we obtained high quality spectra with FEROS
and CO emission spectra with SEST, both at La Silla. The purpose is
to  characterize these evolved objects and to compare them with
 GCVS RV\,Tauri stars. We want to investigate the possible relation between
the specific SED characteristics; the binary nature of the central object
and the presence of chemical peculiarities. The latter is a widespread
phenomenon among GCVS RV\,Tauri stars. Many objects display depletion
patterns and mimic in their photospheric pattern, the abundances of the gas
phase of the Interstellar Medium (ISM): species with a high dust condensation
temperature are underabundant relative to non-refractory elements which have
a low dust condensation temperature.

In our first paper on this sample \citep{irasnulacht} we highlighted one object, IRAS\,08544-4431. 
Its spectral energy distribution shows a broad IR
 excess starting already at $H$. The spectral type
of F3 places it well to the high-temperature side of the instability strip
(Lloyd Evans 1999, and in preparation), and the photometric peak-to-peak
amplitude of $\Delta V$ = 0.17 mag is much smaller than those of recognised
RV Tauri stars.
 Our radial velocity monitoring revealed it
to be a binary with an orbital period of 503 $\pm$ 2 days. Furthermore,
 IRAS\,08544-4431 is detected in both the CO (2-1) and (1-0) mm-wave emission
 lines. The triangular shape of the weak CO profile confirms  that
part of the circumstellar  material is not freely expanding but resides
  probably in a dusty circumbinary disc. Moreover, our  chemical abundance
 analysis of a high resolution spectrum of high S/N revealed that a depletion 
process has modified the photospheric abundances to a moderate extent
([Zn/Fe]=+0.4). All these findings confirm that the F-type star 
IRAS\,08544-4431 is a binary post-AGB star surrounded by a dusty disc.

\begin{figure}
\begin{center}
\caption{\label{sed} The spectral energy distribution (SED) of IRAS\,15469-5311. The SED is 
typical for the sample, showing a large and broad IR-excess.  A total reddening of E(B-V)=1.5 was
deduced minimizing the difference between the deredenned optical fluxes (plus signs) and the model
 atmosphere (full line).}
\resizebox{\hsize}{!}{\includegraphics{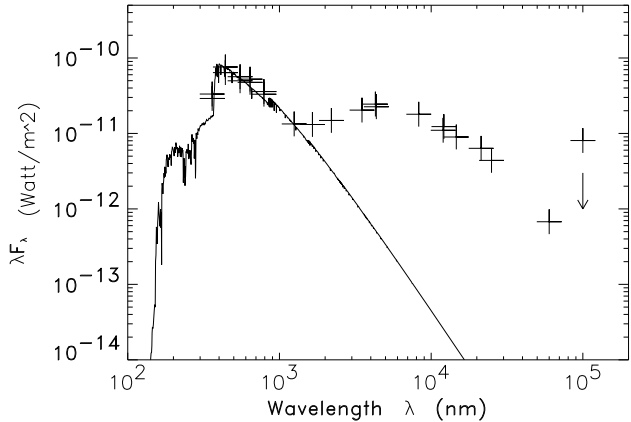}}
\end{center}
\end{figure}

In this paper we report on our chemical abundance analysis of the
high resolution and high signal-to-noise spectra of the stars
of our sample with spectral type F and G. They are listed in 
Table~\ref{tabprogram}. The positions of the optical
sources agree well with the positions of the 2MASS sources 
(Lloyd Evans, in preparation). 
IRAS\,16230-3410, IRAS\,17038-4815,
IRAS\,17233-4330 and IRAS\,17243-4348 were classified
by Lloyd Evans (1999 and in preparation) as genuine RV\,Tauri stars.
Low resolution spectra of IRAS\,17038-4815 and IRAS\,17233-4330
show strong CN and CH bands with respect to the rest of
the spectrum. Both stars belong to the spectroscopic RV\,Tauri
group RVB (for the definition of the different spectroscopic 
classes of RV\,Tauri stars, see \citet{1963ApJ...137..401P,2002A&A...386..504M}).
 The low resolution spectra of IRAS\,16230-3410
and IRAS\,17243-4348 are typical for members of the spectroscopic
group RVA (Lloyd Evans, 1999 and in preparation). Since IRAS\,09538-7622 
has a high peak-to-peak photometric amplitude
and a spectral type G, we also consider it to be a genuine RV\,Tauri (Maas et al., 
in preparation). It belongs to the spectroscopic group RVA 
(Lloyd Evans, 1999 and in preparation).

%\onecolumn

\begin{table*}
\caption{\label{tabprogram}Basic data for the programme stars.
 Besides the IRAS name,
the spectral type (determined on the basis of low resolution spectra (Lloyd 
Evans, in preparation)), the
Equatorial coordinates, ($\alpha_{2000}$,  $\delta_{2000}$), the 
galactic coordinates, (l, b), the visual magnitude, V (Maas et al. in preparation),
and the 2MASS K-magnitude are given. Notes :
 p for peculiar spectral features, (R) for the presence of enhanced
CH and CN bands, De for strong emission in the Na D line (Lloyd Evans, 1999 and in preparation) and
: for poor photometry.}
\label{initiallist}
\begin{center}
\begin{tabular}{l|lrrrrrr}
  
IRAS     & sp. type & \multicolumn{1}{c}{$\alpha_{2000}$} & \multicolumn{1}{c}{$\delta_{2000}$} & \multicolumn{1}{c}{l} & 
 \multicolumn{1}{c}{b}   & \multicolumn{1}{c}{V} & \multicolumn{1}{c}{K} \\
\hline
\rule[0mm]{0mm}{3mm}08544-4431 & F3    & 08 56 14.18 & -44 43 10.7 & 265.51 & +0.39 &   9.1 & 3.52:\\
09060-2807 & F5    & 09 08 10.14 &  -28 19 10.4 & 254.59 & +12.94 &  11.3 & 6.95\\
09144-4933 & G0    & 09 16 09.64 &  -49 46 12.1 & 271.51 & -0.50  &  11.1: & 6.37\\
09538-7622 & G0    & 09 53 57.67 &  -76 36 52.3 & 293.17 & -17.24 &  11.7 & 7.72 \\
15469-5311 & F3    & 15 50 43.80 &  -53 20 43.3 & 327.82 & +0.63  &   10.6 & 4.97 \\
16230-3410 & F8    & 16 26 20.39 &  -34 17 12.9 & 345.54 & +10.26 &    11.4 & 7.81\\
17038-4815 & G2p(R)& 17 07 36.66 &  -48 19 08.5 & 339.79 &  -4.67 &    10.6 & 6.52\\
17233-4330 & G0p(R)& 17 26 58.64 &  -43 33 13.6 & 345.64 & -4.69  &   12.1 & 8.37\\
17243-4348 & G2    & 17 27 53.63 &  -43 50 46.3 & 345.50 & -5.00  &   10.4 & 6.46\\
19125+0343 & F2    & 19 15 01.17 &  +03 48 42.7   &  39.02 & -3.49  &   10.2 & 5.65:\\
19157-0247 & F3    & 19 18 22.71 &  -02 42 10.8 &  33.59 & -7.22  &   10.7 & 7.02\\
20056+1834 & G0 De & 20 07 54.60 &  +18 42 54.8 &  58.44 & -7.46  &   12.5 & 7.47\\

\end{tabular}
\end{center}
\end{table*}

%\twocolumn

Almost all objects listed in Table~\ref{initiallist} are virtually unstudied
prior to our research efforts.
Only for IRAS\,17243-4348 and IRAS\,20056+1834 is there a substantial
literature. IRAS\,17243-4348, also known as LR Sco, was 
misclassified as an R Coronae Borealis star by \citet{1978IBVS.1453....1S},
 and was classified to be a yellow supergiant by
 \citet{1990Obs...110..120G}. In this paper, we classify the object as a low
mass evolved object. 
\citet{1988MNRAS.233..697M}
observed IRAS 20056+1834 (QY Sge) and discovered the near infrared excess
and the Na D emission superimposed on a spectrum of G0 I; they adopted a
model in which the star is seen by reflection from circumstellar matter
while direct light is obscured by dust, as suggested by \citet{1970ApJ...162..557H}.
 \citet{2002MNRAS.334..129K} proposed a disk torus geometry for the circumstellar dust,
with a bipolar flow which is the site for formation of emission lines from
low levels of abundant neutral metal atoms.
 The abundance analysis of high resolution spectra by \citet{2002MNRAS.334..129K}
 revealed that the chemical composition is 
approximately solar except for highly condensable elements which are depleted
by factors of 5-10, on the basis of which it is suggested that QY\,Sge is a
 genuine RV\,Tauri star. In addition, its temperature and luminosity fall in
 the same range as do those of the RV\,Tauri stars.

The outline of this paper is as follows.
First, we give an overview of the observations, after which we make a short
interlude on the \halfa-profile in the spectra. In section~\ref{analysis}
we specify how we determined the photospheric 
model parameters and discuss the errors on the abundances due
to the uncertainties on the model parameters.
In section~\ref{abundances} we present the abundances. We look for traces
of the depletion process and/or nucleosynthesis processes in the deep 
layers of the stars, followed by dredge-up.
In section~\ref{discussion} we discuss the observed abundance patterns
 and compare them with those observed in other depleted stars.
We end with the conclusions.
A detailed modeling of the spectral energy distribution
of these stars will be discussed in a forthcoming paper 
(De Ruyter et al., in preparation).

%*******************************************************************************
%                             DATA
%*******************************************************************************

\section{Observations}
The spectra were almost all obtained by H. Van Winckel with the Fiber-Fed 
Extended Range Optical Spectrograph (FEROS) mounted on the 1.52\,m 
telescope at ESO, La Silla (Chile) in the ESO periods $\#$64
 (22-24 March 2000) and $\#$67 (26-28 March 2001). A spectrum of
IRAS\,08544-4431 was already previously obtained with FEROS by G. Meeus
 during the ESO Period $\#$62. This spectrograph covers the complete
optical wavelength domain (3700-8600 \AA) and has a resolution of 48000.
Two fibers feed simultaneously the spectrograph. While the first fiber
 transmits the light of the object, the second is fed by the background sky.
The spectrograph is equipped with a CCD camera incorporating a monolithic
 thinned and backside-illuminated CCD chip by EEV with 2048 $\times$
4096 pixels of 15 $\times$ 15 $\mu$m size, which is cooled to a temperature
between $-$150 and $-$80 $\degree$ C.
For this analysis we used the pipeline reduced spectra. The pipeline
reduction consists of background subtraction, flat fielding, order extraction and
wavelength calibration. The normalisation of the spectra was done 
by fitting a smoothed spline through interactively chosen continuum points.
For IRAS\,09538-7622 we do not have a FEROS spectrum at our disposal.
Since we are monitoring this
star with the spectrograph CORALIE mounted on the Swiss Euler telescope at La
Silla, we analysed the CORALIE spectrum with the highest S/N.  An overview of 
the obtained spectra is presented in Table~\ref{spectraldata}.
For IRAS\,20056+1834 we do not have a good FEROS spectrum either. We will
use the results of \citet{2002MNRAS.334..129K} throughout the paper.

\begin{table}
\begin{center}
\caption{\label{spectraldata}Overview of the data. The date, the UT and the S/N 
(measured around 5800 \AA) of the analysed spectra are shown. All spectra were
taken with the FEROS spectrograph, except for IRAS\,09538-7622, for 
which we used a CORALIE spectrum.}
\begin{tabular}{lllr}
IRAS       & Date & UT & S/N \\
\hline
08544-4431 & 25/01/1999 & 6:20 & 150 \\ 
09060-2807 & 23/03/2000 & 0:11 & 140 \\ 
09144-4933 & 24/03/2000 & 0:31 & 30 \\ 
09538-7622 & 05/12/1998 & 7:58 & 40 \\ 
15469-5311 & 23/03/2000 & 6:26 & 180 \\ 
           & 27/06/2001 & 2:22 & 140 \\  
16230-3410 & 24/03/2000 & 6:00 & 160 \\ 
17038-4815 & 24/03/2000 & 8:07 & 160 \\ 
17233-4330 & 23/03/2000 & 8:13 & 110 \\
17243-4348 & 24/03/2000 & 9:12 & 140 \\
           & 28/06/2001 & 4:19 & 120 \\
19125+0343 & 28/06/2001 & 7:18 & 190 \\
19157-0247 & 28/06/2001 & 5:52 & 110 \\
\end{tabular}
\end{center}
\end{table}

\section{\halfa-profile}

The \halfa-line at 6562.8 \AA\, of the programme stars shows
 a wide variety of profiles (see Fig.~\ref{overviewhalfa}).
 All profiles, except that of IRAS\,17233-4330 show an emission component 
(the double absorption profile of IRAS\,17243-4348 is very likely due to a central
 emission superimposed on a wider absorption profile).
 In the stars IRAS\,08544-4431, 
IRAS\,15469-5311, (IRAS\,16230-3410), IRAS\,17038-4815, IRAS\,19125+0343
 and IRAS\,19157-0247,  a P-Cygni profile is seen, giving evidence for
 ongoing mass loss. The blue shifted  absorption in these profiles points 
to a very high velocity outflow. The mass-loss mechanism in these stars is,
 however, not well understood : the stars have spectral type F and G
and are expected to be too cool to develop significant mass loss by
 line-driven winds. 
Nevertheless, this mass loss in the early 
post-AGB phase could play an important role in the shaping of post-AGB nebulae
 and planetary nebulae. We refer to a detailed analysis of the profile 
variability of IRAS\,08544-4431 in \citet{irasnulacht}. A more quantitative
analysis of the \halfa\, profile and stellar wind is beyond the scope of this paper
but will certainly be rewarding.

\begin{figure}
\begin{center}
\caption{\label{overviewhalfa}The H$\alpha$-profile for all programme stars.}
\resizebox{\hsize}{!}{\includegraphics{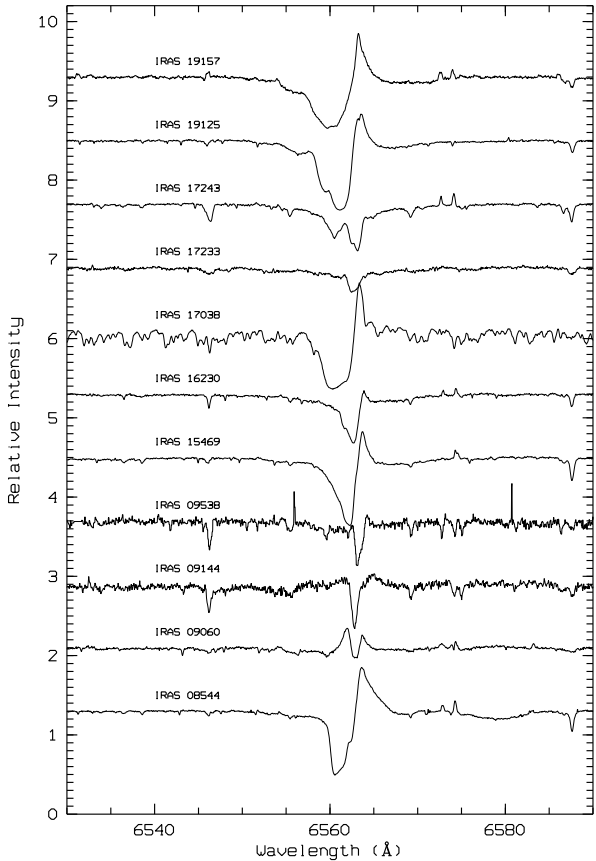}}
\end{center}
\end{figure}

\begin{figure}
\begin{center}
\caption{\label{dib}The spectral domain between 5740-5800 \AA\, with the
Diffuse Interstellar Bands around 5780 \AA\, and 5797 \AA. Note that we
aligned the spectra on the photospheric lines.}
\resizebox{\hsize}{!}{\includegraphics{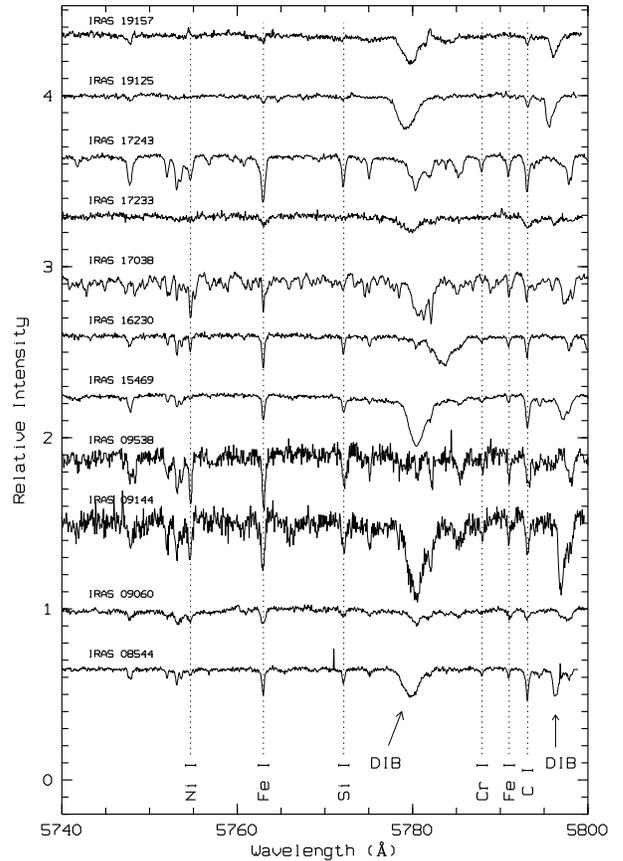}}
\end{center}
\end{figure}

\section{Analysis}
\label{analysis}

%\subsection{Atomic data}
%\label{atomic}

%Atomic data needed for an LTE abundance analysis consist of the
%wavelength, the excitation energy of the lowest level of the
%transition and the log(gf) values or the oscillator strength.
%Especially for the latter, large differences from author to
%author can be found.
%To derive precise abundances, accurate oscillator strengths (log gf-values) 
%are of utmost importance.
%Over many years of research, a considerable list of lines with good 
%oscillator strengths is accumulated at t.
% The values for Fe are taken from \citet{1996ApJS..103..183L},
%together with data from \citet{1980A&A....81..340B} for FeII.
%The data for C, N and O are from 
%\citet{1991A&AS...88..505H,hibbert93} and \citet{biemont93}, 
%and those for s-process elements from the Vienna Atomic Line Database
%(VALD2); the other values originate from
% \citet{fuhr88,1997ApJ...481..452G,martin88,reader80,thevenin89,thevenin90,venn95,wiese66}.

\subsection{Determination of the model parameters and abundances}

For the abundance analysis we used the models of \citet{1993KurCD..13.....K}.
The abundances are calculated with the LTE chemical abundance programme MOOG of C. Sneden 
(version April 2002).

To derive precise abundances, accurate oscillator strengths (log gf-values) 
are of utmost importance. Over many years of research, a considerable list of lines with good 
oscillator strengths has been accumulated at the Instituut voor Sterrenkunde, KULeuven.
 For more details see \citet{2000A&A...354..135V}.

The model parameters : the effective temperature, T$_{\rm eff}$, 
the (logarithm of)  surface gravity, $ \log g$, and the microturbulent
 velocity, $\xi_t$, are 
determined on the basis of the spectral lines of Fe. 
%Iron is the preferential 
%element in our spectral type range since it has many absorption lines with a large range in excitation
% potentials and both lines of neutral and singly ionised iron are present
%in the spectrum of all programme stars.
 The effective temperature is determined by demanding that the iron
 abundance, derived from the Fe~I-lines, is independent of the excitation potential.
This method becomes less accurate due to non-LTE effects for the hotter stars
in our sample (T$_{\rm eff} \geq $ 7500 K) : IRAS\,15469-5311, IRAS\,19125+0343
 and IRAS\,19157-0247. For these stars we used the Fe~II lines to determine
T$_{\rm eff}$. For IRAS\,19157-0247 this was even the only possibility as the 
low excitation Fe~I lines are all in emission making
 the excitation range of Fe~I absorption lines too small to derive a
 temperature.
 As an upper limit indicator for the temperature at a given gravity, the presence
of He-lines in the spectrum can be used. These lines appear in the spectrum
above a temperature of 7750 K (depending on the gravity). Only in IRAS\,19125+0343
the He-line at 5875.6 \AA\, is marginally detected.

 The surface gravity is fixed by postulating an equal abundance derived from
 Fe~I and Fe~II lines  and the microturbulent velocity by 
demanding the independence of the abundance, derived from the Fe~I lines, on the reduced equivalent width (EW/$\lambda$).

The derived parameters are shown in Table~\ref{model}.
The derived temperature and gravities confirm the results of 
\citet{1999IAUS..191..453E} that the programme stars have 
a high luminosity and that their temperatures put them in the
instability strip or just at the blue side of it. 
  Moreover, the low surface gravities of the programme stars are in agreement
with the long pulsation time scales of most of them (Maas et al., in preparation).

Typical errors on the derived model parameters are :
 $\Delta$T$_{\rm eff}$ =250 K,
 $\Delta \log g$= 0.5 , $\Delta \xi_t=$ 1.0 \kms. We study the
influence of these parameter uncertainties on the abundances in the
next section. 

For each ion we looked for unblended lines with an EW $<$ 150 m\AA. 
% Lines with a higher equivalent width are formed higher in the atmosphere, 
%where the assumed LTE is less probable, moreover these lines are saturated.
 We used the VALD2 database \citep{1999A&AS..138..119K} to check for blends. When the abundance of a 
line exceeded the average abundance of
the other lines of the same atom by 0.2 dex, we calculated the expected
central depths of all lines in the neighbourhood, adopting the abundances
determined by other unblended lines and the model parameters of the star. 
Next, we determined the EWs of the lines  of which the central
wavelength differed less than 0.3 \AA\, with that of the considered line,
adopting the abundances determined by other lines. In many cases, unresolved
blends were found in this way.

Because of the presence of strong interstellar bands in the spectra of most 
programme stars (see Fig.~\ref{dib}), we also examined the possible
 contamination of the EWs of the absorption lines, which were used 
in the analysis, by these bands.

\begin{figure}
\begin{center}
\caption{\label{overview}The spectral domain between 6120-6180 \AA.}
\resizebox{\hsize}{!}{\includegraphics{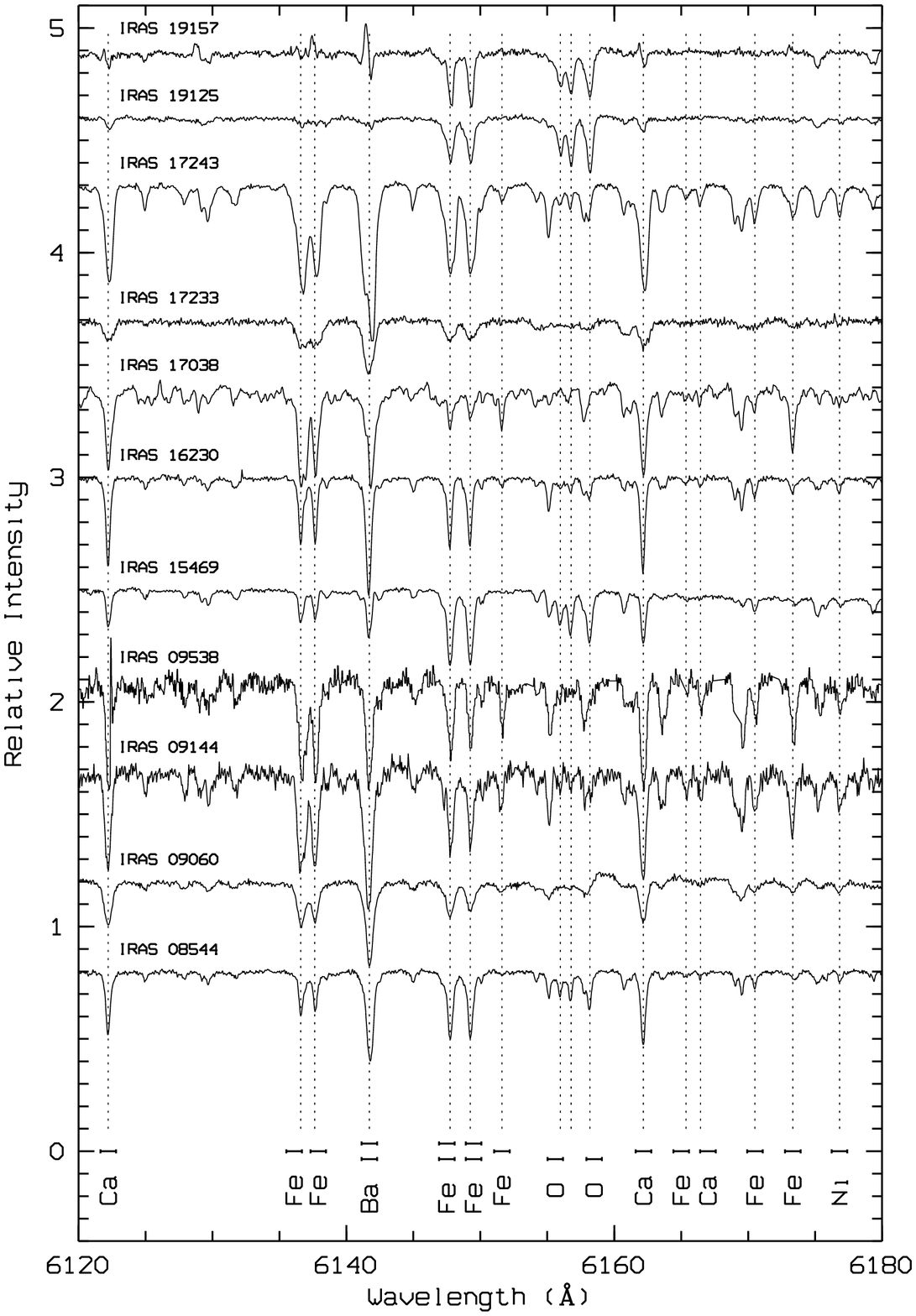}}
\end{center}
\end{figure}

\begin{figure}
\begin{center}
\caption{\label{sulfur}The spectral domain between 6710-6770 \AA\, with the
S-triplet at 6743, 6749 and 6757 \AA.}
\resizebox{\hsize}{!}{\includegraphics{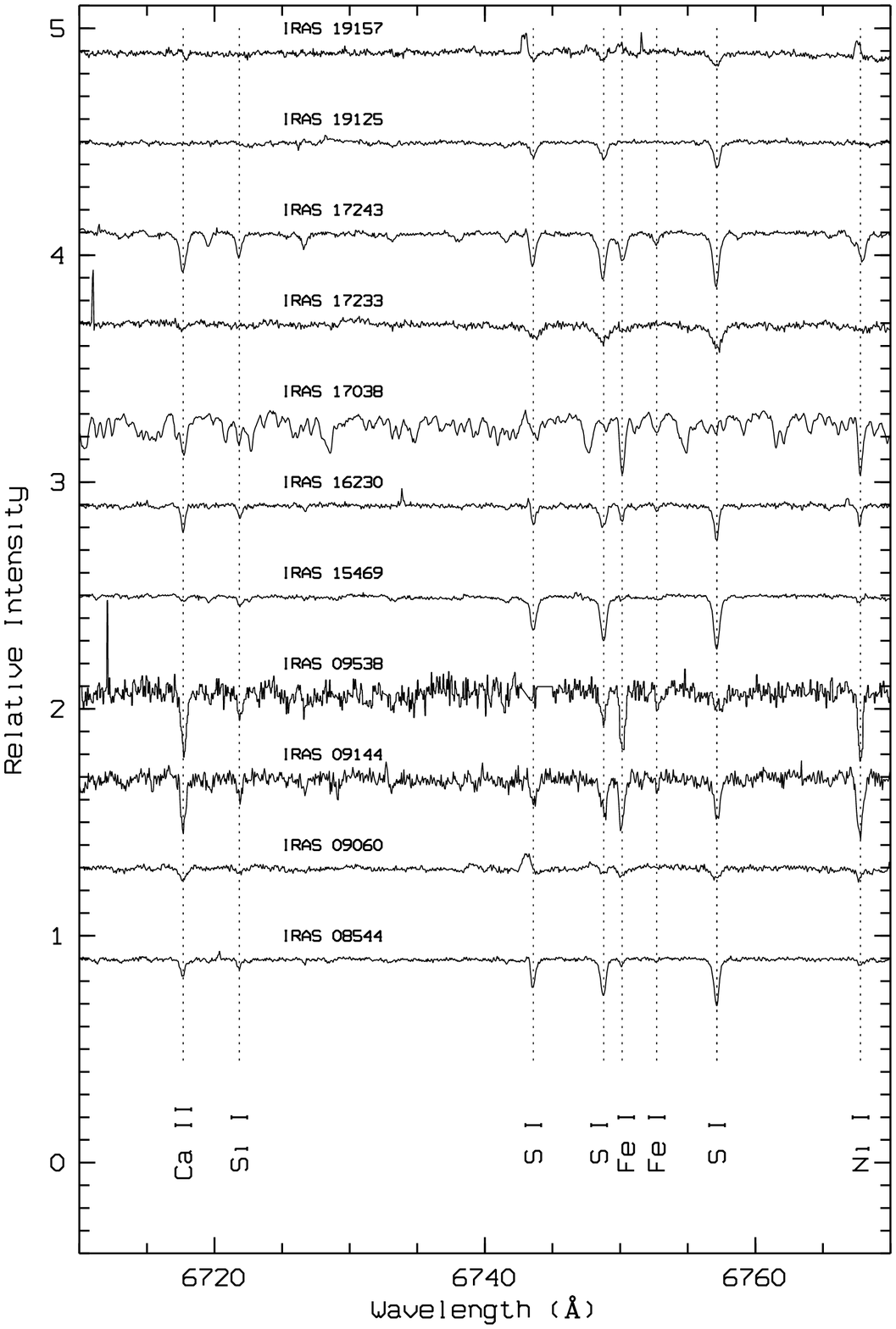}}
\end{center}
\end{figure}

We give a small general overview of the typical problems and
 characteristics of the abundance determination of a given chemical
element.

The carbon(C)-abundance is determined for all stars. This abundance is
 mainly based on the high excitation multiplet at 7100 \AA\,
 (except for IRAS\,09060-2807,
IRAS\,09538-7622 and IRAS\,17038-4815, because of the low C-abundance,
limited available wavelength domain and low temperature, respectively).

Good nitrogen(N)-lines are sparse in the spectra. For only five objects a
 N-abundance was determined, based on an average of 2 lines.
Also for oxygen (O) not many lines are present. The oxygen-triplet with
 excitation temperature of 10.74 eV around 6156 \AA\, shows up only in the
 hotter stars
 of the sample (see Fig.~\ref{overview}). The forbidden oxygen lines 
at 6300 \AA\, and 6364 \AA\, are in emission in all programme stars.
The NLTE sensitive red triplet was never used.

Almost always two low excitation neutral lines are used for the 
sodium(Na)-abundance.

The sulfur(S)-abundance is based on some or all lines of the two triplets
around 4695 \AA\, and 6750 \AA\,(see Fig.~\ref{sulfur}). The S-lines
 in the wavelength range 8660-8700 \AA\,  turned out to be inappropriate
because of the very poor quality of our spectra in this wavelength domain.

One or two lines are used for the zinc (Zn) abundance. In total
4 different Zn-lines are used, of which the lines at 4722 \AA\,
 and 4810 \AA\, are most often available.

Elements with neutral and singly ionised lines offer an extra check for the
reliability of the model parameters. For magnesium (Mg), silicon (Si), calcium 
(Ca), titanium (Ti), manganese (Mn) and nickel (Ni) sometimes neutral and
singly ionised lines are available, but mostly there is only one line 
 present in one of the two ionisation stages. However, for chromium (Cr) a
 considerable number of lines in both ionisation stages is often present.

For all stars a scandium(Sc)-abundance is determined.
 
For copper (Cu), aluminium (Al) and the r-process element (Eu), an 
abundance is sometimes derived.

There are few s-process lines detected. These atoms are expected
to be mainly singly ionised
in the T$_{\rm eff}$ range of our programme stars.
 Often only an upper limit for the abundance could be
determined. This upper limit is derived from the strongest predicted line, 
which is not present in the spectrum, ascribing it an EW of 5 m\AA.

\subsection{Abundance uncertainties}
\label{uncertain}

The line-to-line scatter, which is the standard deviation of the
mean, is given for the abundance derived from
each ion and reflects the uncertainties which result from the errors of the
 determined equivalent widths, the log(gf)-values and the continuum 
placement. The lines of
most ions have excitation temperatures in a small range. Therefore the
 influence of the uncertainties on the model parameters,
is almost nil on the scatter of the ion abundance.

To get an idea of the influence of these uncertainties on the derived
absolute and relative abundances, we computed the abundances for
 IRAS\,16230-3410, using models which differ in effective temperature
 by 250 K, in log gravity by 0.5  and in microturbulent velocity by 
 1.0 \kms, respectively, from the preferred model. These are then
compared with the abundances of the preferred model.
We have chosen IRAS\,16230-3410 because of its temperature,
 lying somewhat in the middle
of the temperature range of the programme stars, the high amount of elements,
for which an abundance is derived, and the many lines used. For the
relative abundances we compared the abundance of the singly ionised
atoms and those of S~I, C~I, N~I and O~I (neutral atoms with high excitation
lines) with that of Fe~II. All other
abundances (i.e. those of neutral atoms with low excitation lines) are
compared with that of Fe~I.

The variations of the absolute and the relative abundances are listed 
in Table~\ref{uncertaineen}.
In general, the uncertainties on the relative abundances are smaller than
those on the absolute abundances. 
The variation of the abundances, due to
 the uncertainty on the temperature, is on average larger for the absolute abundances than for the relative abundances. 

The Si~II and Ba~II abundances are both determined from one strong
 line (EW=137 and 148 m\AA, respectively) and are very sensitive to the
assumed microturbulent velocities ($\sigma_{\xi_t}$=0.26 and 0.33,
respectively for the absolute abundances with 1.0 \kms\, change in 
microturbulent velocity).
These two abundances, together with that of N~I are the most
influenced by the uncertainties on the model parameters. For the
other relative abundances the total uncertainty $\sigma_{rel,tot}$
is less than 0.2 dex.

\begin{table}
\begin{center}
\caption{\label{model} The model parameters for the programme objects.
For IRAS\,15469-5311 we used the listed model parameters for both spectra.
The model parameters for IRAS\,20056+1834 are taken from
 \citet{2002MNRAS.334..129K}.}
\vspace{0.5ex}
\begin{tabular}{c|ccrc}
\rule[-2mm]{0mm}{2mm} IRAS & T$_{\rm eff}$ (K) & $ \log g $ (cgs) & [Fe/H] & $\xi_t$ (\kms)\\
\hline 
\rule[0mm]{0mm}{3mm}08544-4431 & 7250 & 1.5 & $-$0.5 & 4.5 \\ 
09060-2807 & 6500 & 1.5 & $-$0.5 & 5.0 \\ 
09144-4933 & 5750 & 0.5 & $-$0.5 & 4.0 \\ 
09538-7622 & 5500 & 1.0 & $-$0.5 & 5.5 \\ 
15469-5311 & 7500 & 1.5 &    0.0 & 5.0 \\ 
16230-3410 & 6250 & 1.0 & $-$0.5 & 4.0 \\ 
17038-4815 & 4750 & 0.5 & $-$1.5 & 5.0 \\ 
17233-4330 & 6250 & 1.5 & $-$1.0 & 5.0 \\
17243-4348 & 6250 & 0.5 &    0.0 & 5.0 \\
           & 5250 & 0.0 &    0.0 & 5.0 \\
19125+0343 & 7750 & 1.0 & $-$0.5 & 4.0 \\
19157-0247 & 7750 & 1.0 &    0.0 & 3.0 \\
20056+1834 & 5850 & 0.7 & $-$0.4 & 4.5 \\ 
\end{tabular}
\end{center}
\end{table}

\begin{table*}
\begin{center}
\caption{\label{uncertaineen}The influence of the uncertainties in the model
 parameters on the absolute and the relative abundances. For each ion, the absolute
 abundance is shown in the second column for the model of IRAS\,16230-3410:
 T$_{\rm eff}$ = 6250 K, $\log g$ =1.0, $\xi_t=4.0 $\kms.
The following three colums show the variation of the absolute abundance due to a decrease of 250 K in
 effective temperature ($\sigma_{T_{\rm eff}}$), due to a decrease of 0.5 in
 $\log g$ ($\sigma_{g}$) and due to a decrease of 1.0 \kms in
microturbulent velocity ($\sigma_{\xi_t}$), respectively. The sixth column 
shows the total variation of the absolute abundance $\sigma_{\rm abs,tot}$ :
$\sigma_{\rm tot}={{(\sigma_{T_{\rm eff}}}^2 + {\sigma_{g}}^2 + {\sigma_{\xi_t}}^2) }^{1/2}$.
The next columns show the corresponding effects on the relative abundances.
The relative abundance, the variations of the relative abundance 
due to the variation of T$_{\rm eff}$, $\log g$ and
$\xi_t$ and the total variation of the relative abundance ($\sigma_{\rm rel,tot}$)
are shown, respectively.}
\vspace{1.ex}
\begin{tabular}{lrrrrrrrrrr}
 ion  & $\log \epsilon$  & $\Delta T_{\rm eff}$ & $\Delta \log g $ & $\Delta \xi_t$ & $\sigma_{\rm abs,tot}$ 
      & [el/Fe]   &  $\Delta T_{\rm eff}$ & $\Delta \log g $ & $\Delta \xi_t$ & $\sigma_{\rm rel,tot}$  \\
      &      &  $-$250 &  $-$0.5 & $-1.0$  & 
      &      &  $-$250 &  $-$0.5 &  $-1.0$  &      \\
\hline 
\rule[0mm]{0mm}{3mm}C~I   & 8.14 & $+$0.06 & $-$0.08 &  $+$0.05 & $+$0.11 &    $+$0.25  & $+$0.10 &  $+$0.06 &  $-$0.07 & $+$0.13\\ 
N~I   & 7.54 & $+$0.23 & $-$0.13 &  $+$0.07 & $+$0.27 &    $+$0.23  & $+$0.27 &  $+$0.01 &  $-$0.05 & $+$0.27\\ 
O~I   & 8.53 & $+$0.15 & $-$0.12 &  $+$0.03 & $+$0.19 &    $+$0.35  & $+$0.17 &  $+$0.02 &  $-$0.09 & $+$0.19 \\
Na~I  & 6.00 & $-$0.13 & $+$0.05 &  $+$0.01 & $+$0.14 &    $+$0.35  & $+$0.05 &  $+$0.00 &  $-$0.10 & $+$0.11 \\
Si~I  & 7.33 & $-$0.11 & $+$0.06 &  $+$0.04 & $+$0.13 &    $+$0.47  & $+$0.07 &  $+$0.01 &  $-$0.07 & $+$0.10 \\
Si~II & 7.15 & $+$0.12 & $-$0.13 &  $+$0.26 & $+$0.31 &    $+$0.29  & $+$0.16 &  $+$0.01 &  $+$0.14 & $+$0.21 \\
S~I   & 6.97 & $-$0.03 & $-$0.03 &  $+$0.04 & $+$0.06 &    $+$0.32  & $+$0.01 &  $+$0.11 &  $-$0.08 & $+$0.14 \\
Ca~I  & 5.62 & $-$0.16 & $+$0.06 &  $+$0.10 & $+$0.20 &    $-$0.06  & $+$0.02 &  $+$0.11 &  $-$0.01 & $+$0.11 \\
Sc~II & 0.89 & $-$0.11 & $-$0.14 &  $+$0.02 & $+$0.18 &    $-$0.60  & $-$0.07 &  $+$0.00 &  $-$0.10 & $+$0.12 \\
Ti~II & 3.57 & $-$0.09 & $-$0.14 &  $+$0.09 & $+$0.19 &    $-$0.77  & $-$0.05 &  $+$0.00 &  $-$0.03 & $+$0.06 \\
Cr~I  & 4.88 & $-$0.26 & $-$0.01 &  $+$0.00 & $+$0.26 &    $-$0.11  & $-$0.08 &  $-$0.06 &  $-$0.11 & $+$0.15 \\
Cr~II & 5.00 & $-$0.07 & $-$0.19 &  $+$0.07 & $+$0.21 &    $+$0.01  & $+$0.01 &  $-$0.05 &  $-$0.05 & $+$0.07 \\
Mn~I  & 4.66 & $-$0.18 & $+$0.05 &  $+$0.07 & $+$0.20 &    $-$0.05  & $+$0.00 &  $+$0.00 &  $-$0.04 & $+$0.04 \\
Fe~I  & 6.83 & $-$0.18 & $+$0.05 &  $+$0.11 & $+$0.22 &             &      &       &        & \\                
Fe~II & 6.83 & $-$0.04 & $-$0.14 &  $+$0.12 & $+$0.19 &             &      &       &        & \\                
Ni~I  & 5.62 & $-$0.16 & $+$0.05 &  $+$0.05 & $+$0.17 &    $+$0.05  & $+$0.02 &  $+$0.00 &  $-$0.06 & $+$0.06 \\
Zn~I  & 4.18 & $-$0.19 & $+$0.04 &  $+$0.17 & $+$0.26 &    $+$0.26  & $+$0.01 &  $-$0.01 &  $+$0.06 & $+$0.06 \\
Zr~II & 1.83 & $-$0.09 & $-$0.13 &  $+$0.02 & $+$0.16 &    $-$0.09  & $-$0.05 &  $+$0.01 &  $-$0.10 & $+$0.11 \\
Ba~II & 1.45 & $-$0.23 & $-$0.04 &  $+$0.33 & $+$0.40 &    $+$0.00  & $-$0.19 &  $+$0.10 &  $+$0.21 & $+$0.30 \\
Sm~II & 0.23 & $-$0.17 & $-$0.10 &  $+$0.01 & $+$0.20 &    $-$0.10  & $-$0.13 &  $-$0.04 &  $-$0.11 & $+$0.17 \\
\end{tabular}  
\end{center}
\end{table*}

\section{Abundances}
\label{abundances}

The derived abundances are shown in
 Tables~\ref{chembegin}-~\ref{chemeinde}.
 For all ions, the solar abundance, the condensation
 temperature, the number of lines, the mean equivalent width,
 the absolute abundance, the $\sigma$ of the line-to-line scatter
and  the abundance to iron are given.
%Traditionally, abundances are always compared with the Fe-abundance. 
%Fe has many lines and its abundance can be derived for a large range of
% spectral types. Furthermore, iron is thought to scale with the age of the star
%(see Fig.~\ref{feage}).
%We decided to compare our abundances also with that of Fe.
% However, one should
% always keep in mind that (part of) the iron abundance can be set by the
% depletion process and not by the galactic chemical enrichment.

%\onecolumn
\begin{table*}
%\begin{sidewaystable*}
\caption{\label{chembegin} Abundances for programme stars. 
 For all ions, the solar abundance, the condensation temperature, the number of lines, the mean 
equivalent width, the absolute abundance, the $\sigma$ of
 the line-to-line scatter and the abundance relative to iron are given. For the solar iron abundance
we used the meteoric value of 7.51. For the solar C,N and O
abundances we adopted resp. 8.57, 7.99 and 8.86
(C: \citealt{biemont93}, N: \citealt{1991A&AS...88..505H},
O: \citealt{biemont91}). For the solar
 Mg and Si abundances we adopt 7.54 \citep{2001sgc..conf...23H};
 the other solar abundances are taken from \citet{1998SSRv...85..161G}.
The dust condensation temperatures are from \citet{Lodders} and references
 therein.
 Only for Si do we use a different condensation temperature than that 
given in \citet{Lodders}. We use a temperature of 1340 K, which is the temperature
 at which MgSiO$_{3}$ and Mg$_{2}$SiO$_{4}$, which contain most Si, condense
\citep{Lodders}. The abundances shown for IRAS\,15469-5311 are derived from
 the spectrum, which was taken in 2000. The analysis of the spectrum,
 which was taken in 2001, leads to very similar abundances.}
\begin{center}
\resizebox{\hsize}{!}{
\begin{tabular}{|lll|rrrrr|rrrrr|rrrrr|rrrrr|}
\hline
\rule[-0mm]{0mm}{4mm}& &  & \multicolumn{5}{|c|}{\bf IRAS\,09060}
&   \multicolumn{5}{|c|}{\bf IRAS\,09144}
&   \multicolumn{5}{|c|}{\bf IRAS\,09538}
&   \multicolumn{5}{|c|}{\bf IRAS\,15469} \\
\rule[-3mm]{0mm}{3mm}& &  & \multicolumn{5}{|c|}{\bf [Fe/H]$=-0.66$}
&   \multicolumn{5}{|c|}{\bf [Fe/H]$=-0.31$ }
&   \multicolumn{5}{|c|}{\bf [Fe/H]$=-0.65$}
&   \multicolumn{5}{|c|}{\bf [Fe/H]$=+0.04$} \\
\hline
\rule[-3mm]{0mm}{8mm}ion & sun & T$_{\rm cond}$ & N & $\overline{W_{\lambda}}$ & $\log \epsilon$ &$\sigma$   & [el/Fe] &
N & $\overline{W_{\lambda}}$ & $\log \epsilon$ &$\sigma$   & [el/Fe] & N & $\overline{W_{\lambda}}$ & $\log \epsilon$ &$\sigma$   & [el/Fe] & N & $\overline{W_{\lambda}}$ & $\log \epsilon$ &$\sigma$   & [el/Fe] \\
\hline
\rule[-0mm]{0mm}{4mm}C~I   & 8.57    &  78  &  3 & 32 & 7.66 & 0.15 &$-0.25$& 11 & 58 & 8.26 & 0.14& $0$
                      & 2  & 83 & 8.30 & 0.09 &$+$0.38& 14 & 73 & 8.90 & 0.10&$+0.29$ \\
N~I   & 7.99    &  120 &    &    &      &      &       &  2 & 85 & 8.15 & 0.06&$+0.47$
                       &    &    &      &      & &  1 &  19 & 8.62 &       & $+0.59$\\
O~I   & 8.86    &      &    &    &      &      &       &  4 & 43 & 8.74 & 0.17&$-0.19$
                       &  1 & 81 &9.12  &      &$+0.91$&  5 &  56 & 8.93 &0.12&$+0.03$ \\
Na~I  & 6.33    &  970 &    &    &      &      &       &  2 & 48 & 6.23 & 0.01&$+0.22$
                      &  1 & 27 &5.99  &      &$+0.19$&  2 &  63 & 6.97 & 0.07& $+0.74$  \\
Mg~I & 7.54    &  1340 &    &    &      &      &       &    &    &      &     &       
                       &    &    &      &      &  &  1 &  48 & 7.64 &       &$+0.20$\\
Si~I  & 7.54    & 1529 &  5 & 31 & 7.23 & 0.13 &$+0.37$& 11 & 63 & 7.48 & 0.15&$+0.26$ 
                       &  5 &  68& 7.51  & 0.13& $+0.50$& 11 &  29 & 7.96 & 0.10& $+0.52$\\
Si~II & 7.54    & 1529 &  1 & 127& 6.94 &      &$+0.06$&    &    &      &     &      
                       &  1 & 150&7.57  &      &$+0.68$& &    &      &     &  \\
S~I   & 7.33    &  674 &  1 & 42 & 6.61 &      &$-0.06$&  3 & 100& 7.32 & 0.12&$+0.30$ 
                       &  1 &  55 & 7.02  &    &$+0.34$& 6 &  77 & 7.84 & 0.04& $+0.47$\\
Ca~I  & 6.36    & 1634 &  9 & 87 & 5.89 & 0.14 &$+0.21$&  4 &  99& 5.99 & 0.13&$-0.05$ 
&  6 & 110 &    5.77  & 0.17 &$-0.06$&10 & 38   & 6.00     & 0.14    & $-0.26$\\
Ca~II & 6.36    & 1634 &    &    &      &      &       &    &    &      &     &    
                       &    &    &      &      &&  1 &  10 & 5.80 &       &$-0.60$\\
Sc~II & 3.17    & 1652 &  1 & 70 & 2.35 &      &$-0.16$&  4 & 51 & 1.52 & 0.25&$-1.34$
&  1 & 139 &    2.29  &      &$-0.23$&4 &  34 & 1.61 & 0.13  & $-1.60$ \\ 
Ti~I  & 5.02    & 1600 &    &    &      &      &       &  1 & 27 & 3.73 &     &$-0.97$
&  2 &  65 &    4.32  & 0.21 &  $-0.17$&&    &      &     & \\
Ti~II & 5.02    & 1600 &  2 & 102& 4.28 & 0.05 &$-0.08$&  3 & 85 & 3.75 & 0.20&$-0.96$
&  2 &  93 &    4.36  & 0.33 & $-0.01$&  6 &  20 & 3.45 & 0.13 & $-1.61$\\
Cr~I  & 5.67    & 1301 &  4 &  86& 5.05 & 0.18 &$+0.06$&  2 & 104& 5.39 & 0.02&$+0.04$ 
                       &    &    &      &      &&  8 &  32 & 5.65 & 0.10& $+0.08$\\
Cr~II & 5.67    & 1301 &  7 &  70& 5.02 & 0.12 &$+0.01$&  5 & 88 & 5.38 & 0.16&$+0.02$
&  6 & 108 &    5.17  & 0.11 &  $+0.15$&6 & 103 & 5.69 & 0.06 & $-0.02$\\
Mn~I  & 5.39    & 1190 &  2 &  83& 4.84 & 0.09 &$+0.13$&  1 & 66 & 5.05 &     &$+0.02$
&  2 &  79 &    4.63  & 0.00 &$-0.23$&1 &  51 & 5.74 &       &$+0.45$\\
Mn~II & 5.39    & 1190 &    &    &      &      &       &    &    &      &     &        
                       &    &    &      &      &&  3 & 103 & 5.97 & 0.14& $+0.54$ \\
Fe~I  & 7.51    & 1337 & 53 &  87& 6.83 & 0.13 &       & 35 & 81 & 7.19 & 0.15& 
& 38 &  91 &    6.98  & 0.13 &   & 53 &  51 & 7.41 & 0.18 &    \\
Fe~II & 7.51    & 1337 & 13 &  88& 6.85 & 0.14 &       & 15 & 80 & 7.20 & 0.18&
&  8 &  96 &    6.86  & 0.16 & & 24 &  66 & 7.55 & 0.14  & \\
Ni~I  & 6.25    & 1354 &  8 &  67& 5.70 & 0.11 &$+0.13$& 15 & 66 & 6.02 & 0.17&$+0.09$ 
& 17 &  82 &    5.86  & 0.12 & $ +0.14$& 10 &  31 & 6.14 & 0.09  &$-0.01$ \\
Ni~II & 6.25    & 1354 &    &    &      &      &       &    &    &      &     &         
&  1 & 100 &    5.72  &     & $ +0.12$&  1 &  85 & 5.97 &       &$-0.32$ \\
Cu~I  & 4.21    & 1170 &    &    &      &      &       &    &    &      &     &         
                       &    &    &      &    &  & 1 &  26 & 4.77 &       &$+0.66$ \\
Zn~I  & 4.60    & 684  &  2 & 74 & 4.02 & 0.01 & $+0.10$&   &    &      & & 
 &  1 & 145 &    3.98  &      & $ -0.09$ &  2 &  43 & 4.85 & 0.02 & $+0.35$\\
Y~II  & 2.24    & 1622 &  3 & 44 & 1.62 & 0.07 & $+0.04$& 1 & 70 &$<$0.85 && $<-1.08$    
                       &    &    &      &     & &  1 &  49 & 1.74 &       & $-0.54$ \\
Zr~II & 2.60    & 1717 &  2 & 98 & 2.08 & 0.02 & $+0.14$&    &    &      &     &      
                       &    &    &      &      & &    &    &      &      & \\
Ba~II & 2.13    & 1520 &    &    &      &      &        &    &    &      &     &      
                       &    &    &      &      & &  3 &  69 & 1.56 & 0.10& $-0.61$\\
La~II & 1.17    & 1520 &  1 & 18 & 0.70 &      & $+0.19$&  1 &  5 &$-0.38$&&$-1.24$ 
                       &    &    &      &      & & &   &    &      &       \\
Ce~II & 1.58    & 1440 &    &    &      &      &        &    &    &        &&
                       &    &    & $<0.34$&    &  $<-0.59$&   & &    &      &       \\
\rule[-2mm]{0mm}{0mm}Nd~II & 1.50    & 1563 &    &    &      &      &        &  1 &  7 & 0.23 &    & $-0.96$ 
                       &    &    & $< 0.50$&    &$<-0.35$ &    &    &      &      &\\
\hline
\end{tabular}}
%\end{scriptsize}
\end{center}
%\end{sidewaystable*}
\end{table*}

%\twocolumn

\subsection{Individual objects}

The objects are now discussed individually on the basis of the
 Tables~\ref{chembegin}-~\ref{chemeinde} and Figs~\ref{abotemp} and~\ref{abotemptwee}. 
We look for chemical patterns determined by dredge-up processes
from the stellar interior. Alternatively, we look for chemical
signatures pointing to depletion. A 
correlation between the abundance and the condensation temperature is
 taken as evidence for a dust-gas separation. 
However, even if a depletion process is the main effect present
 in the abundances, we do not expect a tight correlation between the
 abundance and the condensation temperature T$_{\rm cond}$,
since it is unlikely that the dust formation near the star occurs
 in thermal equilibrium and since the composition is, 
in general, not solar (The condensation temperatures are computed 
using a solar abundance mix at a pressure of 10$^{-4}$ bar in 
thermal equilibrium.). 
 We will come back to the effects of gas-dust separation and 
dredge-up in the next section.

%\onecolumn

\begin{table*}
\caption{Continuation of table~\ref{chembegin}. Abundances for IRAS\,16230-3410, IRAS\,17038-4815 and IRAS\,17233-4330.}
\begin{center}
\resizebox{\hsize}{!}{
\begin{tabular}{|lll|rrrrr|rrrrr|rrrrr|}
\hline
\rule[-0mm]{0mm}{4mm}& &  & \multicolumn{5}{|c|}{\bf IRAS\,16230}
&   \multicolumn{5}{|c|}{\bf IRAS\,17038}
&   \multicolumn{5}{|c|}{\bf IRAS\,17233} \\
\rule[-3mm]{0mm}{2mm}& &  & \multicolumn{5}{|c|}{\bf [Fe/H]$=-0.68$}
&   \multicolumn{5}{|c|}{\bf [Fe/H]$=-1.54$ }
&   \multicolumn{5}{|c|}{\bf [Fe/H]$=-0.98$}\\
\hline
\rule[-3mm]{0mm}{8mm}ion & sun & T$_{\rm cond}$ & N & $\overline{W_{\lambda}}$ & $\log \epsilon$ &$\sigma$   & [el/Fe] &
N & $\overline{W_{\lambda}}$ & $\log \epsilon$ &$\sigma$   & [el/Fe] & N & $\overline{W_{\lambda}}$ & $\log \epsilon$ &$\sigma$   & [el/Fe] \\
\hline
\rule[-0mm]{0mm}{4mm}C~I   & 8.57    &  78  & 15 & 68 & 8.14 & 0.12  & $+0.25$ & 3 & 64 & 8.79 & 0.14 & $+$1.76
& 10 & 87 & 8.41 & 0.15 & $+0.82$ \\
N~I   & 7.99    &  120 &     2 &  46 & 7.54 & 0.12 & $+0.23$ &  &    &      &       &
& 1 &  25 &    8.23 &      & $+1.22$ \\
O~I   & 8.86    &      & 3 &  30 & 8.53 & 0.10 & $+0.35$ &   &    &      &       &
& 3 &  27 &    8.60 & 0.04& $+0.72$ \\
Na~I  & 6.33    &  970 &  2 &  20 & 6.00 & 0.06 & $+0.35$ &  1 &  73 &   5.31   &&$+0.43$
& 2 & 108 &    6.45 & 0.09  & $+1.05$ \\
Mg~I  & 7.54    &  1340 &   &    &      &       & &  1 & 100 &   6.61   &      &$+0.52$
&   &    &      &       & \\
Mg~II & 7.54    & 1340 &   &    &      &       & &   &    &      &       & 
 & 2 & 141 &    6.26 & 0.16& $-0.30$\\
Si~I  & 7.54    & 1529 & 15 &  38 & 7.33 & 0.11 & $+0.47$ &  3 &  56 &   6.72 & 0.30&$+0.63$
&   &    &      &       & \\
Si~II & 7.54    & 1529 & 1 & 137 & 7.15 &      & $+0.29$  &    &    &&  &       
& 1 &  97 &    6.48 &      &$-0.08$ \\
S~I   & 7.33    &  674 & 5 &  45 & 6.97 & 0.04 & $+0.32$ &   &    &      &       &
& 4 & 108 &    7.47 & 0.05 &$+1.12$  \\
Ca~I  & 6.36    & 1634 & 16 &  70 & 5.62 & 0.13 &  $-0.06$ &  7 &  94 &   4.94   & 0.16&$+0.03$ 
& 5 &  79 &    4.97 & 0.07 &$-0.46$ \\
Ca~II & 6.36    & 1634 &   &    &      &       & &   &    &      &       &
&   &    &      &       & \\
Sc~II & 3.17    & 1652 &  3 &  32 & 0.89 & 0.06 & $-1.60$ &  5 &  73 &   1.18   & 0.14 &$-0.45$
& 2 & 104 &    1.57 & 0.17 &$-0.62$  \\
Ti~I  & 5.02    & 1600 &   &  &    &  & &   &    &      &       & 
&   &    &      &       & \\
Ti~II & 5.02    & 1600 &  21 &  73 & 3.57 & 0.13  & $-0.77$ &  7 &  92 &   3.04   & 0.16 &$-0.44$
& 3 &  48 &    3.40 & 0.22&$-0.64$  \\
Cr~I  & 5.67    & 1301 &  8 &  53 & 4.88 & 0.11 & $-0.11$ &  1 &  37 &   4.35   & &$+0.13$
& 1 &  78 &    4.47 &      &$-0.27$  \\
Cr~II & 5.67    & 1301 & 13 &  82 & 5.00 & 0.10 & $+0.01$ & 3 &  66 &   4.30   & 0.19&$+0.17$   
& 5 &  82 &    4.48 & 0.16 &$-0.21$  \\
Mn~I  & 5.39    & 1190 &  6 &  54 & 4.66 & 0.13  & $-0.05$&  4 &  96 &   4.15   & 0.08 &$+0.21$
& 2 & 119 &    4.70 & 0.03 &$+0.24$     \\
Fe~I  & 7.51    & 1337 & 92 &  70 & 6.83 & 0.13  &         &  33 &  79 &   6.06   & 0.20 & 
&37 & 102 &    6.58 & 0.12 &              \\
Fe~II & 7.51    & 1337 & 24 &  82 & 6.83 & 0.10  &         & 12 & 101 &   5.97   & 0.15 & 
&11 &  99 &    6.53 & 0.10 &               \\
Ni~I  & 6.25    & 1354 &24 &  45 & 5.62 & 0.10 & $+0.05$ & 12 &  65 &   4.88   & 0.15&$+0.08$
& 1 &  67 &    5.12 &      &$-0.20$  \\
Cu~I  & 4.21    & 1170 &   &    &      &       &  &   &    &      &       & 
& 1 &  90 &    4.18 &      &$+0.90$ \\
Zn~I  & 4.60    & 684  & 2 & 100 & 4.18 & 0.05   & $+0.26$   &  1 & 150 &   3.42   &    &$+0.27$
& 2 & 146 &    4.39 & 0.03 &$+0.72$ \\
Y~II  & 2.24    & 1622 & & & $<$1.29  & &  $<-0.27$ &  1 &   6 &   0.27   &      & $-0.43$
& 1 &   9 &    0.84 &      &$-0.42$  \\
Zr~II & 2.60 & 1717 & 1 & 20 & 1.83 & & $-0.09$ & & &&&&&&&& \\
Ba~II & 2.13    & 1520 & 1  & 148   & 1.45 &    & 0.00  &  &    &      &       & 
& 1 & 134 &    1.28 &      &$+0.13$ \\
La~II & 1.17    & 1520&    &     &$<-$0.17 &   & $<-0.66$    &   &    &      &       & 
&   &     & $<-$0.06&      &$<-0.25$ \\
Ce~II & 1.58    & 1440 &    &     & $< -0.15$ & &$<-1.05$&        &      &       & 
&   &  &   & $<-$0.01&      &$<-0.61 $  \\
Pr~II & 0.71 & 1557  & &    &$<-$0.37 &     & $<-0.40$ &   &    &      &       &  
&   &    &      &       & \\
Nd~II & 1.50    & 1563 &    &     & $< 0.16$ & &$<-0.66$ & 1 &  39 &$-$0.33   &      &$-0.29$
&    &    & $ <0.16$&      &$<-0.36$  \\
\rule[-2mm]{0mm}{0mm}Sm~II & 1.01    & 1560 & 2  & 14   & 0.23    & 0.03 & $-0.10$    &   &    &      &       & 
&   &     & $<-0.56$& &$<-0.59$  \\
\hline
\end{tabular}}
\end{center}
\end{table*}

\begin{table*}
\caption{\label{chemeinde}Continuation of Table~\ref{chembegin}. Abundances for IRAS\,17243-4348, IRAS\,19125+0343
and IRAS\,19157-0247  For IRAS\,17243-4348
we show the abundances, derived from both spectra.}
\begin{center}
\resizebox{\hsize}{!}{
\begin{tabular}{|lll|rrrrr|rrrrr|rrrrr|rrrrr|}
\hline
\rule[-0mm]{0mm}{4mm}& & &   \multicolumn{5}{|c|}{\bf IRAS\,17243}
&   \multicolumn{5}{|c|}{\bf IRAS\,17243} 
& \multicolumn{5}{|c|}{\bf IRAS\,19125}
&   \multicolumn{5}{|c|}{\bf IRAS\,19157} \\
& & & \multicolumn{5}{|c|}{24/03/2000} & \multicolumn{5}{|c|}{28/06/2001} &
&&&&&&&&&  \\
\rule[-3mm]{0mm}{2mm}& &  & \multicolumn{5}{|c|}{\bf [Fe/H]$=-0.05$}
&   \multicolumn{5}{|c|}{\bf [Fe/H]$=-0.10$ }
&   \multicolumn{5}{|c|}{\bf [Fe/H]$=-0.35$}
&   \multicolumn{5}{|c|}{\bf [Fe/H]$=+0.07$} \\
\hline
\rule[-3mm]{0mm}{8mm}ion & sun & T$_{\rm cond}$ & N & $\overline{W_{\lambda}}$ & $\log \epsilon$ &$\sigma$   & [el/Fe] &
N & $\overline{W_{\lambda}}$ & $\log \epsilon$ &$\sigma$   & [el/Fe] & N & $\overline{W_{\lambda}}$ & $\log \epsilon$ &$\sigma$   & [el/Fe] & N & $\overline{W_{\lambda}}$ & $\log \epsilon$ &$\sigma$   & [el/Fe] \\
\hline
\rule[-0mm]{0mm}{4mm}C~I   & 8.57    &  78 & 23 & 72 &  8.35 & 0.12 & $-$0.17 & 10 & 86 & 8.63 & 0.14 & 0.16 
 & 24 & 57 & 8.72 & 0.13 & $+0.50$ & 13 & 45 & 8.61 & 0.13 & $-0.03$ \\
N~I   & 7.99    &  120 &  5 & 58  &    7.78 & 0.15 & $-$0.16  & &   &         &      &    
&&&& &&&&&&\\
O~I   & 8.86    &      &  3 & 53  &    8.73 & 0.01 & $-$0.08  & &   &         &      &         
&  8 & 76 & 9.06 & 0.12 & $+0.55$ &  5 & 85 & 9.15 & 0.12 & $+0.22$ \\
Na~I  & 6.33    &  970 &  2 & 54  &    6.57 & 0.00 &    0.26  &  2 & 77 &   6.31 &0.09&0.14
 &  3 & 35 & 6.70 & 0.11 & $+0.66$ &   &  &  &      &  \\
Mg~I  & 7.54    & 1340&  1  & 137 &    7.65 &      &    0.13  &    &    &     &      &        
 &  1 & 125 & 7.51 &     & $+0.26$ &&&&& \\
Mg~II & 7.54    & 1340&     &&&&&&&&&& 1 & 71 & 7.29 & & +0.10 && &&& \\ 
Al~I  & 6.47    & 1670 &&&&&&&&&& 
 &  1 &  57 &   4.87 &   & $-1.31$ &&&&&  \\
Si~I  & 7.54    & 1529 & 16  & 68 &    7.81 & 0.12 &    0.29  & 11 & 101 &   7.67 & 0.12 &    0.29
 &  1 &   9 &   7.36 &   & $+0.11$ &&&&&  \\
Si~II & 7.54    & 1529&  1 & 137  &    7.92 &      &    0.43  &    & &        &      &        
 &&&&& &&&&&  \\
S~I   & 7.33    &  674 &  6 & 88  &    7.37 & 0.09 &    0.09  &  6 & 91 &   7.41 & 0.13 &    0.18
&  6 &  37 & 7.81 & 0.08& $ +0.83$ &  3 & 30 & 7.50 & 0.21 & $+0.10$  \\
Ca~I  & 6.36    & 1634 &  9 & 91  &    6.19 & 0.10 & $-$0.15  &  5 & 113 &  6.08 & 0.18 & $-$0.12
 &  3 &  27 & 5.85 & 0.14& $ -0.22$ &&&&&   \\
Sc~II & 3.17    & 1652 &  1  &  119 &    2.06 &      & $-$1.06  &  4 & 90 &   1.71 & 0.18 & $-$1.36 
&    &     &$<$0.72 &   & $<-2.10$ &  3 & 40 & 2.73 & 0.05& $-0.51$  \\
Ti~I  & 5.02    & 1600  &    &   &         &      &          &  2 & 23 &  4.11 & 0.15 & $-$0.75
&&&&&  &&&&&   \\
Ti~II & 5.02    & 1600 &  2  & 108 &    4.41 & 0.21 & $-$0.56  &  1 & 127 &   4.33 &      & $-$0.59
  &    &     &$<$2.87 &   & $<-1.80$ & 12 & 83 & 4.80 & 0.12& $-0.29$  \\
V~II  & 4.00    & 1455 &&&&&  &&&&& 
 &  1 &  77 &$<$3.18 &   & $<-0.47$ &&&&&  \\
Cr~I  & 5.67    & 1301&   &    &         &      &          &  6 & 92&   5.49 & 0.12 & $-$0.02 
 &  4 &  63 &   5.48 & 0.07& $ +0.10$ &&&&&  \\
Cr~II & 5.67    & 1301 &  1 & 80  &    5.60 &      & $-$0.02  &  2 & 87 &   5.50 & 0.05 & $-$0.07 
 & 16 &  81 &   5.56 & 0.12& $ +0.24$ & 10 & 95 & 5.74&0.16&$+0.00$   \\
Mn~I  & 5.39    & 1190 &  2 & 75  &    5.30 & 0.02 & $-$0.07  &  3 & 88  &  5.14 & 0.13 & $-$0.09 
 &  3 &  35 &   5.66 & 0.13& $ +0.56$ &  1 & 17 & 5.26&    &$-0.15$  \\
Fe~I  & 7.51    & 1337 & 53  & 66 &    7.49 & 0.12 &          & 43 & 88 &   7.35 & 0.07 &         
& 45 &  47 &   7.22 & 0.16  &&19 & 55 & 7.53 & 0.13 &   \\
Fe~II & 7.51    & 1337 & 13  & 82 &    7.46 & 0.12 &          & 13 & 78 &   7.41 & 0.16 &        
& 18 &  53 &   7.16 & 0.12  &&23 & 71 & 7.58 & 0.15 &   \\
Ni~I  & 6.25    & 1354 & 25 & 71  &    6.26 & 0.11 &    0.03  & 16 & 111 &  6.13 & 0.12 &    0.04
&  1 &  12 &   5.56 & &$-0.40$       & &&&&  \\
Ni~II & 6.25    & 1354 &  &&&&& &&&&&
 1 &  24 &   5.27 &       &$-0.63$& &&&&  \\
Cu~I  & 4.21    & 1170 & 1 & 113   &    4.49 &      &    0.30  &    &    &     &      &  
 &  1 &   7 &   4.64 &       &+0.72& &&&&  \\
Zn~I  & 4.60    & 684 & 1   & 65 &    4.82 &      &    0.24  &  1 & 92 &   4.53 &      &    0.09 
 &  2 &  24 &   4.73 & 0.11  & +0.42& &&&&  \\
Y~II  & 2.24    & 1622 & 1 & 6   &    0.44 &      & $-$1.75  &  3 & 49 &  0.61 & 0.15 & $-$1.53 
 &  1 &  17 &$<$1.53 &       &$<-0.36$& &&&&  \\
Zr~II & 2.60    & 1717 &   &    &         &      &          &  2 & 73 &   1.25 & 0.03 & $-$1.25
&  1 &  14 &$<$1.94 &       &$<-0.31$& &&&&  \\
La~II & 1.17    & 1520 &  &     &         &      &          &  2 & 9 &$-$0.34 & 0.13 & $-$1.41 
 &&&&&  &&&&& \\
Ce~II & 1.58    & 1440 &   &    &         &      &          &  3 & 66 &   0.29 & 0.13 & $-$1.19
 &&&&& &&&&&  \\
Nd~II & 1.50    & 1563&  &     &         &      &          &  2 & 7 & $-$0.34 & 0.02 & $-$1.74 
 &&&&&  &&&&&  \\
\rule[-2mm]{0mm}{0mm}Eu~II & 0.51    & 1338 & 1 & 50    &    0.75 &   &    0.29  &  1 & 85 &   0.49 &      &    0.08
  &&&&&  &&&&&  \\
\hline
\end{tabular}}
\end{center}
\end{table*}

\begin{figure*}
\begin{center}
\caption{\label{abotemp} The absolute abundances of the elements versus their
condensation temperature for all the programme stars.}
\resizebox{6.5cm}{6.5cm}{\includegraphics{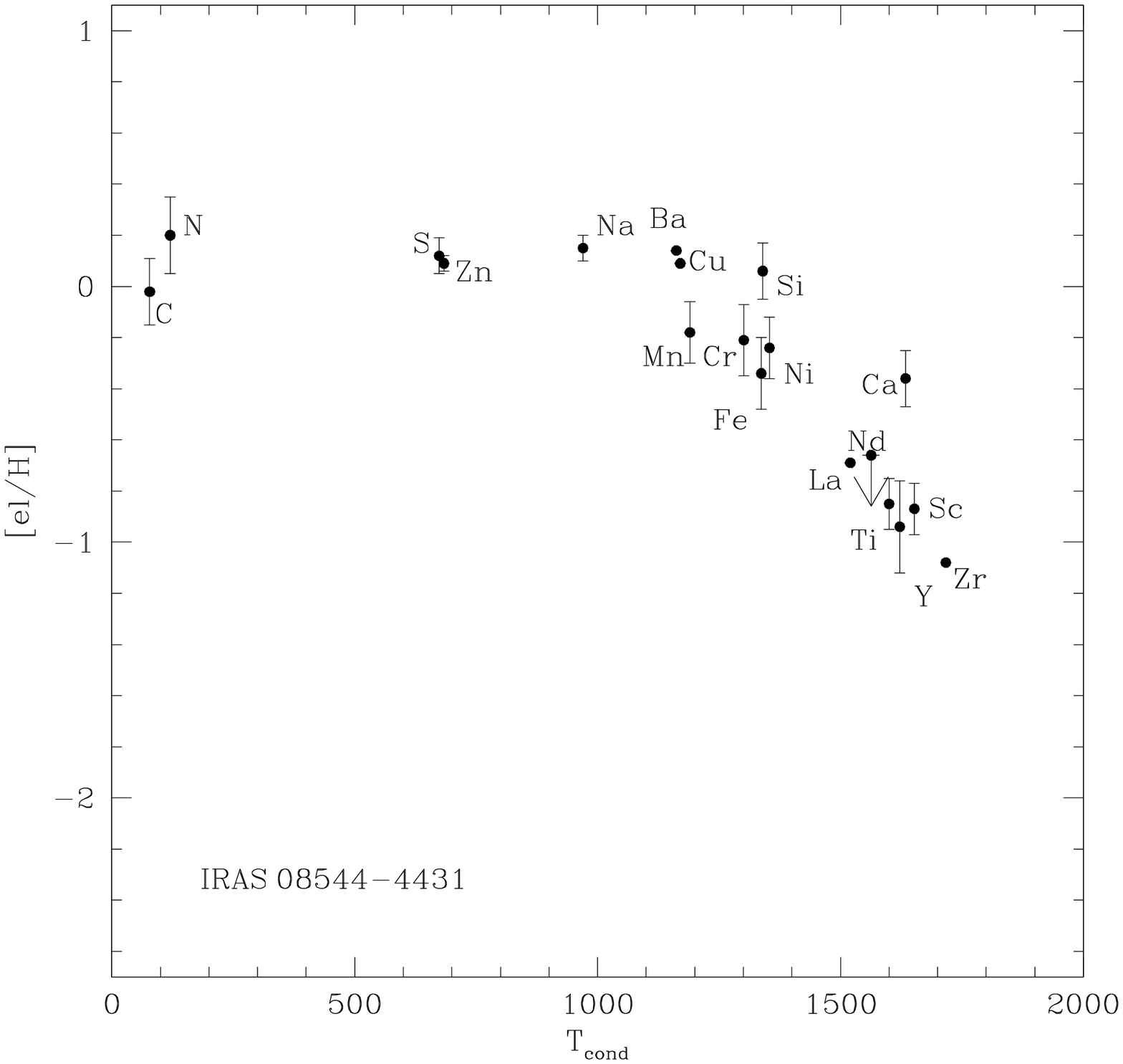}}
\resizebox{6.5cm}{6.5cm}{\includegraphics{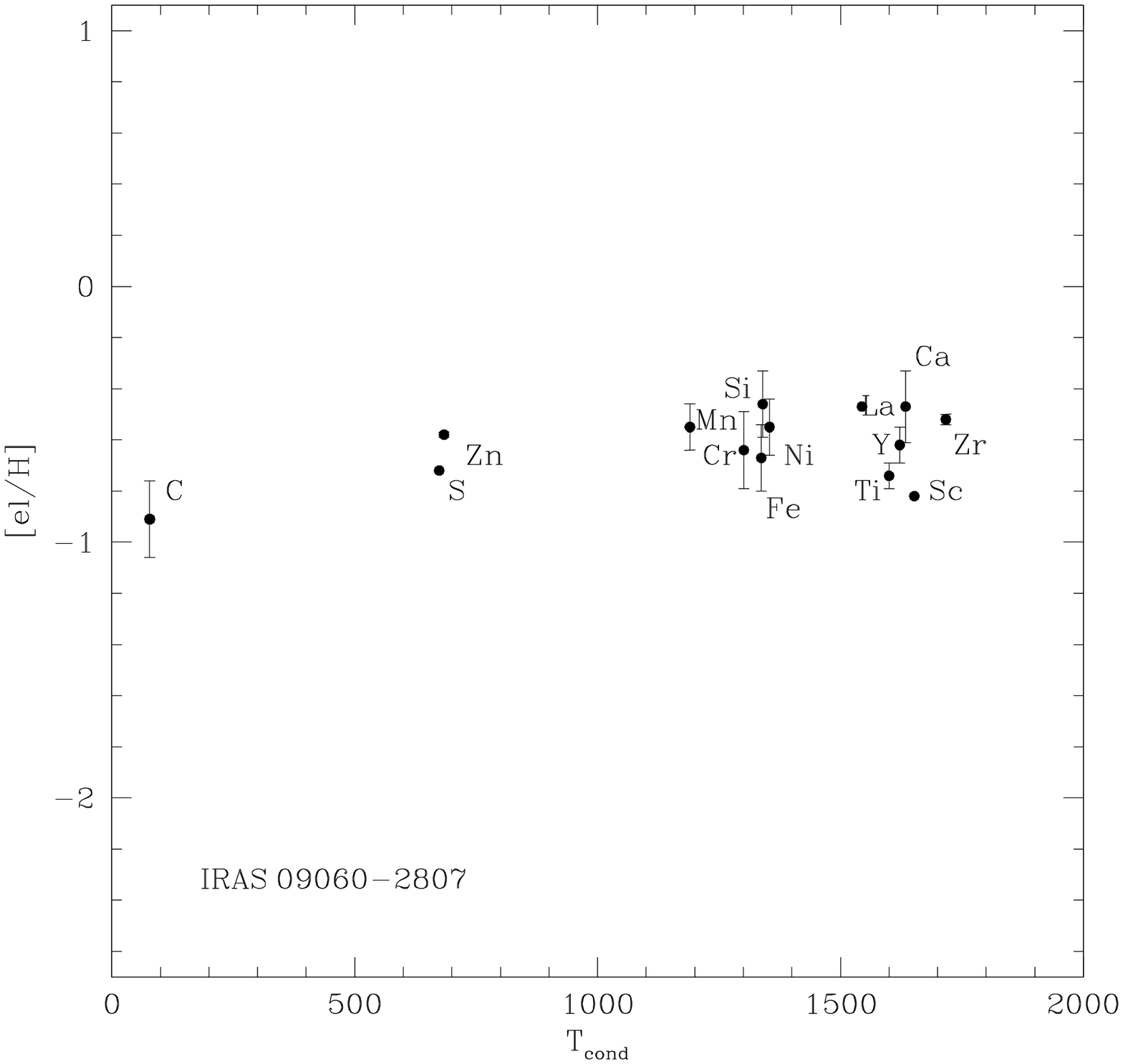}}
\resizebox{6.5cm}{6.5cm}{\includegraphics{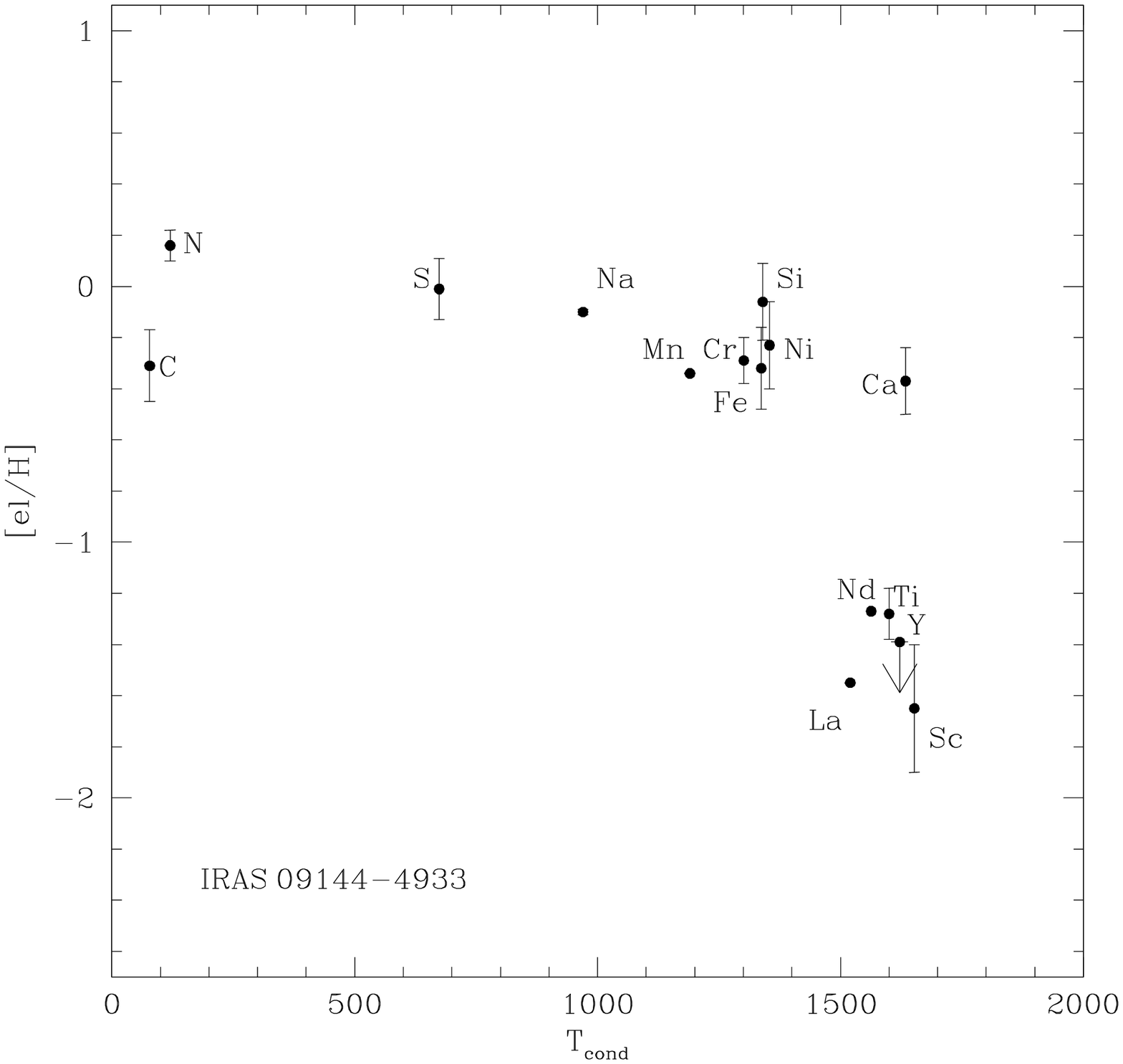}}
\resizebox{6.5cm}{6.5cm}{\includegraphics{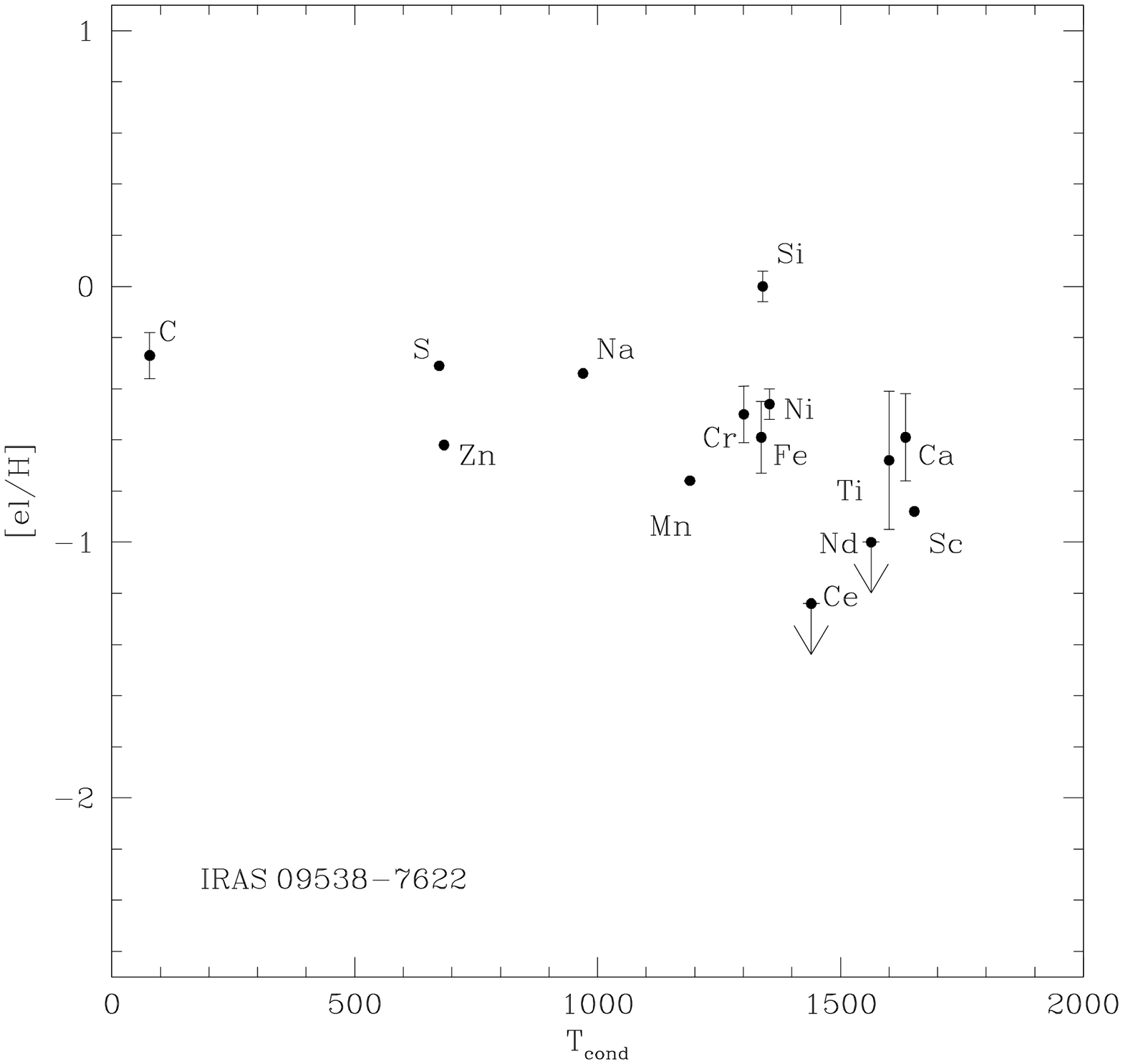}}
\resizebox{6.5cm}{6.5cm}{\includegraphics{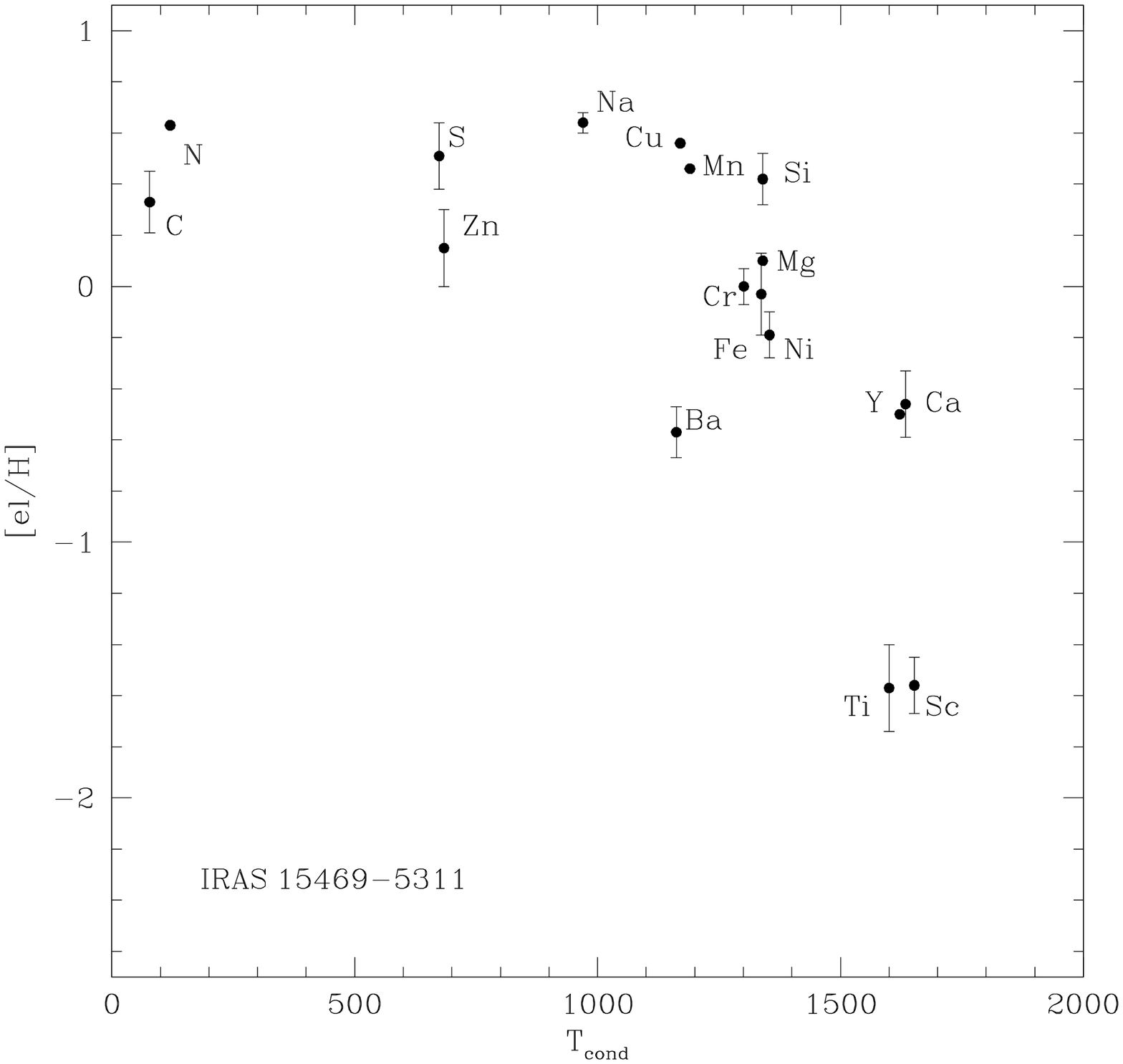}}
\resizebox{6.5cm}{6.5cm}{\includegraphics{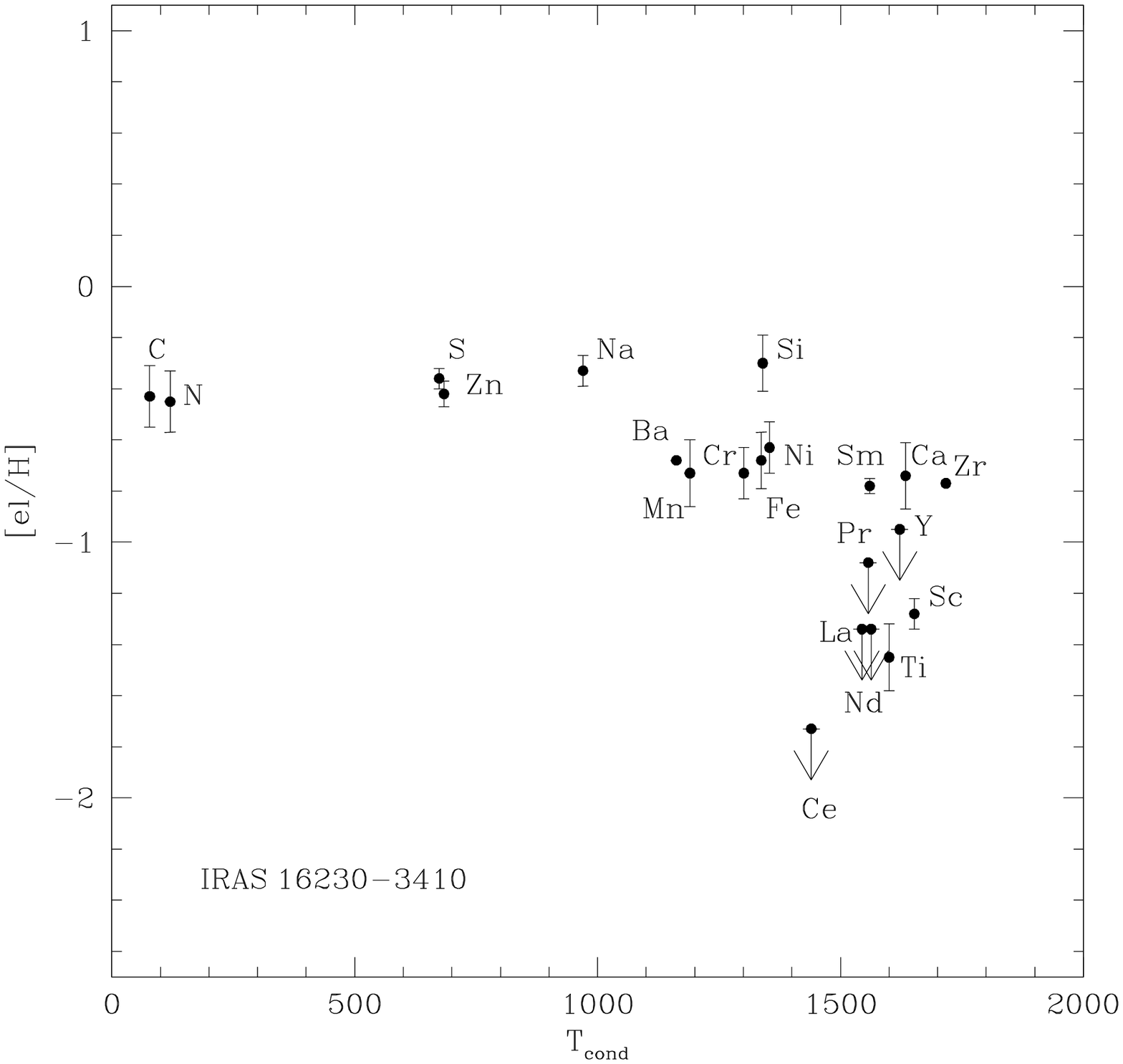}}
\end{center}
\end{figure*}

\begin{figure*}
\begin{center}
\caption{\label{abotemptwee} Continuation of Fig.~\ref{abotemp}.}
\resizebox{6.5cm}{6.5cm}{\includegraphics{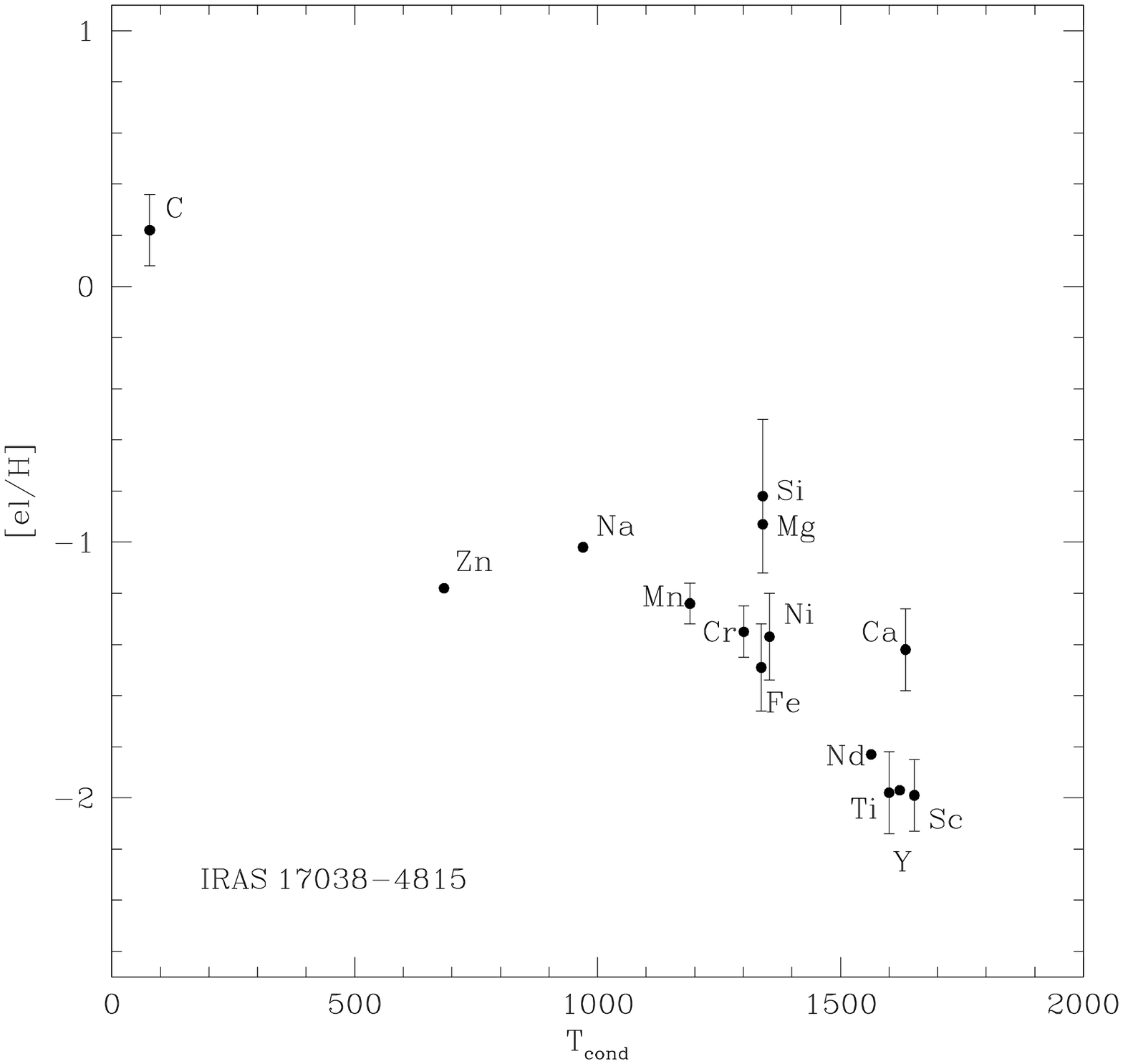}}
\resizebox{6.5cm}{6.5cm}{\includegraphics{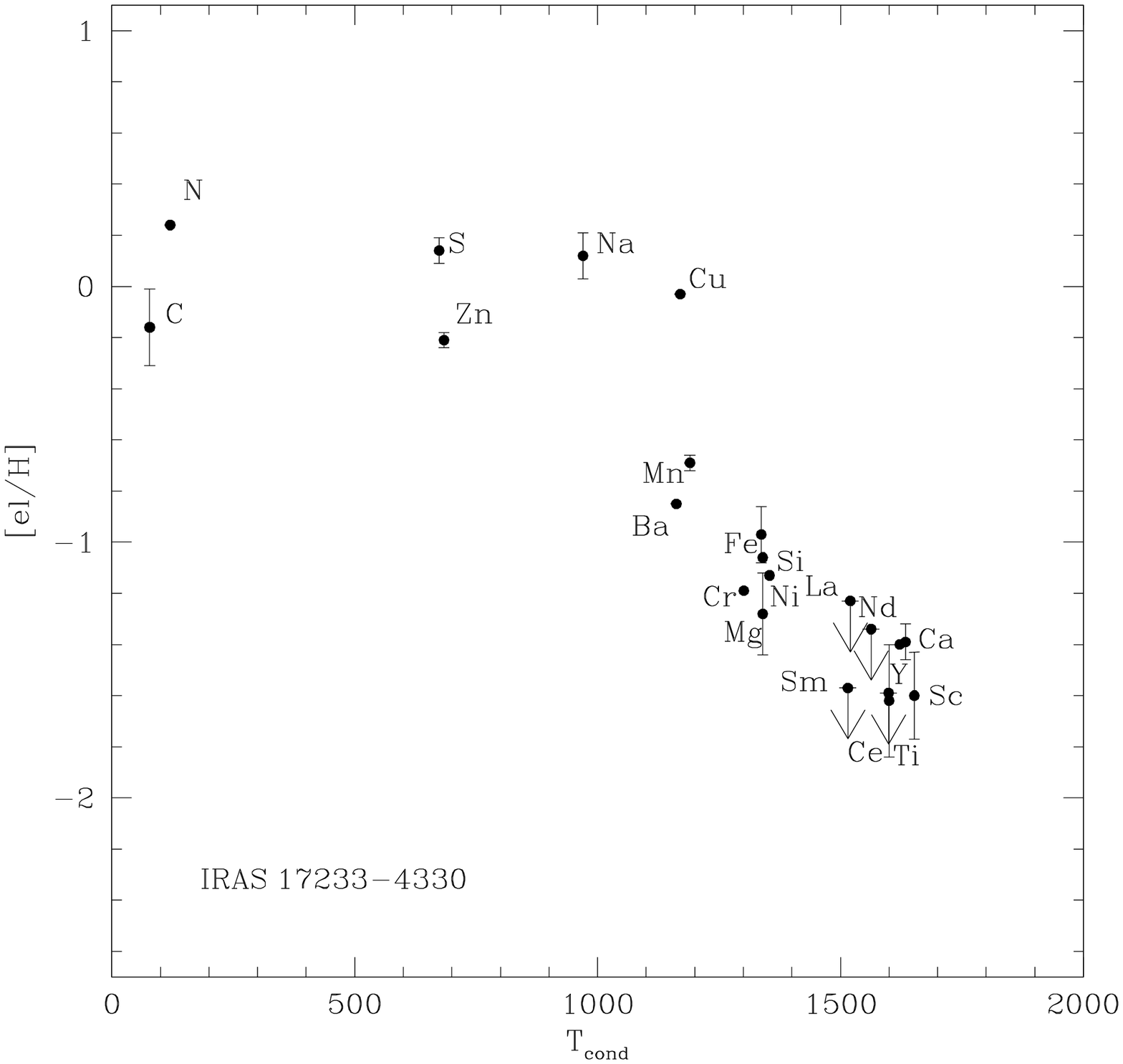}}
\resizebox{6.5cm}{6.5cm}{\includegraphics{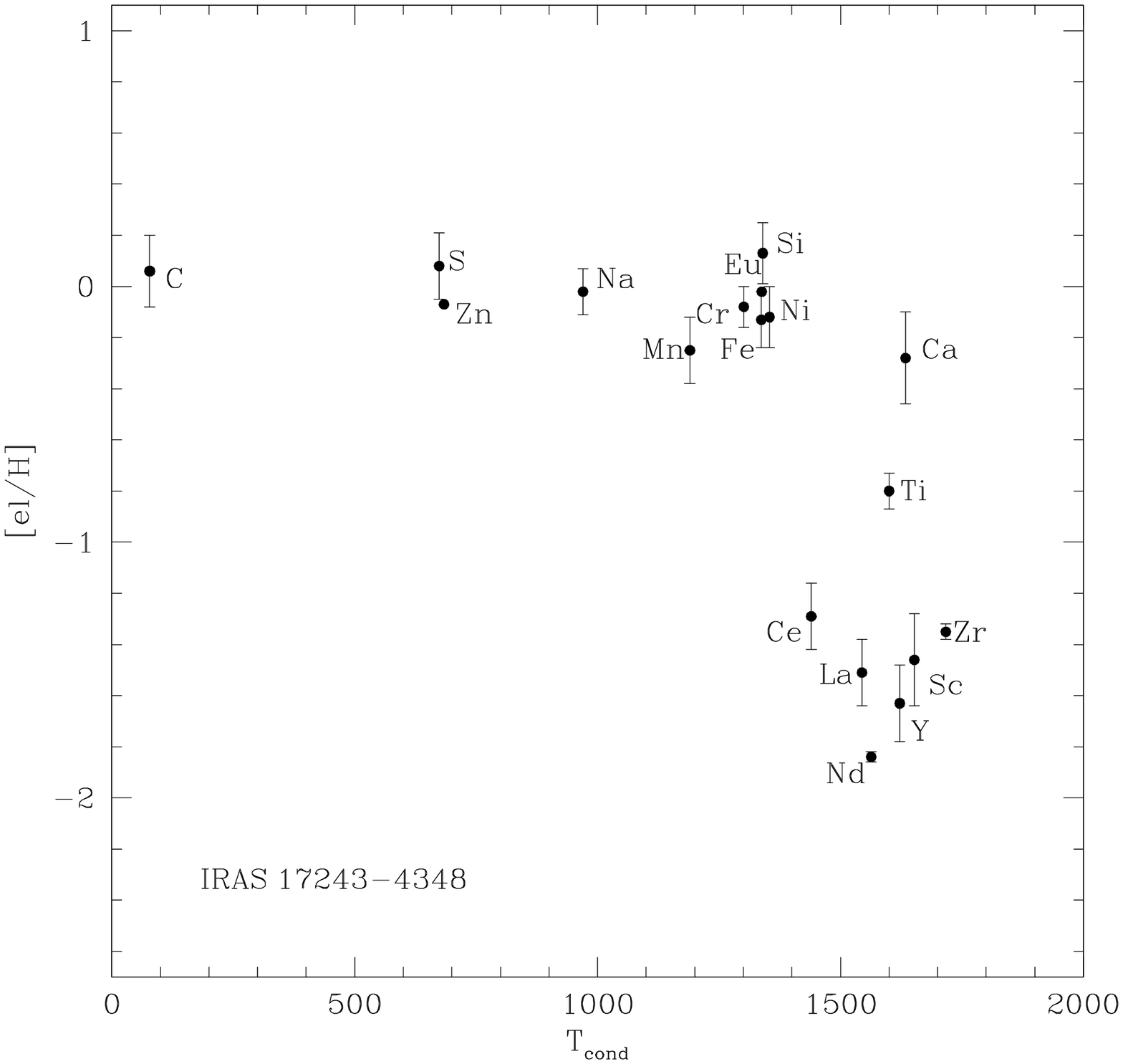}}
\resizebox{6.5cm}{6.5cm}{\includegraphics{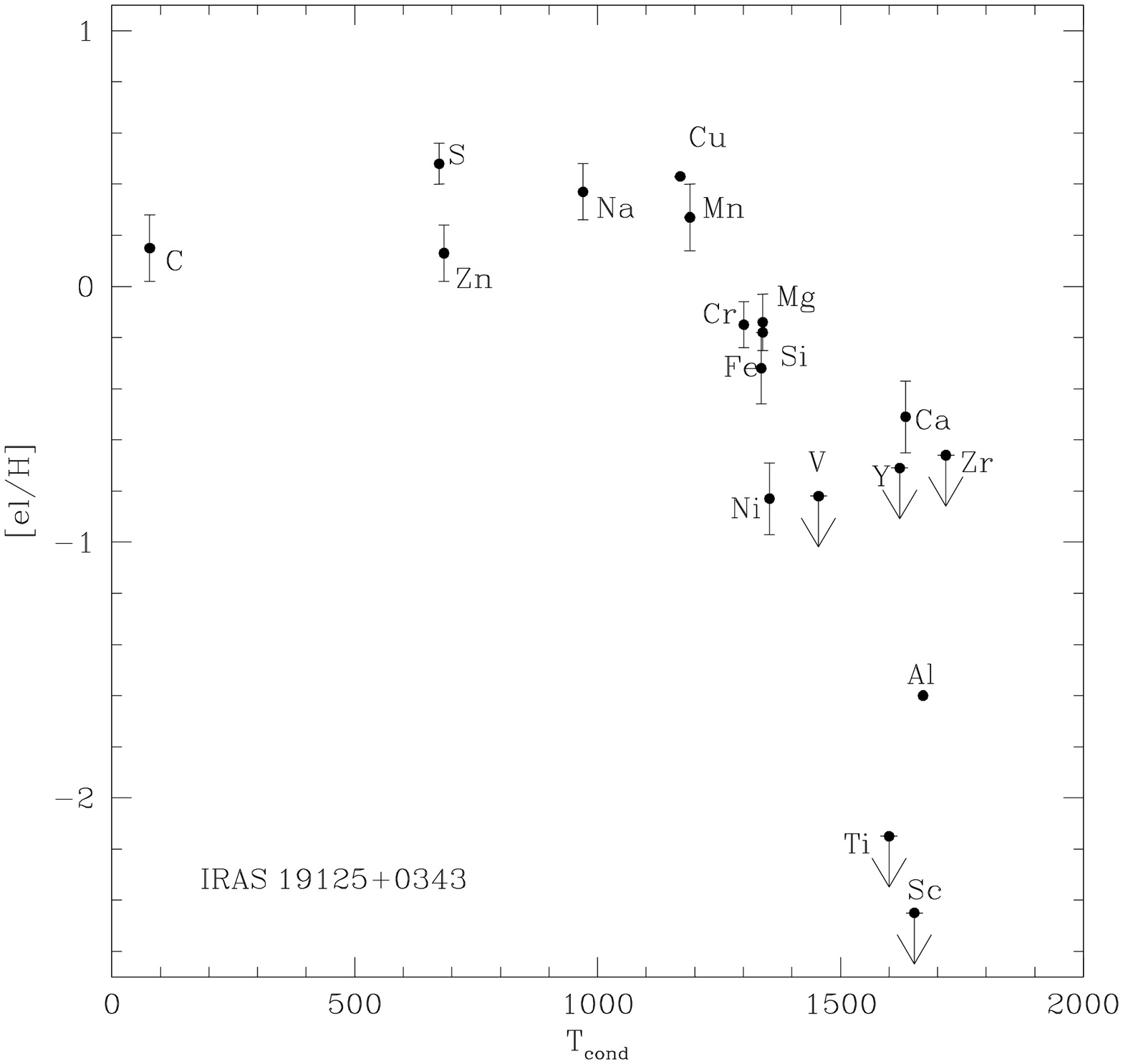}}
\resizebox{6.5cm}{6.5cm}{\includegraphics{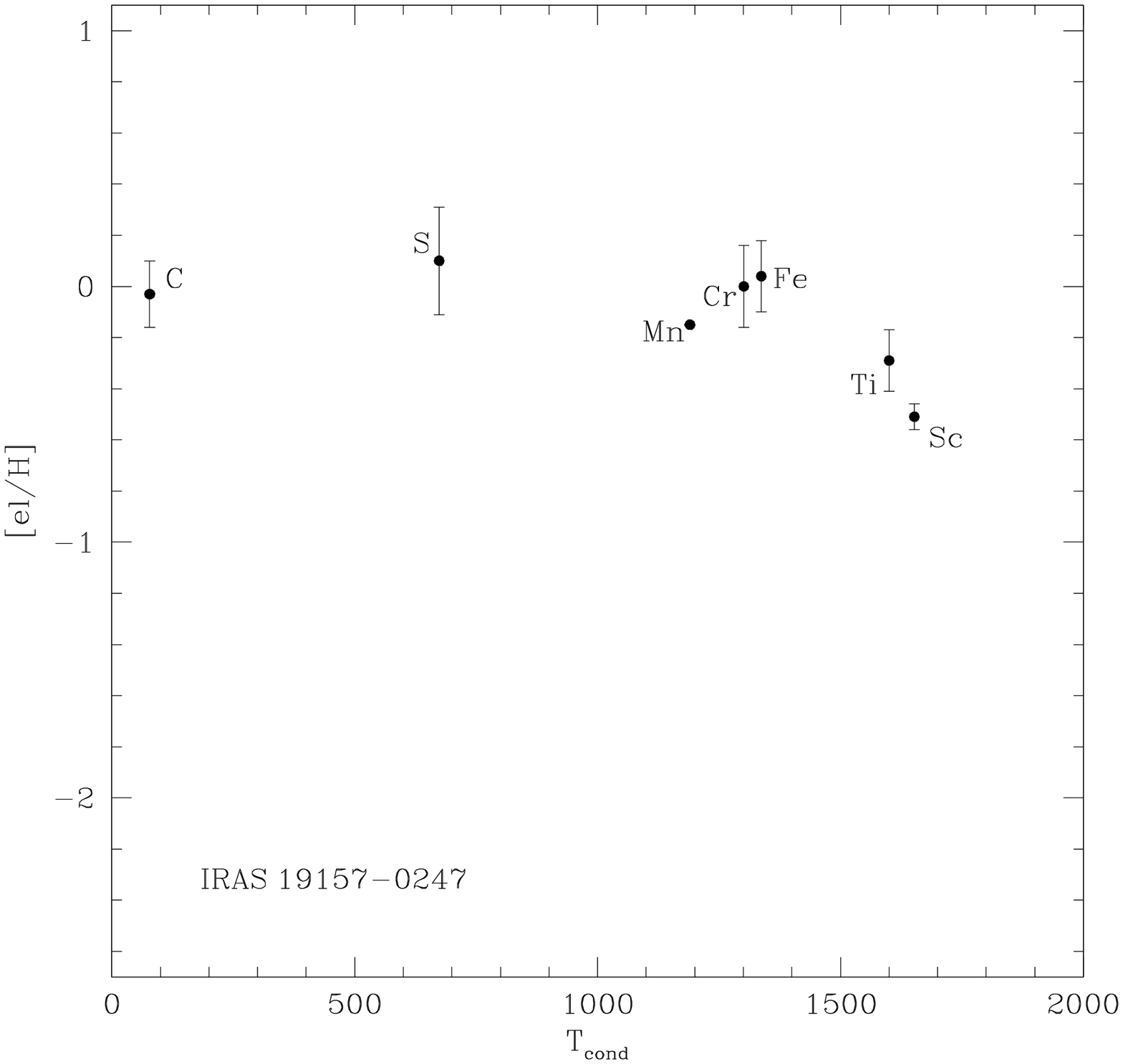}}
\resizebox{6.5cm}{6.5cm}{\includegraphics{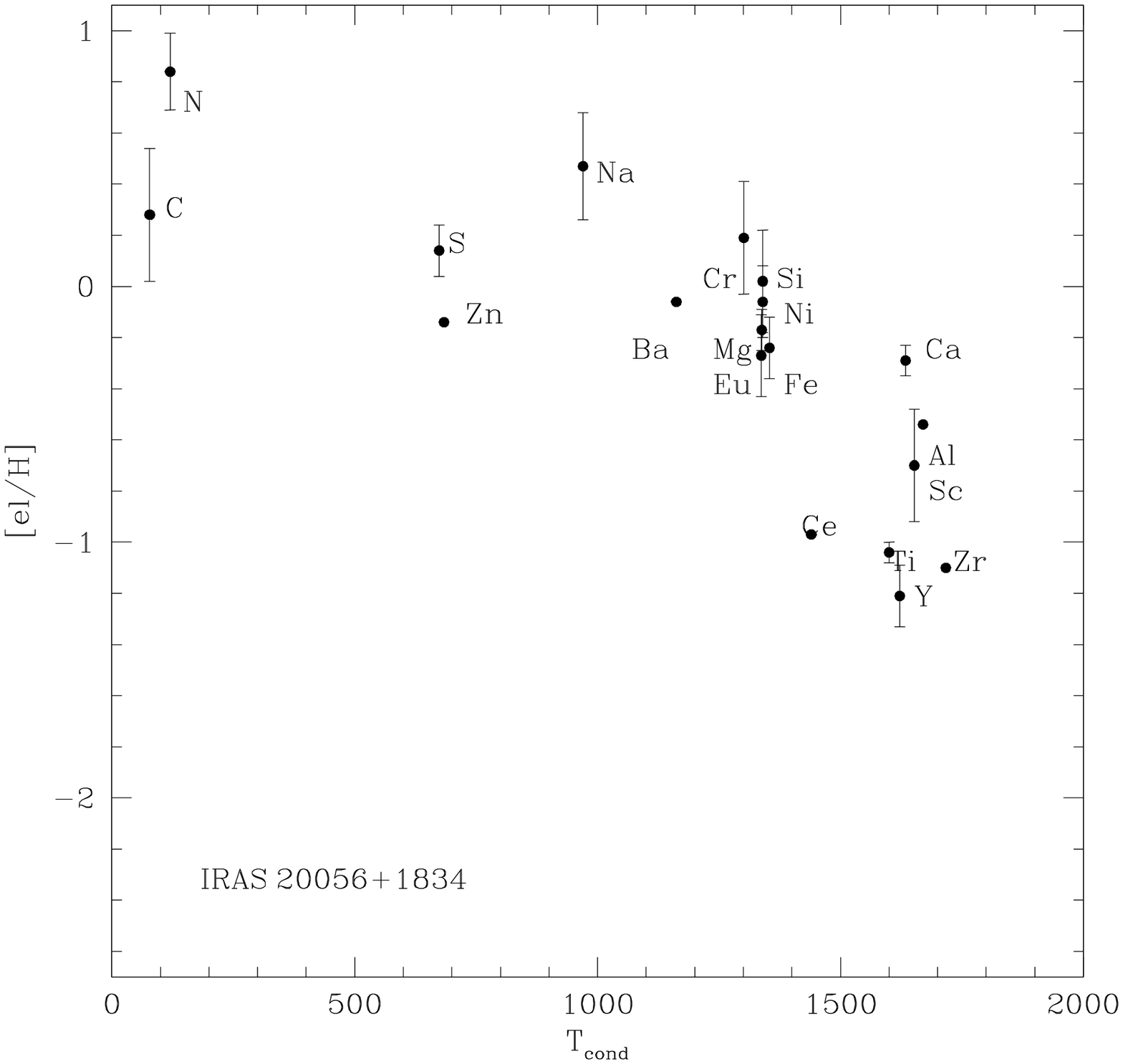}}
\end{center}
\end{figure*}

%\twocolumn

All stars, for which we performed a chemical analysis, are pulsating variables. 
In this analysis we assume that we can approximate the atmosphere of the star at a
certain pulsational phase by a hydrostatic model.
For a part of the programme stars, the photometric variation
is small (Lloyd Evans 1999) and this is very likely
a good assumption. For the stars with a larger pulsational amplitude, it should be checked
that the results of the chemical analysis at different pulsational phases are consistent.
For one object, IRAS\,17243-4348 ($\Delta V= 0.69$, Maas et al. in preparation), we have two spectra, taken at different
pulsational phases, at our disposal. Although the derived model parameters are very 
different (a T$_{\rm eff}$ of 6250 K and a $\log g$ of 0.5 
for one phase and a T$_{\rm eff}$ of 5250 K
 and a $\log g$ of 0.0 for the other phase), the results for the chemical composition are in 
good agreement. \citet{1997ApJ...481..452G} performed a chemical analysis
of a number of RV\,Tauri stars. Several stars were observed on more than one occasion.
Spectra with absorption lines showing line-doubling or remarkable asymmetries or with 
strong emission in H$\beta$\, were rejected for analysis. Spectra, not showing these
characteristics represent the atmosphere at its stablest time.
Their analysis of three spectra of SS\,Gem, obtained at three different pulsational
phases,  is consistent. The analysed spectra of our stars do not show line-splitting,
 severe line-asymmetries or emission at H$\beta$.\\

{\bf IRAS\,08544-4431}\\

The abundances of this object, of which the composition is moderately
changed by the depletion process, are thoroughly
discussed in \citet{irasnulacht}. \\

{\bf IRAS\,09060-2807}\\

Fig.~\ref{abotemp} shows that the abundances are independent
 of the condensation temperature and that all elements scale with iron.
 No depletion process has altered the chemical composition and the derived 
metallicity of  $-0.7$ reflects the initial composition,
indicating that it is an older object than the other stars in the sample.
The old nature of this object is supported by its high galactic latitude 
of 12.94 deg. For a star of this metallicity one expects a small
 overabundance of $\alpha$-elements [$\alpha$/Fe] = 0.2. 
For IRAS\,09060-2807 we find
 [$\alpha$/Fe]=0.1 (with $\alpha$ = Si, S, Ca, Ti). 
The C-abundance of [C/Fe]=$-0.25$ is rather low. This could be caused by 
CNO-cycling on the first giant branch, leading to an N-enhancement. Unfortunately,
no good N-lines are present. Neither are good O-lines available.
For three s-process elements Y, Zr and La we determined abundances with a mean
of +0.12 relative to iron. Since we found no carbon or s-process enhancement, 
there is no chemical evidence for third dredge-up in IRAS\,09060-2807.
\\

{\bf IRAS\,09144-4933}\\

This star is faint and highly reddened which results in a spectrum with a
 low S/N of 30. As a consequence the number of used lines is low and we did not
 use lines in the domain until 4900 \AA. Unfortunately, no 
Zn-abundance could be derived. Looking at Fig.~\ref{abotemp},
 we note that up to a certain T$_{\rm cond}$ no correlation between the 
abundances and T$_{\rm cond}$ is present. However, La, Nd, Ti, Y, Sc, all
 having T$_{\rm cond} > 1500 $ K, are clearly underabundant by at
 least one order of magnitude. Ca complicates the picture, since it has a
 T$_{\rm cond}$ of 1624 K and is not underabundant. \\

{\bf IRAS\,09538-7622}\\

The abundances of this object are based on a CORALIE spectrum, of which
the S/N (40) is considerably less than the other spectra.
 Because also the wavelength domain covered by CORALIE is smaller than that by FEROS, 
the derived abundances rely on fewer lines and are not as accurate as
those of the other stars.
 A higher S/N spectrum over a larger spectral range is needed to
 confirm the  abundances obtained in this work.
Maas et al. (in preparation) identify IRAS\,09538-7622 as a genuine RV\,Tauri
star. Its RV\,Tauri spectroscopic class is A (Lloyd Evans, 1999 and in preparation). 
Its abundances are not affected by the depletion process and scale roughly with iron.
Only Nd and Ce seem to be underabundant, but it is not clear with which strength. 
Since [Zn/Fe]=$-0.1$, the low metallicity $-0.6$ is initial and points to an old nature. This
is also indicated by the high measured radial velocity of 97 \kms
of this CORALIE spectrum and the high galactic latitude of $-$17.24 deg. \\

%The high O abundance coming from the forbidden 6363 
%line is maybe circumstellar.

{\bf IRAS\,15469-5311}\\

 The temperature for IRAS\,15469-5311 was determined on the basis of Fe~II-lines.
 We believe that the model parameters are not as precise as for the other stars
(note the difference in abundance for Fe~I and Fe~II)
so the absolute abundances (certainly the super solar abundances of S, Mn, Si, Na, Cu)
should be handled with caution, the relative abundances
are considered to be more precise.

For this star we have two spectra, one taken on 23/03/2000,
 the other on 27/06/2001.
For both spectra we used the same model parameters, which
resulted in similar abundances. We show the abundances
of the spectrum taken in 2000 in Fig.~\ref{abotemp}, which has the highest signal-to-noise.

The high [S/$\alpha$] of 0.9 indicates that a depletion process has
changed a part of the chemical composition. However, we note that there is a large
spread among the underabundances : [Y/Fe]=$-0.5$ and [Ca/Fe]=$-0.4$ versus
[Ti/Fe] = $-1.6$ and [Sc/Fe]=$-1.6$. Also Ba has a low abundance ([Ba/Fe]=$-0.6$),
despite its 'lower' T$_{\rm cond}$ (=1162 K). \\

{\bf IRAS\,16230-3410}\\

[Zn/Fe] = $+$0.3 suggest that the iron abundance ([Fe/H]=$-$0.7) is slightly affected.
Abundances of elements up to a T$_{\rm cond}$ of about 1500 K are at most 
mildly
changed by the depletion process. Above that temperature, there is a large
spread in underabundances going from $-$0.1 for [Ca,Zr,Sm/Fe] to $-0.8$ for [Ti/Fe]
 and the upperlimit of $-1.0$ for [Ce/Fe]. IRAS\,16230-3410 is identified as
a genuine RV\,Tauri star (Lloyd Evans, 1999 and in preparation).
 It belongs to the spectroscopic group A. The depletion process
 has, at most, effected a moderate change in
the chemical composition of the RV\,Tauri stars of this class
 \citep{2000ApJ...531..521G}.
In this way the chemical composition of IRAS\,16230-3410 is in agreement with
its membership of group A. \\

{\bf IRAS\,17038-4815}\\

IRAS\,17038-4815 was noticed by Lloyd Evans (1999 and in preparation)
to be an extreme example of an RV\,Tauri star of spectroscopic group B with
by far the strongest C$_2$ bands of any  B-group star. A spectral
classification based on the low-resolution data would indicate a CH star
or possibly a regular R star. Many CH stars have even shallower C$_2$ bands.
Our analysis based on high-resolution spectra indicate, however, that
the object is strongly depleted (Table 6, Fig. 6).
Since the s-process elements are not enhanced relative to the species
with similar condensation temperature,
we have no evidence for 3rd dredge-up enrichment. Unfortunately, good O lines
are not available but despite the fact that the object has a very strong
[C/Fe] ratio of $+$ 1.8, it is not at all clear if this object is a genuine
post-carbon star. This star illustrates the finding that strong depletion
can mimic intrinsic carbon enhancement by increasing the C/O ratio. \\

{\bf IRAS\,17233-4330}\\

[Zn/Fe]=$+0.7$ and [S/$\alpha$]=$+$1.5 leaves no doubt about the depleted photosphere
of IRAS\,17233-4330, which is a genuine RV\,Tauri star of group B
 (Lloyd Evans, 1999 and in preparation). \\

{\bf IRAS\,17243-4348 (LR\,Sco)}\\

\begin{table}
\caption{\label{compar} Comparison between the abundances of IRAS\,17243-4348 derived from
the spectrum taken on 24/03/2000 and that taken on 28/06/2001.
  $\Delta$ [el/H]= [el/H]$_{2001}-$[el/H]$_{2000}$, $\Delta$ 
[el/Fe]=[el/Fe]$_{2001}-$[el/Fe]$_{2000}$. The last column shows the difference
of the 2001 relative abundances with those derived by \citet{1992JApA...13..307G} :
$\Delta$[el/Fe]$_{Gir}$ = [el/Fe]$_{Giridhar}-$[el/Fe]$_{2001}$.}
\vspace{0.5ex}
\begin{center}
\begin{tabular}{l|rrr}
\rule[-3mm]{0mm}{3mm} ion  & $\Delta$ [el/H] & $\Delta$ [el/Fe] &  $\Delta$[el/Fe]$_{Gir}$ \\
\hline 
\rule[0mm]{0mm}{3mm} C~I    &    0.28  &    0.33 &    0.08 \\
Na~I   & $-$0.26  & $-$0.12 &    0.03 \\
Si~I   & $-$0.14  &    0.00 &    0.09 \\
S~I    &    0.04  &    0.09 &    0.24 \\
Ca~I   & $-$0.11  &    0.03 & $-$0.17 \\
Sc~II  & $-$0.35  &    0.30 &    0.71 \\
Ti~I   &          &         &    0.76 \\
Ti~II  & $-$0.08  &    0.03 &    0.58 \\
Cr~I   &          &         &    0.08 \\
Cr~II  & $-$0.10  &    0.05 &    0.16 \\
Mn~I   & $-$0.16  &    0.02 &    0.34 \\
Fe~I   & $-$0.14  &         & $-$0.11 \\
Fe~II  & $-$0.05  &         & $-$0.19 \\
Ni~I   & $-$0.13  &    0.01 &    0.00 \\
Zn~I   & $-$0.29  & $-$0.21 &    0.37 \\
Y~II   &    0.17  &    0.22 &    1.67 \\
Eu~II  & $-$0.26  & $-$0.21 &         \\
\end{tabular}
\end{center}
\end{table}

For this star, which is considered as a genuine RV\,Tauri star by
 \citet{1999IAUS..191..453E}, we have two spectra at our disposal.
 The first is taken on 24/03/2000,
 the second on 28/06/2001, corresponding to different phases in the pulsation cycle. 
For the first model we find a T$_{\rm eff}$ of 6250 K and a $\log g$ of 0.5
for the second a T$_{\rm eff}$ of 5250 K and a $\log g$ of 0.0. To compare the abundances
 derived from the different spectra, we first intended to use only the same lines
 in both analyses. However, we dropped this restriction, since due to the large
 differences in model parameters, the lines in common are too few to derive accurate
 abundances.
  
The differences in absolute and relative abundances between
 both spectra are shown in Table~\ref{compar}.
We note that the differences in relative abundances are smaller than
the absolute differences (except for C, S and Y).
The agreement between the different relative abundances is for most
elements good ($<$ 0.2). For C, Sc, Zn, Y and Eu the differences
 are larger.
 However the Sc, Zn, Y and Eu abundances are only determined
on the basis of one line. For C we prefer the abundance of 
the hot model, since this abundance is based on many more lines.

The ratios [Zn/Fe] , 0.24 and 0.09, and [S/$\alpha$], 0.21 and 0.36,
 for the 2000 spectrum and 2001 spectrum respectively,
 are also in good agreement. The underabundances of the s-process
 elements are also confirmed by the 2000 spectrum.
However, the lines, used for the 2001 spectrum, are blended and can not 
be used for accurate abundance determination.

Fig.~\ref{abotemp} shows the abundance versus the
condensation temperature for the 2001 analysis.
The diagram for the 2000 analysis shows the same trends and would lead
 to the same conclusions. Up to a condensation temperature of
 about 1500 K, the abundances are approximately solar.
The elements with a higher condensation temperature, like the
 s-process elements, (Ca), Ti and Sc are clearly underabundant. There is,
 however, a large spread in underabundances.
IRAS\,17243-4348 belongs to group A (Lloyd Evans, 1999 and in preparation), which is in agreement
 with its chemical composition.

\citet{1992JApA...13..307G} also made an abundance analysis of
 IRAS\,17243-4348. They used a T$_{\rm eff}$ of 5500 K, a $\log g$ of 
0.5 and a $\xi_t$ of 5.5  \kms. Lines with an EW up to 450 m\AA\, are
 used in their analysis. We compared their relative abundances with ours
 derived from the spectrum taken in 2001 (see Table~\ref{compar}),
 because the used temperatures differ only by 250 K. For C, Na, Si, Ca,
 Cr, Fe and Ni the agreement is very good. Worse is the agreement
for Sc, Ti, Mn, Zn and Y. For the first two the uncertainty on their
 abundances is very large (0.82 for Sc, 0.29 for Ti~I and 0.38 for Ti~II).
 Their Y-abundance is based on one line. \\ \\

{\bf IRAS\,19125+0343}\\

[Zn/Fe]=$+$0.4 and [S/$\alpha$]=$+0.8$ point to a depleted photosphere. \\

{\bf IRAS\,19157-0247}\\

Many low excitation lines appear in emission in this object. 
These were not taken into account for the abundance analysis. As a
consequence, abundances are determined for few elements in 
 IRAS\,19157-0247. Fe is approximately solar, no clear trend is seen in 
the abundance-condensation temperature diagram and the underabundances
 are not large ([Ti/Fe]=$-$0.3, [Sc/Fe]=$-$0.5). In this way the evidence
for a depleted photosphere for IRAS\,19157-0247 is lacking.  \\

{\bf IRAS\,20056+1834 (QY\,Sge)}\\

The abundances of this object are taken from ~\citet{2002MNRAS.334..129K}.
They do not limit their lines towards a low EW. For Ba, for example, they
 use the 5853 \AA\, line with an EW of 288 m\AA.
We can see in Fig.~\ref{abotemp} that they obtain very similar trends
to those we observe for IRAS\,09144-4933, IRAS\,15469-5311, 
IRAS\,16230-3410, IRAS\,17243-4348 and IRAS\,19125+0343.
They note that `there is a clear tendency for the elements with a high
 T$_{\rm cond}$ to be underabundant relative to those of low 
T$_{\rm cond}$'. Furthermore, they remark that `the abundances anomalies
 are reminiscent of those observed in stars of group B,
but that, in contrast, iron and other elements are only slightly, if at
 all, depleted'.

%\onecolumn

\begin{table*}
\begin{center}
\caption{\label{synopsis} The synopsis of the abundance results. 
[Fe/H]$_{0}$ is the estimated initial metallicity on the
 basis of the Zn-abundance, if this abundance was believed to be 
unaffected. For the initial metallicity determination we made use of the
 relation : [Zn/Fe] =+0.04 from \citet{1991A&A...246..354S}.}
\vspace{0.5ex}
\begin{tabular}{rrrrrrrrrrrr} 
IRAS  & [Fe/H] & [Fe/H]$_{0}$ & [C/Fe] & [N/Fe] & [O/Fe] & C/O &  [Zn/Fe] & [S/$\alpha$] & [S/Ti] & [Zn/Ti] & [s/Fe]  \\	 
\hline 							 									 
\rule[0mm]{0mm}{3mm}					 							 
08544 & $-0.3$ &    0.0      & $ 0.0$ & $+0.2$ & $-0.2$   & /  & $+0.4$ & $+0.5$ & $+1.0$ & +0.9 & $<-0.3$  \\		 
09060 & $-0.7$ & $-$0.7      & $-0.2$ &   /    &  /       & /  & $+0.1$ & $-0.2$ & $ 0.0$ & +0.2 & $+0.1$  \\ 		 
09144 & $-0.3$ &    0.0      & $ 0.0$ & $+0.5$ & $-0.2$   & 0.3 & /      & $+0.6$ & $+1.3$ & /    & $<-1.1$  \\		 
09538 & $-0.6$ & $-$0.6      & $+0.4$ &  /     & $+0.9$   & 0.1 & $-0.1$ & $+0.1$ & $+0.3$ & +0.1 & $<-0.5$  \\ 		 
15469 & $ 0.0$ &    0.2      & $+0.3$ & $+0.6$ &   0.0    & 0.9 & $+0.3$ & $+0.9$ & $+2.1$ & +1.8 & $-0.6$  \\  		 
16230 & $-0.7$ & $-$0.4      & $+0.2$ & $+0.2$ & $+0.3$   & 0.4 & $+0.3$ & $+0.4$ & $+1.1$ & +1.0 & $-0.4$ \\  		 
17038 & $-1.5$ & /           & $+1.8$ &   /    &   /      &  /  & $+0.3$ &    /   &   /    & /    & $-0.4$  \\  		 
17233 & $-1.0$ & $-$0.2      & $+0.8$ & $+1.2$ & $+0.7$   & 0.6 & $+0.7$ & $+1.5$ & $+1.8$ & +1.4 & $<-0.3$  \\ 		 
17243 & $ 0.0$ &  0.2        & $-0.2$ & $-0.2$ & $-0.1$   & 0.4 & $+0.2$ & $+0.2$ & $+0.6$ & +0.8 & $-1.4$  \\ 		 
19125 & $-0.3$ &  0.1        & $+0.5$ &   /    & $+0.5$   & 0.5 & $+0.4$ & $+0.8$ & $>2.6$ & $>$2.3 & $<-0.3$  \\ 	 
19157 & $+0.1$ & $+$0.1      & $0.0$ &   /    & $+0.2$   & 0.3 &    /   & $+0.4$ & $+0.4$ &  /     & /  \\  	 
20056 & $-0.3$ & $-$0.2      & $+0.6$ & $+1.1$ & $+0.6$   & 0.5 & $+0.1$ & $+0.5$ & $+1.2$ & +1.2   & $-0.54$  \\ 	 
\hline							 									 
subdwarf &$-0.2$&$-0.2$      &  0.0   &   0.0  & $+0.1$   & 0.4 &   0.0   &  0.0   &  0.0   & 0.0    & 0.0  \\           
\\
\\
\end{tabular}
\end{center}
\end{table*}

%\twocolumn

\subsection{Abundance synopsis} 

In this section we discuss the synopsis of the abundance results,
 listed in Table~\ref{synopsis}. Note that the values for IRAS\,17243-4348
are from the spectrum, taken in 2000, except for [s/Fe].
In this table we include also the abundances
for an unevolved mildly metal-poor dwarf ([Fe/H]=$-$0.2)
\citep{1985pdcn.conf..131N}. These 
abundances reflect the chemical composition of the ISM out of which the 
stars (with an initial metallicity of $-0.2$) were formed. Differences
 in abundances between the programme
stars and these abundances are the consequence of the changes in 
the chemical composition of the photosphere during the star's evolution.
However, conclusions are not easily drawn on the basis of this comparison.
First of all, it is possible that the observed metallicity 
does not correspond with
the initial abundance of the star, because the depletion process has
 lowered the Fe-abundance. Secondly, abundance changes can be due to two 
processes : the depletion process and/or nucleosynthesis processes in 
the deep layers of the star, followed by deep mixing.  \\ 

{\bf Metallicities} \\

The observed metallicities range between $-1.5$ and $+0.1$. 
We estimated the initial metallicity [Fe/H]$_{0}$ on the
 basis of the Zn-abundance, if this abundance was believed to be 
unaffected. For the initial metallicity determination we made use of the
 relation :
 [Zn/Fe] =+0.04 from \citet{1991A&A...246..354S}.
Only for IRAS\,17038-4815 and IRAS\,17233-4330 ([Zn/Fe]=+0.7),
  was the metallicity greatly affected. Since for IRAS\,17038-4815 also 
the Zn-abundance is (probably) affected, it is difficult to trace the initial
 metallicity. The estimated initial metallicity
 of IRAS\,17233-4330 is $-0.2$.
For IRAS\,09060-2807 ([Fe/H]=$-0.7$), IRAS\,09538-7622 ([Fe/H]=$-0.6$) and
 IRAS\,19157-0247 ([Fe/H]=$+0.1$) the metallicity reflects the initial
 composition, because the abundances were not affected by a depletion 
process.
 For all the other stars, the influence of the depletion process on the
 metallicity was rather small. \\ 

{\bf Depletion} \\

Based on the depletion process, discussed in the previous section,
we can divide the programme stars in three groups. 

In a first group, which consists of three objects, IRAS\,09060-2807,
IRAS\,09538-7622 and IRAS\,19157-0247, there is no evidence that a depletion
process has taken place. IRAS\,09538-7622 is a genuine RV\,Tauri star
of spectroscopic group A.

The second group contains IRAS\,08544-4431, IRAS\,09144-4933,
IRAS\,15469-5311, IRAS\,16230-3410, IRAS\,17243-4348, IRAS\,19125+0343
and IRAS\,20056+1834. In these stars the elements with a condensation 
temperature up to about 1500 K (like Fe) are not or only moderately
 depleted. Above that temperature most elements (the s-process elements, 
except for Ba, Ti and Sc) are clearly underabundant. We call this pattern in 
the abundance versus condensation temperature diagram,
the 'flat-steep' pattern in the following.
This group contains also the genuine RV\,Tauri stars of 
spectroscopic group A, IRAS\,16230-3410
and IRAS\,17243-4348 (Lloyd Evans 1999 and in preparation).

The last group exists of the two genuine RV\,Tauri stars of spectroscopic
group B:
 IRAS\,17038-4815 and IRAS\,17233-4330. A depletion process
 has strongly changed their abundances.
For IRAS\,17233-4330 the observed iron abundance differs 
 from the estimated initial metallicity by $-0.8$.
This was expected as all RV\,Tauri stars of spectroscopic
group B, studied till now, are depleted. \\

{\bf CNO abundances } \\

In all stars (except IRAS\,17243-4348) [N/Fe] is positive, but only in 
IRAS\,15469-5311, IRAS\,17233-4330 and IRAS\,20056+1834 there is evidence
that it is not only
 due to the decrease of the Fe-abundance but also the result of an 
enhancement in N.
The enhancement of the N abundance is most likely due to CNO-cycling 
and first dredge-up on the giant branch.

All C/O ratios are smaller than 1, so all stars are oxygen rich.
The observed C/O ratios are, within the errors, solar, except for 
IRAS\,15469-5311 (C/O =0.9) and IRAS\,09538-7622 (C/O = 0.1).
 A higher than solar C/O ratio
 can be caused by C-nucleosynthesis on the AGB and subsequent dredge-up.
 Re-accretion of CO molecules will also change the C/O ratio but it
is unlikely that a C/O ratio $> 1$ can be expected : CO is the most
abundant molecule after H$_{2}$ and after efficient re-accretion C$\sim$O
is expected.
The almost solar C/O ratio for most of the stars can best be explained 
by the fact that no C-nucleosynthesis, followed by deep mixing, took place
and that the depletion process operated only marginally in most of these 
stars. Together with the already mentioned small decrease of the iron
 abundance, also the amount of accreted material was low.
  Consequently, the effect on C/O is small.

Since the depletion process has also been marginally active
 in IRAS\,15469-5311, it is not likely that it is the cause of the high C/O
 ratio. C-nucleosynthesis on the AGB is mostly accompanied by the
 production of s-process elements.
However, Ba and Y are the only s-process elements for which we could 
determine  abundances, and both are underabundant.

The low C/O for IRAS\,09538-7622 is probably caused by the high
 O-abundance, which is based on only one line. \\

{\bf Iron peak elements} \\

Both Cr and Ni have a condensation temperature close to that of Fe 
(T$_{\rm cond}$ = 1337 K), so the effects of the depletion process on 
the abundances of these
elements, should be the same. This is confirmed in all stars in which
 the depletion process was active. Also in the other stars Cr and
 Ni follow the Fe abundance. Zn (T$_{\rm cond}$ = 684 K)
 has a much lower condensation temperature 
than the other iron peak elements. For this reason [Zn/Fe] is often used 
as a prime indicator for the depletion process.
In the stars of the first group the low [Zn/Fe] ($-$0.1 or +0.1)
 points to the inactivity of the depletion process. In the second
group [Zn/Fe] varies between +0.1 and +0.4, which is rather low.
This can be explained as follows.
In these stars elements with a 
condensation temperature up to 1500 K, are  slightly or not at all
 affected by the depletion process. Only elements with a condensation 
temperature higher than 1500 K are underabundant. As a consequence
the 'flat-steep' pattern in these stars is not traced by [Zn/Fe] but
more by [Zn/Ti] or [S/Ti] for instance.
The low [Zn/Fe] of +0.3 for IRAS\,17038-4815 is due to the fact that
also the Zn-abundance is likely affected. \\

{\bf $\alpha$ elements} \\

Metal-poor unevolved dwarfs show $\alpha$-overabundances
 ([$\alpha$/Fe]=+0.2 for [Fe/H]=$-0.7$ \citealt{1989ARA&A..27..279W}), reflecting the chemical history of our Galaxy.
 As the lowest initial metallicity in our programme stars
is probably $-0.7$ (for IRAS\,17038-4815 we can not estimate an initial
 metallicity), this effect is small. Moreover, the depletion process 
dilutes this effect, as $\alpha$ elements have different condensation
 temperatures: S : 674 K, Mg : 1340 K, Si : 1340 K,
  Ca : 1634 K, Ti : 1600 K. [S/$\alpha$] is used to trace the 
depletion process. For stars of the second group, [S/$\alpha$] compares 
the abundance of S with that of two unaffected elements (Mg, Si) and two
 affected elements (Ca, Ti). Therefore it
is not characteristic for  this kinds of depletion.
[S/Ti] seems therefore a better indicator for this 'flat-steep' pattern. \\

{\bf s-process elements} \\

We do not detect s-process enhancements in any of our objects. However, since
s-process elements have a high  T$_{\rm cond}$ and the depletion process
 was  active in most of the stars, possible s-process overabundance
 could be  diluted.
Nevertheless, if we compare the s-process abundances with the abundances
 of elements with the same condensation temperature, the s-process 
elements are not enriched. In this way we can conclude that there is no 
s-process evidence for AGB-nucleosynthesis
and third dredge-up. In contrast, the s-process underabundances support
in most stars the idea that a depletion process has taken place.

\section{Discussion}
\label{discussion}

\subsection{Abundance patterns}

The stars of our new sample have been selected on the basis of a large and broad IR excess.
We limited our chemical abundance analysis to the stars with spectral
type F and G. Clear chemical patterns which could be due to dredge-up
processes during stellar evolution were not observed and we focus our
discussion on depletion.

We observe three kind of depletion patterns. In the first
group no depletion is observed, the metallicity corresponds to
 the initial metallicity. In the second
group the picture is more complicated. Elements with a T$_{\rm cond}$ below
1500 K are not or only slightly affected. Large underabundances are
measured only for most elements with a T$_{\rm cond}$ above 1500 K. 
The third group contains two stars, in which the depletion process has
 considerably changed the chemical composition.

The uncertainties in the relative abundances, induced by the
 uncertainties in the model parameters, derived in
 Section~\ref{uncertain}, can not
explain the low relative abundance of the elements with a high
condensation temperature versus those with a low condensation temperature,
observed in group 2 and 3 of the programme stars.
Non-LTE effects are, however, not excluded and their impact on the
abundances is difficult to assess.
Such a non-LTE effect was put forward to explain 
s-process deficiencies in low-mass supergiants, which included type II
 Cepheids, 
RV\,Tauri stars and supergiants at a high latitude \citep{1989ApJ...342..476L}. 
They argued that in low mass variables a shock can be present, causing the production
 of Lyman-continuum photons. These photons lead to an overionisation of 
elements with a low second ionisation potential (IP)
into the second-ionised state. This results in an underabundance for those
 elements in an LTE-analysis, which are mainly based on single ionised
lines.
All s-process elements together with Sc, Ca and Eu have low second IPs
 ($<$ 13.6 eV). Ti with a second IP of 13.58 eV is right at the Lyman edge.

\begin{figure}
\begin{center}
\caption{\label{ionabo} The abundances of the elements 
versus their second ionisation potential for IRAS\,16230-3410 ($\Delta V=0.46 $),
 IRAS\,17233-4330 ($\Delta V=0.99 $) and IRAS\,09538-7622 ($\Delta V=1.35 $) (Maas et al., in 
preparation). The dashed line represents the Lyman limit.}
\resizebox{6.5cm}{6.5cm}{\includegraphics{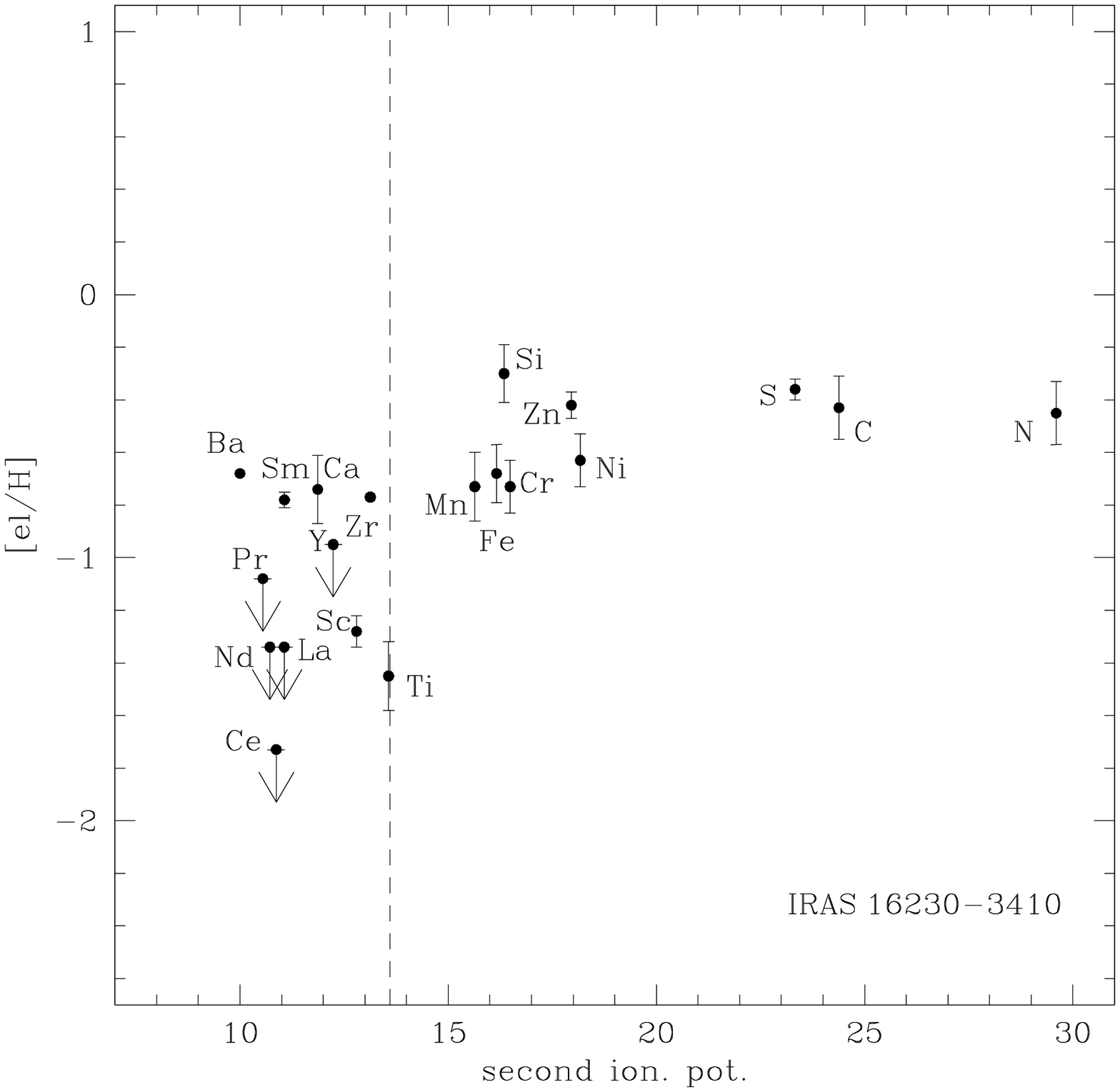}}
\resizebox{6.5cm}{6.5cm}{\includegraphics{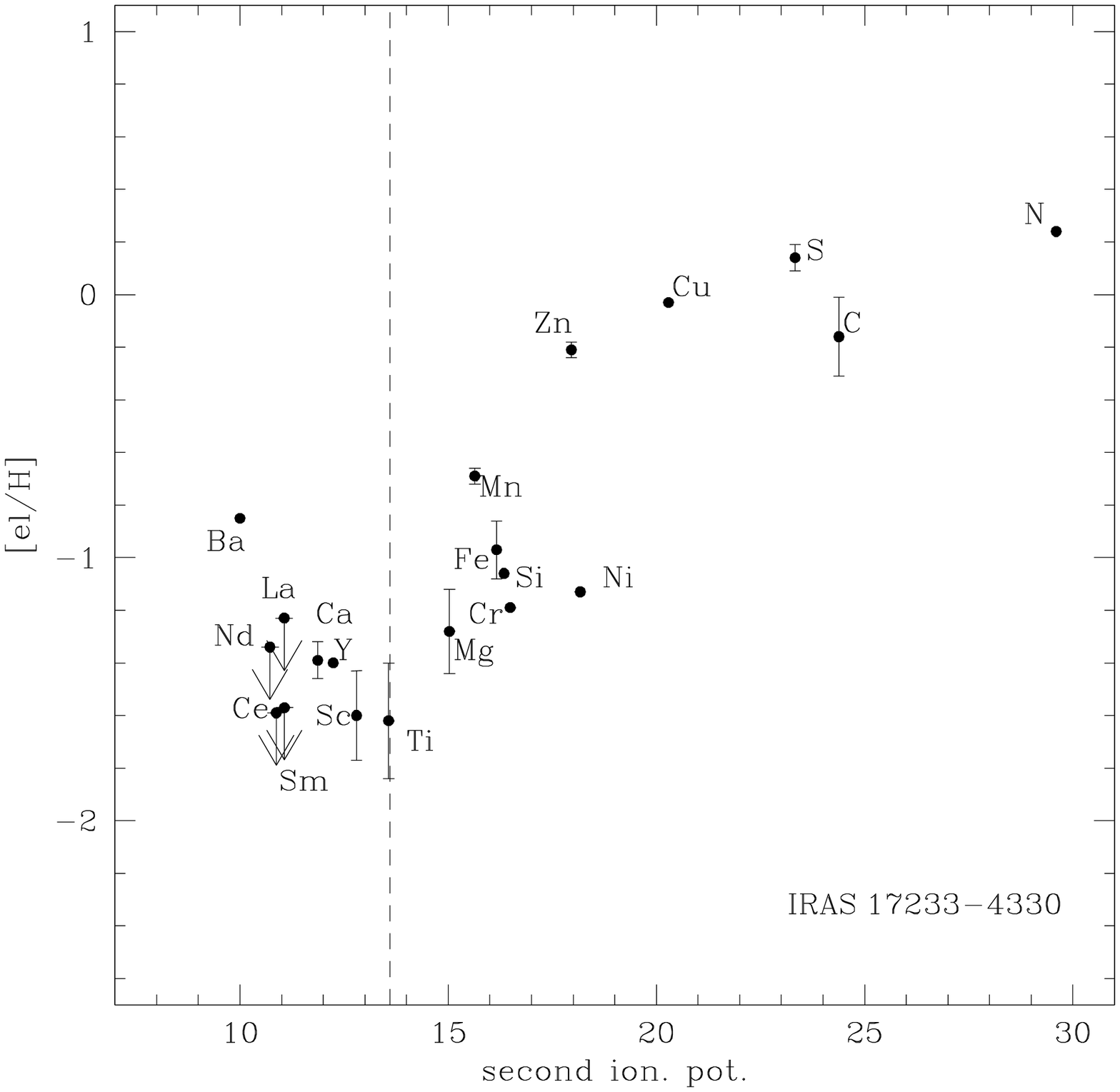}}
\resizebox{6.5cm}{6.5cm}{\includegraphics{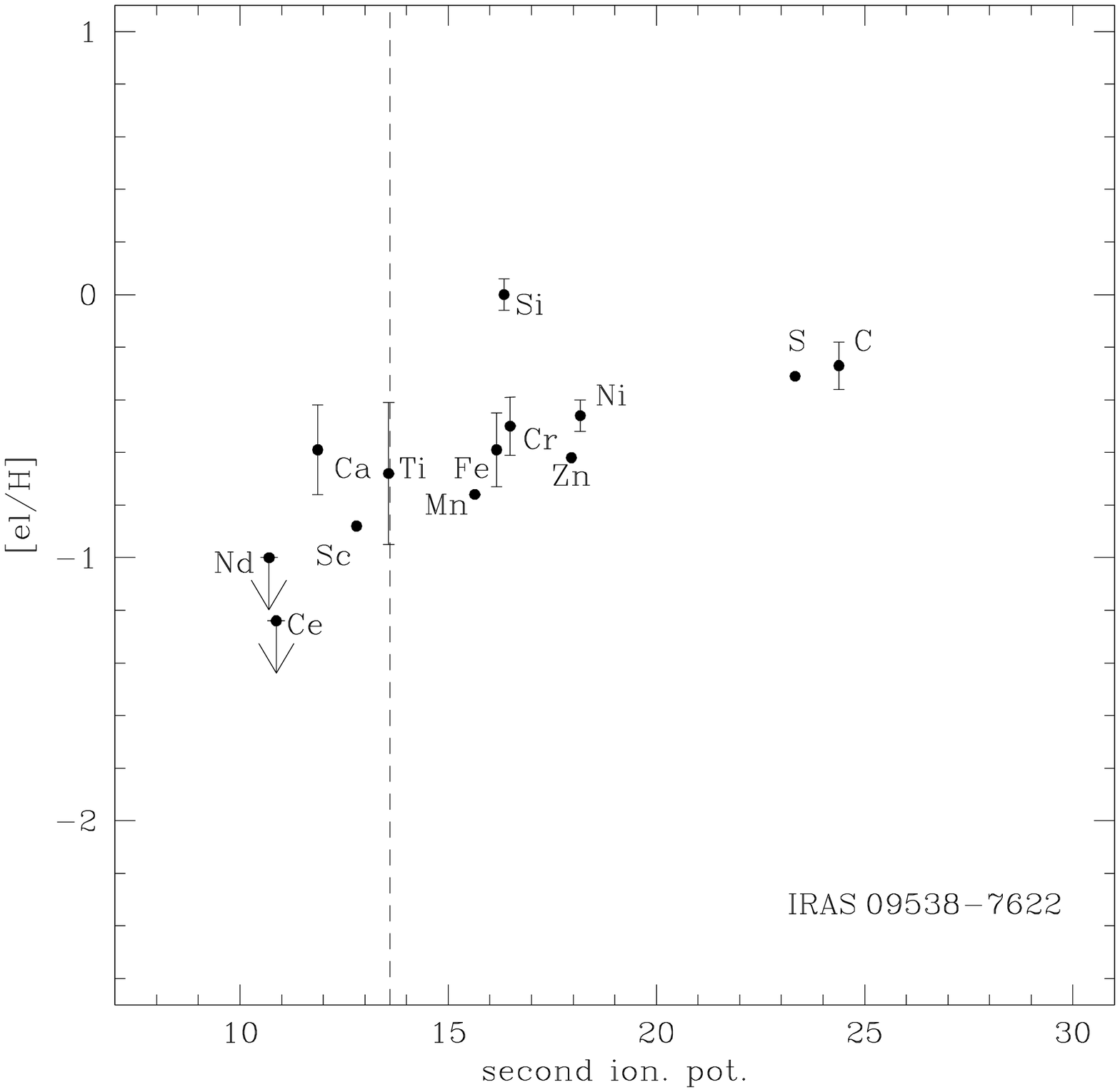}}
\end{center}
\end{figure}

In Fig.~\ref{ionabo} we plotted for three of our programme stars, 
IRAS\,16230-3410, IRAS\,17233-4330 and IRAS\,09538-7622, the abundances of the elements
versus their second IP. For IRAS\,16230-3410 the strong underabundances
are only observed for the elements with a second IP below 13.6 eV. However,
for elements with a second IP below 13.6 eV, there is a large scatter in
underabundances. This is also the case if one plots
the abundances versus condensation temperature (see Fig.~\ref{abotemp}).
This non-LTE effect can, however, not explain the small but significant
[Zn/Fe]=+0.3. Both Zn and Fe have a second IP larger than 13.6 eV and
their abundances can not be affected by this non-LTE effect. The same is,
to a larger extent, true for [Zn/Fe] in IRAS\,17233-4330. Also the 
underabundances of Mn, Si, Ni, Cr, Mg can not be explained, since their second
IP is higher than 13.6 eV.

If one looks only at the abundance patterns, this non-LTE effect can
explain at least a part of the observed abundances for the programme stars 
with a flat-steep pattern.  However,
essential  to affect the population of the different
ionisation stages of the elements is of course the production of
Lyman-continuum photons. The presence of a shock, creating these photons, is
 doubtful in many of our programme stars, in for example IRAS\,15469-5311, 
for which the photometric variation is small ($\Delta V$ = 0.17, Maas et al. in 
preparation) .
If this non-LTE effect would apply for IRAS\,15469-5311, it is then surprising
that it does not lead to underabundances in IRAS\,09538-7622 (see Fig.~\ref{ionabo}),
 for which the peak-to-peak photometric variation in V is high ($\Delta V$ =1.35,
Maas et al. in preparation) and 
shock caused line-deformation is observed in the cross-correlation profiles,
which is not seen in IRAS\,15469-5311 (Maas et al., in preparation).

Another way to discriminate between both explanations are the
elements Ba and the r-process element Eu. They have both a low second 
ionization potential, but their condensation temperature (T$_{\rm cond}$ =1162 
and 1338 K, respectively) is not as high as for the other s-process elements. 
In IRAS\,08544-4431, IRAS\,16230-3410, IRAS\,17233-4330 and IRAS\,20056+1834,
Ba scales with iron while it is underabundant in IRAS\,15469-5311
 ([Ba/Fe]=$-0.6$). Eu is solar in IRAS\,17243-4348 and IRAS\,20056+1834. 
Thus, the abundances of Ba and Eu are, in general,
 in favour of the depletion process.
However, it must be stated that the Ba and Eu abundance are based on only one
line and that the Eu line can be affected by hyper-fine splitting. Moreover, 
in the abundance versus condensation temperature diagram the scatter at a
 certain condensation temperature is sometimes also large.

Finally, we point to the fact that the underabundances found here are much
 larger than those found by \citet{1989ApJ...342..476L}, which report 
underabundances, by a factor of about 2 (0.3 in logarithmic units). 
The underabundances for our stars attain values less than  $-2.1$ for
 IRAS\,19125+0343.

We conclude that we found no observational evidence for this strong non-LTE effect
to be active.

%More in general, if a shock produces Lyman photons, which change the 
%population of hte ionisation levels, this would also affect the abundances
%of the  RV\,Tauri stars. The low observed Ca, Sc and Ti abundances (see
%table~\ref{overdep}) are then not real and not caused by the gas-dust separation
%process. Note, however, that not all abundances are sensible for this effect :
%for instance, the for depletion characteristic [Zn/Fe]; both elements 
%have a high second ionisation potential.
% Since the scale of the depletion is not the same in
%all RV\,Tauri stars, this implicates that the shock is stronger in the  
%strongly depleted RVB than in the RVA and RVC stars. Shock intensity
%variations from cycle to cycle are noted 
%\citet{1997MNRAS.286....1P,1990A&A...237..159G,1989A&A...215..316G},
% but comparing studies in shock intensity between RV\,Tauri stars,
%are not undertaken and not straight forward.
%\citet{2000ApJ...531..521G} remark for the s-process underabundances in
%R\,Scuti, that a relation with condensation temperature is much more
%obvious than one with a second ionisation potential. The five heavy
%elements in their analysis (La, Ce, Nd, Sm and Eu)
% span a range of 1 dex in [X/Fe], but differ in
%ionisation potential only 0.5 eV. In condensation temperature they differ
%more than 200 K.

\subsection{Accretion scenarios}

Abundance anomalies resembling the chemical composition of the gas phase
of the ISM are also observed in $\lambda$ Bootis. These are stars
with effective temperature between 8000-10000 K and population I kinematics.
They are often surrounded by the remains of their formation disc. The scale
of the depletion in these stars is comparable with that in 
RV\,Tauri stars of spectroscopic group B.
\citet{1991IAUS..145..341B} proposed that selective removal of
metals from the photosphere through grain formation 
and mass loss could account for the observed abundance anomalies in the 
strongly depleted post-AGB stars. The photosphere is, however, too hot
for dust formation, so the gas-dust separation has to take place in the CS
environment. \citet{1992A&A...259L..39M} pointed out two different
possible accretion scenarios and \citet{1992A&A...262L..37W} stressed
that the most likely place for the depletion process to take place is
a stationary disc. In this scenario, the gas falls back on the star at a low 
accretion rate.
The radiation pressure exerts an outward force on the dust and keeps it in the 
disc; the gas is, in contrast, not hampered by the radiation pressure 
under the  condition that the density is low enough so that the gas
is not dragged along with the dust.
The observational evidence for the scenario of \citet{1992A&A...262L..37W} at
 that time
were the discs surrounding the $\lambda$ Boo stars  and the binarity of
the two peculiar post-AGB binaries HR\,4049 and BD+39$\degree$4926.
 In the latter systems it
 is likely a disc has formed during mass-transfer at the AGB. Later on,
the finding that all severely depleted post-AGB stars are binaries 
\citep{1995A&A...293L..25V} and the observed dust discs around some of them
\citep{1991A&A...242..433W,1995ApJ...443..249R,1997IAUS..180..211B},
 enlarged the observational arguments for this scenario.
This scenario became less clear after the detection of many more depleted
systems \citep{2000ApJ...531..521G,1994ApJ...437..476G,1998ApJ...509..366G,1997AJ....114..341G,1996MNRAS.280..515G,1997ApJ...479..427G,1998A&A...336L..17V}.
\citet{2000ApJ...531..521G} considers two possible sites for the dust
formation and dust-gas separation in RV\,Tauri stars : the dusty
stellar wind in a single star or a dusty circumbinary disc.

Different kinds of data revealed that
 the prototypical RV\,Tauri star AC\,Her
strengthens the link between the RV\,Tauri stars and the strongly 
depleted post-AGB stars and, as a consequence, the evidence for
a circumbinary disc to be the most likely place for the gas-dust separation.
Radial velocity monitoring shows that AC\,Her is a spectroscopic binary.
Its photosphere is depleted and the SED of AC\,Her points to the presence
of hot as well as cold dust \citep{1998A&A...336L..17V}.
 Moreover, ISO-SWS spectra threw light on the 
oxygen rich nature  of CS material of which a high amount is crystalline
 \citep{1999A&A...350..163M}. Also the similarity with the ISO-SWS spectrum
of the comet Hale-Bopp is striking. The CO-emission profile is 
different from that observed for post-AGB stars with an outflow 
\citep{1988A&A...206L..17B}. Last but not least, the disc around AC\,Her
is resolved in the mid-IR \citep{2000ApJ...541..264J}. Recent observations
could, however, not confirm this \citep{biller2003}.

In this paper we analysed the chemical composition of stars with spectral
type F and G, which were selected on the basis of a large IR excess : the object
is surrounded by hot and cold dust. The most probable geometry of the dust
around a post-AGB star with spectral type F and G is a disc.
This work reveals that a high fraction (9 out of 12  or 75 $\%$) of the sample 
shows the signature of depletion, indicating that a disc is a preferential 
site for the depletion process to take place. This strengthens the scenario
of \citet{1992A&A...262L..37W}.  However, a disc is clearly not a
 sufficient condition for the dust-gas 
separation to take place, since three objects are not affected.
In 89\,Her there is also evidence for a disc \citep{1993A&A...269..242W}, but no
 depletion is observed \citep{1990ApJ...357..188L}.

\citet{2003A&A...397..595D} outline that grain settling in the disc may play
an essential role for accretion to take place.
The larger grains will eventually settle in the midplane of
the disc. The gas will then have a higher scale height than the dust. The dust free
 gas layers cease to experience an outward drag by the dust and, subsequently, 
fall back on the star.

Since the temperature in the disc decreases with increasing radius,
 at the inner radius of the disc only the elements
 with a high condensation temperature are caught in grains,
 given that the inner temperature of the disc
 is high. This could possibly give an explanation for the `flat-steep'
  depletion pattern observed in group 2 of the programme stars.
 When grains settle in the midplane, gas falls back on the star.
 This gas contains all elements except 
those with the highest condensation temperature. The accretion of this gas
gives rise to the flat-steep pattern. 

%wide range of severity is observed.
%? function of convection, amount of material that falls back,
%amount of refractory elements that condense onto dust grains, mass loss.
%not well known

\subsection{Depletion and RV\,Tauri stars}

In what follows we compare our results concerning the observed chemical patterns
with what is known from the chemical composition of RV\,Tauri stars and other
depleted objects.

Table~\ref{overdep} shows the abundances for all analysed RV\,Tauri stars, the
stars analysed in this paper, the strongly and moderately depleted post-AGB binaries,
the type II Cepheid ST\,Pup and $\lambda$ Boo.

\citet{2000ApJ...531..521G} studied the chemical abundances of the RV\,Tauri
 stars and clearly established with this last paper in a series that depletion 
is widely observed in RV\,Tauri stars and is not just limited to a select
 club of
 peculiar objects (i.e. the severely depleted post-AGB binaries). In this publication
nine of the twelve studied objects are affected by the depletion process.
 In this way, we subscribe that the depletion process is a widespread 
phenomenon.

\citet{2000ApJ...531..521G} pointed out a difference in chemical composition between the 
spectroscopic classes : the RV\,Tauri stars of spectroscopic
group A are not or are only
 moderately affected, the stars of group B are severely affected and
those of group C turn out to be immune for the depletion process.

%\onecolumn

\begin{table*} 
\caption{\label{overdep} Abundances of the RV\,Tauri stars, our programme stars,
the severely and mildly depleted post-AGB binaries, the Population II Cepheid
ST\,Pup and $\lambda$ Boo. The second column refers to the spectroscopic class
of the RV\,Tauri stars. We also included the T$_{\rm eff}$. For the genuine
RV\,Tauri stars we took the mean temperature, if more than one temperature was measured
during the pulsation cycle.}
\begin{center}
%\resizebox{\hsize}{!}{
\begin{tabular}{lrrrrrrrrrr}
        & class   & [S/H]  & [Zn/H]  & [Sc/H]  & [Ti/H] & [Fe/H] & [Fe/H]$_{0}$ & T$_{\rm eff}$(K) & ref. \\
\hline
\rule[0mm]{0mm}{3mm}DY\,Aql & A &        &       &  $-$2.1 && $-$1.0 &&4250 & 1     \\
TW\,Cam & A &    0.0 & $-$0.4  &  $-$0.5 & $-$0.7 & $-$0.5 &  $-$0.4 & 4800 & 2\\
SS\,Gem & A & $-$0.4 &    0.0  &  $-$1.9 & $-$2.0 & $-$0.9 &  $-$0.2 & 5600 & 1\\
U\,Mon  & A & $-$0.1 & $-$0.6  &  $-$1.0 & $-$0.6 & $-$0.8 &  $-$0.5 & 5000 & 2\\
TT\,Oph & A &    0.0 & $-$0.8  &  $-$1.1 & $-$0.8 & $-$0.8 &  $-$0.8 & 4800 & 2\\
R\,Sct  & A &        & $-$0.3  &  $-$1.4 & $-$0.4 & $-$0.4 &  $-$0.3 & 4500 & 2\\
R\,Sge  & A &    0.4 & $-$0.2  &  $-$1.5 & $-$1.3 & $-$0.5 &     0.1 & 5100 & 3\\
RV\,Tau & A &        &    0.0  &  $-$0.3 & $-$0.5 & $-$0.4 &  $-$0.4 & 4500 & 2\\
CE\,Vir & A &        & $-$0.7  &  $-$2.6 & $-$1.3 & $-$1.2 &  $-$0.7 & 4400 & 1\\
AD\,Aql & B &    0.0 & $-$0.1  &  $-$1.8 & $-$2.6 & $-$2.1 &  $-$0.1 & 6300 & 4\\
UY\,Ara & B &    0.0 & $-$0.4  &  $-$1.7 &        & $-$1.0 &  $-$0.3 & 5500 & 2\\     
IW\,Car & B &    0.4 & $-$0.1  &         &        &        &  $+$0.2 & 6700 & 5\\
RU\,Cen & B & $-$0.7 & $-$1.0  &  $-$1.9 & $-$2.0 & $-$1.9 &         & 6000 & 6\\	     
SX\,Cen & B & $-$0.1 & $-$0.5  &  $-$2.0 & $-$2.0 & $-$1.1 &	    &  6250 & 6\\
AC\,Her & B & $-$0.4 & $-$0.9  &  $-$1.7 & $-$1.6 & $-$1.4 & $-$0.7 &  5900 & 4\\
EP\,Lyr & B & $-$0.6 & $-$0.7  &  $-$2.1 & $-$2.0 & $-$1.8 & $-$0.7 &  6100 & 3\\
CT\,Ori & B & $-$0.5 & $-$0.6  &  $-$2.6 & $-$2.5 & $-$1.9 & $-$0.6 &  5750 & 1\\
DY\,Ori & B &    0.2 &    0.2  & $<-$2.7 &        & 	   & $+$0.2 &  6000 & 3\\
AR\,Pup & B &    0.4 &         &  $-$2.2 &        & 	   & $+$0.2 &  6000 & 3\\
DS\,Aqr & C & $-$0.8 & $-$1.1  &  $-$1.2 &        & $-$1.1 & $-$1.1 &  6500 & 2\\
V360\,Cyg&C & $-$0.9 & $-$1.4  &         & $-$1.3 & $-$1.4 & $-$1.3 &  5300 & 4\\
BT\,Lib & C & $-$0.8 & $-$1.1  &  $-$0.7 &        & $-$1.1 & $-$1.1 &  5800 & 2\\
V453\,Oph&C &    &         &  $-$2.2 & $-$1.9 & $-$2.2 & $-$2.2 & 5800 & 4\\
\hline
\rule[0mm]{0mm}{3mm}IRAS\,08544 &   &    0.1 &    0.1  & $-$0.9 & $-$0.9 & $-$0.3 &    0.0 & 7250 & 6\\
IRAS\,09060 &   & $-$0.7 & $-$0.5  & $-$0.8 & $-$0.7 & $-$0.7 & $-$0.7 & 6500 & 6\\
IRAS\,09144 &   &    0.0 &         & $-$1.6 & $-$1.3 & $-$0.3 &    0.0 & 5750 & 6\\
IRAS\,09538 & A & $-$0.3 & $-$0.6  & $-$0.9 & $-$0.7 & $-$0.6 & $-$0.6 & 5500 & 6\\
IRAS\,15469 &   &    0.5 &    0.2  & $-$1.6 & $-$1.6 &    0.0 &    0.2 & 7500 & 6\\
IRAS\,16230 & A & $-$0.4 & $-$0.4  & $-$1.3 & $-$1.4 & $-$0.7 & $-$0.4 & 6250 & 6\\
IRAS\,17038 & B &        & $-$1.2  & $-$2.0 & $-$1.4 & $-$1.5 & /      & 4750 & 6\\
IRAS\,17233 & B &    0.1 & $-$0.2  & $-$1.6 & $-$1.6 & $-$1.0 & $-$0.2 & 6250 & 6\\
IRAS\,17243 & A &    0.0 &    0.2  & $-$1.1 & $-$0.6 &    0.0 &  0.2   & 5750 & 6\\  
IRAS\,19125 &   &    0.5 &    0.1  &$<-$1.4 &$<-$2.1 & $-$0.3 &  0.1   & 7750 & 6\\   
IRAS\,19157 &   &    0.2 &         & $-$0.4 & $-$0.2 & $+$0.1 & $+$0.1 & 7750 & 6\\ 
IRAS\,20056 &   &    0.1 & $-$0.1  & $-$0.7 & $-$1.0 & $-$0.3 & $-$0.2 & 5850 & 7\\
\hline
\rule[0mm]{0mm}{3mm}HR\,4049 &  & $-$0.4 & $-$1.3  &&& $-$4.8 & $-$0.4 & 7600 & 8,9\\
HD\,52961 & & $-$1.0 & $-$1.4 &          &        &              $-$4.8 & / & 6000 & 10,11\\
HD\,44179 & & $-$0.3 & $-$0.6 &&& $-$3.3 & $-$0.6 & 7500 & 8,12\\
BD+39$\degree$4926 & & $-$0.1 &         & $-$3.0 & $-$3.3 &  $-$2.9 & $-$0.1 & 7500 & 13\\
HD\,46703 & & $-$0.4 & $-$1.4  &&& $-$1.6 & $-$0.4 & 6000 & 14,15 \\
HD\,213985 & &   0.4 &        &    $-$1.3 & $-$1.5& $-$0.9 & / & 8200 & 8 \\
\hline
\rule[0mm]{0mm}{3mm}ST\,Pup       & & $-$0.2 & $-$0.1 & $-$2.2 & $-$2.2 & $-$1.5 & $-$0.1 & 5500 & 16 \\
$\lambda$ Boo & &   &  &  & $-$2.0&   $-$2.0     & / & 8650 & 17\\
\hline
\multicolumn{10}{l}{\small{(1): \citealt{1997ApJ...481..452G}; (2): \citealt{2000ApJ...531..521G}; (3): \citealt{1997ApJ...479..427G}; (4): \citealt{1998ApJ...509..366G}}} \\
\multicolumn{10}{l}{\small{(5): \citealt{1994ApJ...437..476G} (6) : this work; 
 (7) \citealt{2002MNRAS.334..129K}; (8) : \citealt{hansthesis}}} \\
\multicolumn{10}{l}{\small{(9): \citealt{2002PASJ...54..765T}; 
(10) : \citealt{1991A&A...251..495W}; (11) : \citealt{1992Natur.356..500V}}} \\
\multicolumn{10}{l}{\small{(12) : \citealt{1996A&A...314L..17W} (13) : \citep{1973A&A....22..273K}
; (14) : \citealt{1984ApJ...279..729L}}} \\
\multicolumn{10}{l}{\small{(15) : \citealt{1987ApJ...312..203B}; (16) : \citealt{1996MNRAS.280..515G}; (17) : \citealt{1990ApJ...363..234V}}}
\end{tabular}
\end{center}
\end{table*}

%\twocolumn

In our sample, five genuine RV\,Tauri stars are present.
In the stars of group A IRAS\,16230-3410 and IRAS\,17243-4348 strong
 underabundances
are observed only for elements with a condensation temperature above 1500 K,
while no underabundances are observed in IRAS\,09538-7622.
The chemical compositions of the group B stars
 IRAS\,17038-4815 and IRAS\,17233-4330 shows signatures of severe depletion. 
These findings confirm their result.

Another of their results is that not only the scale of the depletion differs
 from star to star, but also the pattern in the abundance versus condensation
temperature diagram. The RV\,Tauri stars
 were divided in several subgroups, according to the pattern observed.
The `flat-steep' pattern we found in a large number of stars, is very similar
 to the one observed in R\,Scuti. DY\,Aql and CE\,Vir are possible members of
 this subgroup of R\,Scuti , but additional element abundances are needed.
In the abundance synopsis of our stars we concluded that the
 abundance ratio [S/Ti] is a good tracer of the `flat-steep' pattern.

The difference in effective temperature between the RV\,Tauri stars
of groups A and B is, 
according to \citet{2000ApJ...531..521G}, the reason why stars of group B
 are heavily affected 
and stars of group A only moderately or not at all.
 The lower effective temperature of RV\,Tauri stars of group A
results in a deeper convective envelope and dilutes the possible
 abundance anomalies. However, it is then not clear why some elements
like Ti and Sc in some stars of group A are as underabundant as in 
RV\,Tauri stars of group B
(see Table~\ref{overdep}).
The stars IRAS\,08544-4431, IRAS\,15469-5311 and IRAS\,19125+0343, which
 show the `flat-steep' depletion pattern, are, on
the basis of this, chemically similar to some RV\,Tauri stars of group A.
 However, with their T$_{\rm eff}$ of 7250 K, 7500 K and 7750 K
they are clearly hotter than the RV\,Tauri stars of group B, 
so convection is not a
 possible explanation for the lower depletion scale. The ongoing mass-loss,
established by the P-Cygni profile of \halfa\, in all three stars, maybe plays
 a role. 

The minor role of the effective temperature on the strength
of the depletion process is further established in Fig.~\ref{tempdep}. 
In this figure we plot 
for all depleted evolved stars the ratios  [Zn/Fe] and [S/Ti] versus T$_{\rm eff}$.
The open boxes correspond to stars with large pulsational amplitudes,
while the full circles are stars with small pulsational amplitudes.
For the former a mean effective temperature is difficult to determine due to
its large variation during the pulsation cycle.
No correlation is present between [Zn/Fe] and T$_{\rm eff}$ , while 
the variation in [S/Ti] around a T$_{\rm eff}$ of 6000 K is 2.
The strongly depleted post-AGB binary HD\,52961 illustrates this. With a
 T$_{\rm eff}$ of 6000 K and a pulsation period of 70.8 days \citep{1991A&A...251..495W,1995AJ....110.3010F},
 it is very similar to the RV\,Tauri stars. However, with a
 [Zn/Fe]=+3.4 it is much more depleted than any RV\,Tauri star 
(the highest [Zn/Fe]=+2.0 for AD\,Aql).
A chemical analysis of the cooler objects in our sample in the future
will allow us to check these findings. 

\begin{figure}
\begin{center}
\caption{\label{tempdep}The ratios [Zn/Fe] and [S/Ti] versus T$_{\rm eff}$ for all
depleted evolved stars, listed in Table~\ref{overdep}. The open boxes indicate
the stars with large pulsational amplitudes, while the full circles are stars
with small pulsational amplitudes.}
\resizebox{6cm}{6cm}{\includegraphics{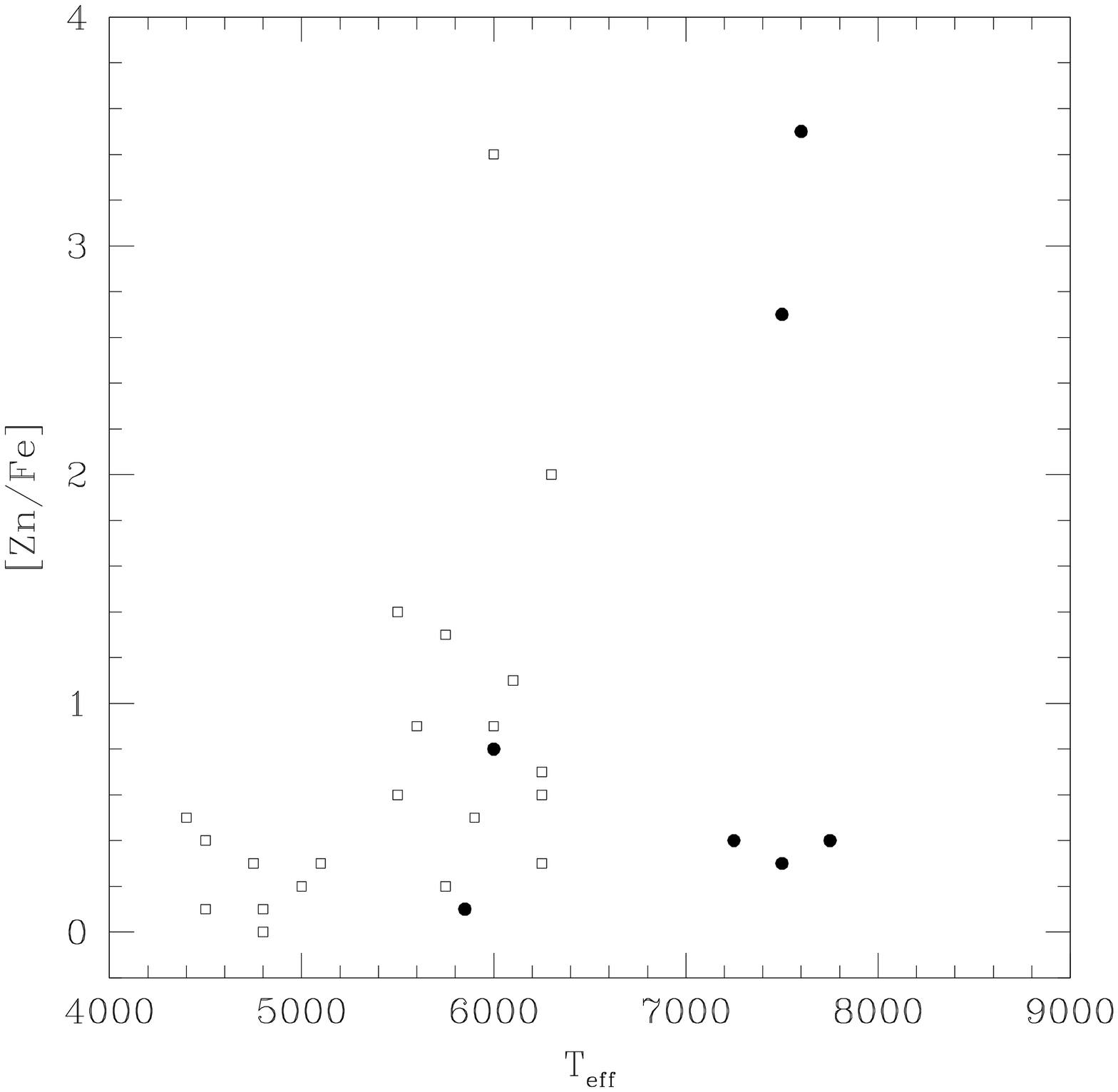}}
\resizebox{6cm}{6cm}{\includegraphics{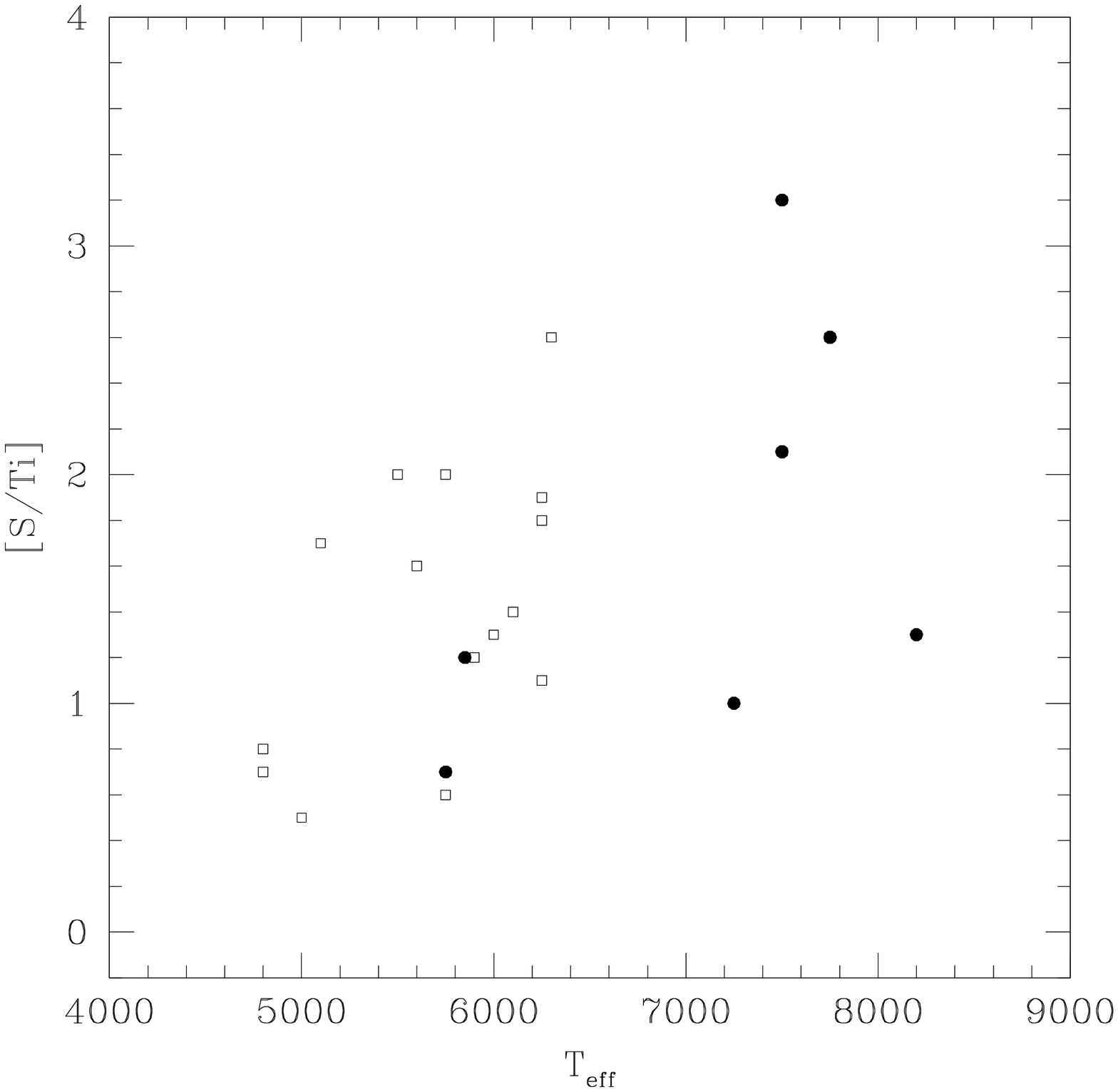}}
\end{center}
\end{figure}

For the unaffected RV\,Tauri stars of group C another reason for their immunity for the
 depletion process is suggested : their metallicity. The stars 
of group C have
a low initial metallicity. The critical metallicity, below which the depletion
process is not active, is estimated at [Fe/H]=$-1.0$ \citep{2000ApJ...531..521G}.
 How this low metallicity makes the depletion
 process inefficient is not yet clear.  In the case of a storage site around a 
binary, a low metallicity can lead to a too slow dust formation so that the
gas falls back on the star before the gas-dust separation has set in
\citep{2000ApJ...531..521G}. 
 We estimated for the objects listed in
Table~\ref{overdep}, the initial metallicity. For the RV\,Tauri stars
we have taken the values from \citet{2000ApJ...531..521G} . For the other
 objects in which depletion occurred, we estimated the initial Fe-metallicity
 on the
 basis of the Zn-abundance, if this abundance was adopted to be unaffected.
In the few cases the Zn-abundance did not reflect the initial composition,
we used the S-abundance with the relation [$\alpha$/Fe]$\sim 0$ for
$0<$ [Fe/H]$<-0.5$ from \citet{1985pdcn.conf..131N}. 
The initial metalicities of the affected objects, listed in Table~\ref{overdep}
range between $-0.8$ and +0.2.
The objects in our sample with the lowest initial metallicity are
 IRAS\,09060-2807 and the genuine RV\,Tauri star of group A IRAS\,09538-7622
 ($-0.7$ and $-0.6$). Both are 
unaffected by the depletion process. The reason of this could be the same 
as for the RV\,C RV\,Tauri
stars. However, their metallicity is higher than the critical metallicity 
[Fe/H]=$-1.0$.

 A simple model would predict a correlation between the IR excess and the
 abundance anomalies. 
This is not the case for the RV\,Tauri stars \citep{2000ApJ...531..521G}
 and neither for the strongly depleted
 post-AGB stars \citep{2002A&A...386..504M}. This is confirmed in our sample.
They all have a very large IR excess, but some are not affected by the
depletion process (like IRAS\,09060-2807), others rather severely 
(IRAS\,17038-4815 and IRAS\,17233-4330).

%The range of the depletion scale in the stars, listed in 
%table~\ref{overdep}, is wide. The severely depleted post-AGB stars
% didn't get their abundance anomalies at once so an evolution from 
%normal to the observed abundances has thus occurred.
%Progenitors of the severely depleted post-AGB stars should thus exist. An
% obvious choice are the RV\,Tauri stars. For post-AGB stars the 
%temperature is the first indication on how long the star is already
%on the post-AGB track and maybe for the time in which the depletion process
%could be operational. IRAS\,08544-4431, IRAS\,15469-5311 and 
%IRAS\,19125+0343, for which we show that they are binaries in the
%next chapter, are only moderately depleted. However, their effective
%temperature of respectively 7250, 7500 and 7750 K, is similar to those
%of the strongly depleted binaries HR\,4049, HD\,44179 and 
%BD+39$\degree$4926 (all have temperatures of about 7500 K). The circumstellar
%material of the strongly depleted post-AGB binaries is C-rich, which
%is not the case for the above mentioned binary programme stars. The chemical
%properties of the dust grains may ,thus, be also important for the gas-dust 
%separation.
%Also HD\,52961 illustrates with a [Fe/H]=$-4.8$ and a T$_{eff}$ of
% 6000 K that there is no simple relation of increase of the depletion
% scale with increasing temperature. 

\subsection{Concluding remarks}

In this paper we performed a detailed analysis of high quality spectra
of 12 stars with spectral
type F and G of our new sample of RV\,Tauri like objects. 
%These stars were selected on the basis of
%a broad IR-excess. We limited our analysis in this paper to the
%stars with spectral type F and G.
 Since nine of the twelve stars show depletion 
patterns, we confirm that depletion is a widespread phenomenon among post-AGB stars
with a near IR excess.
This depletion process takes place in the circumstellar environment.
Elements with a high condensation temperature get locked in dust grains,
while elements with a lower condensation temperature remain in the gas phase
 and are reaccreted. The radiation pressure keeps the dust from falling back on
the star.
We affirm that the efficiency of this process, together with the resulting
abundance pattern, is different from star to star but that there is no
correlation with the strength of the IR-excess \citep{2000ApJ...531..521G}.
However, the presence of a broad 
IR-excess was the criterion on which the stars of our new sample were selected
and in this paper we only analysed the stars with spectral type F and G.
For these stars the most obvious geometry for the surrounding dust is a
 circumbinary disc. The fact that a high fraction of $75 \%$ of the studied 
objects is depleted, indicates that the presence of a disc and the presence of 
depletion are linked. We find, however, that while a disc is a preferential site for the
gas-dust separation to take place, its presence is not a sufficient condition. 

In our following paper we will deal with the suspected binarity of the stars of our new
sample and the relation with the presence of a disc and depletion.

\begin{acknowledgements}
The authors want to acknowledge the Geneva Observatory and its staff for the generous
 time allocation on the Swiss Euler telescope.
TM acknowledges financial support from the Robert A Welch Foundation
 and from the Fund for Scientific Research of
 Flanders (FWO) under the grant G.0178.02. 
This paper is the publication of one
chapter of the PhD thesis ``A study of post-AGB stars
with a dusty disc'' by T. Maas, 2003, Het
Instituut voor Sterrenkunde, KULeuven. The full
PhD thesis is downloadable at
http://www.ster.kuleuven.ac.be/pub/maas\_phd/

\end{acknowledgements}

\aareferences

\begin{thebibliography}{60}
\expandafter\ifx\csname natexlab\endcsname\relax\def\natexlab#1{#1}\fi

\bibitem[{{Bi{\' e}mont} {et~al.}(1993){Bi{\' e}mont}, {Hibbert}, {Godefroid},
  \& {Vaeck}}]{biemont93}
{Bi{\' e}mont}, E., {Hibbert}, A., {Godefroid}, M., \& {Vaeck}, N. 1993, \apj,
  412, 431

\bibitem[{{Bi{\' e}mont} {et~al.}(1991){Bi{\' e}mont}, {Hibbert}, {Godefroid},
  {Vaeck}, \& {Fawcett}}]{biemont91}
{Bi{\' e}mont}, E., {Hibbert}, A., {Godefroid}, M., {Vaeck}, N., \& {Fawcett},
  B.~C. 1991, \apj, 375, 818

\bibitem[{{Biller} {et~al.}(2003){Biller}, {Close}, {Potter}, {Bieging},
  {Hoffman}, {Hinz}, \& {Oppenheimer}}]{biller2003}
{Biller}, B., {Close}, L., {Potter}, D., {et~al.} 2003, in Asymmetric Planetary
  Nebulae III, 0

\bibitem[{{Bond}(1991)}]{1991IAUS..145..341B}
{Bond}, H.~E. 1991, in IAU Symp. 145: Evolution of Stars: the Photospheric
  Abundance Connection, 341

\bibitem[{{Bond} {et~al.}(1997){Bond}, {Fulton}, {Schaefer}, {Ciardullo}, \&
  {Sipior}}]{1997IAUS..180..211B}
{Bond}, H.~E., {Fulton}, L.~K., {Schaefer}, K.~G., {Ciardullo}, R., \&
  {Sipior}, M. 1997, in IAU Symp. 180: Planetary Nebulae, 211

\bibitem[{{Bond} \& {Luck}(1987)}]{1987ApJ...312..203B}
{Bond}, H.~E. \& {Luck}, R.~E. 1987, \apj, 312, 203

\bibitem[{{Bujarrabal} {et~al.}(1988){Bujarrabal}, {Bachiller}, {Alcolea}, \&
  {Martin-Pintado}}]{1988A&A...206L..17B}
{Bujarrabal}, V., {Bachiller}, R., {Alcolea}, J., \& {Martin-Pintado}, J. 1988,
  \aap, 206, L17

\bibitem[{{Dominik} {et~al.}(2003){Dominik}, {Dullemond}, {Cami}, \& {van
  Winckel}}]{2003A&A...397..595D}
{Dominik}, C., {Dullemond}, C.~P., {Cami}, J., \& {van Winckel}, H. 2003, \aap,
  397, 595

\bibitem[{{Fernie}(1995)}]{1995AJ....110.3010F}
{Fernie}, J.~D. 1995, \aj, 110, 3010

\bibitem[{{Gehrz}(1972)}]{1972ApJ...178..715G}
{Gehrz}, R.~D. 1972, \apj, 178, 715

\bibitem[{{Giridhar} {et~al.}(1998){Giridhar}, {Lambert}, \&
  {Gonzalez}}]{1998ApJ...509..366G}
{Giridhar}, S., {Lambert}, D.~L., \& {Gonzalez}, G. 1998, \apj, 509, 366

\bibitem[{{Giridhar} {et~al.}(2000){Giridhar}, {Lambert}, \&
  {Gonzalez}}]{2000ApJ...531..521G}
---. 2000, \apj, 531, 521

\bibitem[{{Giridhar} {et~al.}(1990){Giridhar}, {Rao}, \&
  {Lambert}}]{1990Obs...110..120G}
{Giridhar}, S., {Rao}, N.~K., \& {Lambert}, D.~L. 1990, The Observatory, 110,
  120

\bibitem[{{Giridhar} {et~al.}(1992){Giridhar}, {Rao}, \&
  {Lambert}}]{1992JApA...13..307G}
---. 1992, Journal of Astrophysics and Astronomy, 13, 307

\bibitem[{{Giridhar} {et~al.}(1994){Giridhar}, {Rao}, \&
  {Lambert}}]{1994ApJ...437..476G}
---. 1994, \apj, 437, 476

\bibitem[{{Gonzalez} \& {Lambert}(1997)}]{1997AJ....114..341G}
{Gonzalez}, G. \& {Lambert}, D.~L. 1997, \aj, 114, 341

\bibitem[{{Gonzalez} {et~al.}(1997{\natexlab{a}}){Gonzalez}, {Lambert}, \&
  {Giridhar}}]{1997ApJ...481..452G}
{Gonzalez}, G., {Lambert}, D.~L., \& {Giridhar}, S. 1997{\natexlab{a}}, \apj,
  481, 452

\bibitem[{{Gonzalez} {et~al.}(1997{\natexlab{b}}){Gonzalez}, {Lambert}, \&
  {Giridhar}}]{1997ApJ...479..427G}
---. 1997{\natexlab{b}}, \apj, 479, 427

\bibitem[{{Gonzalez} \& {Wallerstein}(1996)}]{1996MNRAS.280..515G}
{Gonzalez}, G. \& {Wallerstein}, G. 1996, \mnras, 280, 515

\bibitem[{{Grevesse} \& {Sauval}(1998)}]{1998SSRv...85..161G}
{Grevesse}, N. \& {Sauval}, A.~J. 1998, Space Science Reviews, 85, 161

\bibitem[{{Herbig}(1970)}]{1970ApJ...162..557H}
{Herbig}, G.~H. 1970, \apj, 162, 557

\bibitem[{{Hibbert} {et~al.}(1991){Hibbert}, {Bi{\'e}mont}, {Godefroid}, \&
  {Vaeck}}]{1991A&AS...88..505H}
{Hibbert}, A., {Bi{\'e}mont}, E., {Godefroid}, M., \& {Vaeck}, N. 1991, \aaps,
  88, 505

\bibitem[{{Holweger}(2001)}]{2001sgc..conf...23H}
{Holweger}, H. 2001, in AIP Conf. Proc. 598: Joint SOHO/ACE workshop "Solar and
  Galactic Composition", ed. R.~F. {Wimmer-Schweingruber}, 23

\bibitem[{{Jura}(1986)}]{1986ApJ...309..732J}
{Jura}, M. 1986, \apj, 309, 732

\bibitem[{{Jura} {et~al.}(2000){Jura}, {Chen}, \&
  {Werner}}]{2000ApJ...541..264J}
{Jura}, M., {Chen}, C., \& {Werner}, M.~W. 2000, \apj, 541, 264

\bibitem[{{Kodaira}(1973)}]{1973A&A....22..273K}
{Kodaira}, K. 1973, \aap, 22, 273

\bibitem[{{Kupka} {et~al.}(1999){Kupka}, {Piskunov}, {Ryabchikova}, {Stempels},
  \& {Weiss}}]{1999A&AS..138..119K}
{Kupka}, F., {Piskunov}, N., {Ryabchikova}, T.~A., {Stempels}, H.~C., \&
  {Weiss}, W.~W. 1999, \aaps, 138, 119

\bibitem[{{Kurucz}(1993)}]{1993KurCD..13.....K}
{Kurucz}, R. 1993, ATLAS9 Stellar Atmosphere Programs and 2 km/s grid.~Kurucz
  CD-ROM No.~13.~ Cambridge, Mass.: Smithsonian Astrophysical Observatory,
  1993., 13

\bibitem[{{Lloyd Evans}(1985)}]{1985MNRAS.217..493E}
{Lloyd Evans}, T. 1985, \mnras, 217, 493

\bibitem[{{Lloyd Evans}(1997)}]{1997Ap&SS.251..239L}
---. 1997, \apss, 251, 239

\bibitem[{{Lloyd Evans}(1999)}]{1999IAUS..191..453E}
{Lloyd Evans}, T. 1999, in IAU Symp. 191: Asymptotic Giant Branch Stars, ed.
  T.~{Le Bertre}, A.~{L{\` e}bre}, \& C.~{Waelkens}, Vol. 191, 453

\bibitem[{{Lodders} \& {Fegley}(1988)}]{Lodders}
{Lodders}, K. \& {Fegley}, B. 1988, {The Planetary Scientist's Companion}
  (Oxford University Press), 83

\bibitem[{{Luck} \& {Bond}(1984)}]{1984ApJ...279..729L}
{Luck}, R.~E. \& {Bond}, H.~E. 1984, \apj, 279, 729

\bibitem[{{Luck} \& {Bond}(1989)}]{1989ApJ...342..476L}
---. 1989, \apj, 342, 476

\bibitem[{{Luck} {et~al.}(1990){Luck}, {Bond}, \&
  {Lambert}}]{1990ApJ...357..188L}
{Luck}, R.~E., {Bond}, H.~E., \& {Lambert}, D.~L. 1990, \apj, 357, 188

\bibitem[{{Maas} {et~al.}(2003){Maas}, {Van Winckel}, {Lloyd Evans}, {Nyman},
  {Kilkenny}, {Martinez}, {Marang}, \& {van Wyk}}]{irasnulacht}
{Maas}, T., {Van Winckel}, H., {Lloyd Evans}, T., {et~al.} 2003, \aap, 405,
  271:(Paper I)

\bibitem[{{Maas} {et~al.}(2002){Maas}, {Van Winckel}, \&
  {Waelkens}}]{2002A&A...386..504M}
{Maas}, T., {Van Winckel}, H., \& {Waelkens}, C. 2002, \aap, 386, 504

\bibitem[{{Mathis} \& {Lamers}(1992)}]{1992A&A...259L..39M}
{Mathis}, J.~S. \& {Lamers}, H.~J.~G.~L.~M. 1992, \aap, 259, L39

\bibitem[{{Menzies} \& {Whitelock}(1988)}]{1988MNRAS.233..697M}
{Menzies}, J.~W. \& {Whitelock}, P.~A. 1988, \mnras, 233, 697

\bibitem[{{Molster} {et~al.}(1999){Molster}, {Waters}, {Trams}, {Van Winckel},
  {Decin}, {van Loon}, {J{\" a}ger}, {Henning}, {K{\" a}ufl}, {de Koter}, \&
  {Bouwman}}]{1999A&A...350..163M}
{Molster}, F.~J., {Waters}, L.~B.~F.~M., {Trams}, N.~R., {et~al.} 1999, \aap,
  350, 163

\bibitem[{{Nissen} {et~al.}(1985){Nissen}, {Edvardsson}, \&
  {Gustafsson}}]{1985pdcn.conf..131N}
{Nissen}, P.~E., {Edvardsson}, B., \& {Gustafsson}, B. 1985, in Production and
  distribution of C,N,O elements, ESO Conference and Workshop Proceedings, held
  in Garching/Munich, May 13-15, 1985, Garching: European Southern Observatory
  (ESO), 1985, edited by I.J. Danziger, F. Mateucci, and K. Kj{\" a}r., p.131,
  131

\bibitem[{{Preston} {et~al.}(1963){Preston}, {Krzeminski}, {Smak}, \&
  {Williams}}]{1963ApJ...137..401P}
{Preston}, G.~W., {Krzeminski}, W., {Smak}, J., \& {Williams}, J.~A. 1963,
  \apj, 137, 401

\bibitem[{{Rao} {et~al.}(2002){Rao}, {Goswami}, \&
  {Lambert}}]{2002MNRAS.334..129K}
{Rao}, N.~K., {Goswami}, A., \& {Lambert}, D.~L. 2002, \mnras, 334, 129

\bibitem[{{Raveendran}(1989)}]{1989MNRAS.238..945R}
{Raveendran}, A.~V. 1989, \mnras, 238, 945

\bibitem[{{Roddier} {et~al.}(1995){Roddier}, {Roddier}, {Graves}, \&
  {Northcott}}]{1995ApJ...443..249R}
{Roddier}, F., {Roddier}, C., {Graves}, J.~E., \& {Northcott}, M.~J. 1995,
  \apj, 443, 249

\bibitem[{{Sneden} {et~al.}(1991){Sneden}, {Gratton}, \&
  {Crocker}}]{1991A&A...246..354S}
{Sneden}, C., {Gratton}, R.~G., \& {Crocker}, D.~A. 1991, \aap, 246, 354

\bibitem[{{Stephenson}(1978)}]{1978IBVS.1453....1S}
{Stephenson}, C.~B. 1978, Information Bulletin on Variable Stars, 1453, 1

\bibitem[{{Takeda} {et~al.}(2002){Takeda}, {Parthasarathy}, {Aoki}, {Ita},
  {Nakada}, {Izumiura}, {Noguchi}, {Takada-Hidai}, {Sato}, {Tajitsu}, {Honda},
  {Kawanomoto}, {Ando}, \& {Karoji}}]{2002PASJ...54..765T}
{Takeda}, Y., {Parthasarathy}, M., {Aoki}, W., {et~al.} 2002, \pasj, 54, 765

\bibitem[{{Van Winckel}(1995)}]{hansthesis}
{Van Winckel}, H. 1995, {The chemical composition of optically bright post-AGB
  stars} (PhD thesis)

\bibitem[{{Van Winckel} {et~al.}(1992){Van Winckel}, {Mathis}, \&
  {Waelkens}}]{1992Natur.356..500V}
{Van Winckel}, H., {Mathis}, J.~S., \& {Waelkens}, C. 1992, \nat, 356, 500

\bibitem[{{Van Winckel} \& {Reyniers}(2000)}]{2000A&A...354..135V}
{Van Winckel}, H. \& {Reyniers}, M. 2000, \aap, 354, 135

\bibitem[{{Van Winckel} {et~al.}(1995){Van Winckel}, {Waelkens}, \&
  {Waters}}]{1995A&A...293L..25V}
{Van Winckel}, H., {Waelkens}, C., \& {Waters}, L.~B.~F.~M. 1995, \aap, 293,
  L25

\bibitem[{{Van Winckel} {et~al.}(1998){Van Winckel}, {Waelkens}, {Waters},
  {Molster}, {Udry}, \& {Bakker}}]{1998A&A...336L..17V}
{Van Winckel}, H., {Waelkens}, C., {Waters}, L.~B.~F.~M., {et~al.} 1998, \aap,
  336, L17

\bibitem[{{Venn} \& {Lambert}(1990)}]{1990ApJ...363..234V}
{Venn}, K.~A. \& {Lambert}, D.~L. 1990, \apj, 363, 234

\bibitem[{{Waelkens} {et~al.}(1991{\natexlab{a}}){Waelkens}, {Lamers},
  {Waters}, {Rufener}, {Trams}, {Le Bertre}, {Ferlet}, \&
  {Vidal-Madjar}}]{1991A&A...242..433W}
{Waelkens}, C., {Lamers}, H.~J.~G.~L.~M., {Waters}, L.~B.~F.~M., {et~al.}
  1991{\natexlab{a}}, \aap, 242, 433

\bibitem[{{Waelkens} {et~al.}(1991{\natexlab{b}}){Waelkens}, {Van Winckel},
  {Bogaert}, \& {Trams}}]{1991A&A...251..495W}
{Waelkens}, C., {Van Winckel}, H., {Bogaert}, E., \& {Trams}, N.~R.
  1991{\natexlab{b}}, \aap, 251, 495

\bibitem[{{Waelkens} {et~al.}(1996){Waelkens}, {Van Winckel}, {Waters}, \&
  {Bakker}}]{1996A&A...314L..17W}
{Waelkens}, C., {Van Winckel}, H., {Waters}, L.~B.~F.~M., \& {Bakker}, E.~J.
  1996, \aap, 314, L17

\bibitem[{{Waters} {et~al.}(1992){Waters}, {Trams}, \&
  {Waelkens}}]{1992A&A...262L..37W}
{Waters}, L.~B.~F.~M., {Trams}, N.~R., \& {Waelkens}, C. 1992, \aap, 262, L37

\bibitem[{{Waters} {et~al.}(1993){Waters}, {Waelkens}, {Mayor}, \&
  {Trams}}]{1993A&A...269..242W}
{Waters}, L.~B.~F.~M., {Waelkens}, C., {Mayor}, M., \& {Trams}, N.~R. 1993,
  \aap, 269, 242

\bibitem[{{Wheeler} {et~al.}(1989){Wheeler}, {Sneden}, \&
  {Truran}}]{1989ARA&A..27..279W}
{Wheeler}, J.~C., {Sneden}, C., \& {Truran}, J.~W. 1989, \araa, 27, 279

\end{thebibliography}
\end{document}